\documentclass[10pt-default,11pt]{article}
\usepackage{amsmath}
\usepackage{amsfonts}
\usepackage{amssymb}
\usepackage{geometry}

\input{tcilatex}
\begin{document}

\bigskip \medskip

{\Large Huygens triviality of the time-independent Schr\"{o}dinger equation.}

{\Large Applications to atomic and high energy physics}

{\Large \ \ \ \ \ \ \ \ \ \ \ \ }

\ \ \ 

\textbf{Arkady L.Kholodenko}$^{1}$\textbf{\medskip , Louis H. Kauffman}$^{2}$

$^{1}$\textsl{375 H.L.Hunter Laboratories, Clemson University, Clemson,SC
29634-0973,USA}

$^{2}$\textsl{Department of Mathematics, Statistics and Computer
Science,University of }

\ \textsl{Illinois at Chicago, 851 South Morgan Street, Chicago, IL,
60607-7045}

\bigskip

\bigskip

\ \ \ \textbf{Abstract}

Huygens triviality-a concept invented by Jacques Hadamard-describes an
equivalence class connecting those 2nd order partial differential equations
which are transformable into the wave equation. In this work it is
demonstrated, that the Schr\"{o}dinger equation with the time-independent
Hamiltonian belongs to such an equivalence class. The wave equation is the
equation for which Huygens' principle (HP) holds. The HP was a subject of
confusion in both physics and mathematics literature for a long time. Not
surprisingly, the role of this principle was obscured from the beginnings of
quantum mechanics causing some theoretical and experimental
misunderstandings. The purpose of this work is to bring the full clarity
into this topic. By doing so, we obtained a large amount of new results
related to uses of Lie sphere geometry, of twistors, of Dupin cyclides, of
null electromagnetic fields, of AdS-CFT correspondence, of Penrose limits,
of geometric algebra, etc. in physical problems ranging from the atomic to
high energy physics and cosmology.

\bigskip

\bigskip

\bigskip


\ \ Time-independent Schr\"{o}dinger equation

\ \ Lie sphere geometry

\ \ Dupin cyclides

\ \ \ 

\bigskip

\ \textbf{\ 1. \ Motivation and background}

\bigskip

It is well documented $[1]$ that Schr\"{o}dinger's equation in its known
form emerged only in the 4th installment of Schr\"{o}dinger's papers on
quantum mechanics -all published in 1926.The Huygens principle was
introduced in the 2nd installment. In the 1st installment the
Hamilton-Jacobi (H-J) equation \ was used as point of departure. This
equation was then used as an input in his variational derivation of the
stationary Schr\"{o}dinger's equation. Although the variational way of
obtaining this equation was subsequently endorsed by Courant and Hilbert
[2], neither Schr\"{o}dinger himself nor the rest of physics community were
using this (variational) way for obtaining the stationary Schr\"{o}dinger's
equation. Instead, in the 2nd installment, being guided by \ results of De
Broglie (his PhD thesis was completed in 1924), Schr\"{o}dinger presented
the following arguments. "Hamilton's variational principle can be shown to
correspond to the Fermat principle for wave propagation in configuration
(q)-space, and the H-J equation expresses Huygens' principle for this wave
propagation." In the same paper he also writes "The H-J equation corresponds
to Huygens' principle (in its old simple form, not in the form due to
Kirchhoff)." Kirchhoff's way of dealing with Huygens's principle is
discussed, for example, in \ the book by Baker and Copson [3]. Thus, for Schr%
\"{o}dinger the Huygens principle is synonymous with the H-J equation. Later
on, in 1948, Feynman [4] made the following \ comment about Huygens'
principle. In section 7, page 377, of [4] we read that the equation 
\begin{equation}
\psi (x_{k+1},t+\epsilon )=\dint \exp [\frac{i}{\hbar }S(x_{k+1},x_{k})]\psi
(x_{k},t)dx_{k}/A  \tag{1.1}
\end{equation}%
"is easily interpreted physically as the expression of Huygens' principle."
Here the index $k=0,1,2,...,$ represents time ticks, $\psi (x_{k},t)$ is the
wave function, $A$ and $\hbar $ are known constants, $S(x_{k+1},x_{k})$ is
the classical action between space-time points $x_{k+1}$ and $x_{k},\epsilon
\rightarrow 0^{+}$. Further down he writes:\ " Actually Huygens' principle
is not correct in optics. It is replaced by Kirchhoff's modification which
requires that both the amplitude and its derivative must be known on the
adjacent surface. This is a consequence of the fact that the wave equation
in optics is second order in the time. The wave equation of quantum
mechanics is first order in time; therefore, Huygens' principle\textsl{\ is}
correct for matter waves, action replacing time." \ From these quotations
the question emerges: \ \textsl{How the Huygens principle (HP) by Schr\"{o}%
dinger is related to that by Feynman}? \ The first attempt to provide \
mathematically satisfactory answer to this question was made by Gutzwiller
[5]. His work is incomplete though as he acknowledges himself. Thus, this
paper (and those which will follow in the sequel) is aimed at providing
missing details. By doing so, it will become obvious why this topic is still
of such profound importance. To demonstrate the importance, it is sufficient
to recall some facts from Feynman's lectures on physics [6] as well as from
his book on path integrals [7]. They both begin with the discussion of the
two-slit experiment. For the light this experiment was set up originally by
Young in 1801 [8]. Its explanation involves uses of HP [4]. For electrons it
was performed initially in 1961[9]. \ After this, use of other heavier
particles, including C60 fullerens, \ showed the same pattern as that
observed by Young for light [10,11]. In view of these experimental results,
it is appropriate to make some comments on Gutzwiller's paper [5]. On page
54 we find the following statement: "The wave equation%
\begin{equation}
\left( \frac{\partial ^{2}}{\partial t^{2}}\frac{1}{c^{2}}-\frac{\partial
^{2}}{\partial x^{2}}-\frac{\partial ^{2}}{\partial y^{2}}-\frac{\partial
^{2}}{\partial z^{2}}\right) \Psi =0  \tag{1.2}
\end{equation}%
implies Huygens's principle which is the relativistic version of Feynman's
path integral\footnote{%
Perhaps Gutzviller had in mind that iteration of Eq.(1.1) (which is the
epitome of Huygens' principle for Feynman) is leading to the nonrelativistic
path integral satisfying \ standard Schr\"{o}dinger's equation. This result,
apparently, can be generalized to the relativistic case where Huygens'
principle in its conventional form holds. Such path integral is expected to
satisfy the wave Eq.(1.2). This program was left unfulfilled by Gutzwiller
[5].}, valid in its usual form for Schr\"{o}dinger's equation%
\begin{equation}
i\hbar \frac{\partial \varphi }{\partial t}=-\frac{\hbar ^{2}}{2m}\left( 
\frac{\partial ^{2}}{\partial x^{2}}+\frac{\partial ^{2}}{\partial y^{2}}+%
\frac{\partial ^{2}}{\partial z^{2}}\right) \varphi +V\varphi =0,  \tag{1.3}
\end{equation}%
but \textsl{neither Huygens' principle nor the path -integral applies to the
sationary waves} which are the solutions of Eq.s (1.4) and (1.5)". \ In
numeration of this paper these equations are respectively given by 
\begin{equation}
\left( \frac{\partial ^{2}}{\partial x^{2}}+\frac{\partial ^{2}}{\partial
y^{2}}+\frac{\partial ^{2}}{\partial z^{2}}\right) \psi +\frac{8\pi ^{2}m}{%
h^{2}}(E-V)\psi =0  \tag{1.4}
\end{equation}%
and 
\begin{equation}
\left( \nabla ^{2}+k^{2}\right) \Psi =0.  \tag{1.5}
\end{equation}%
By making such a claim Gutzwiller contradicts Schr\"{o}dinger who clearly
stated that the H-J equation is mathematically restated Huygens' principle.
Eq.(1.4) is directly obtainable from the H-J equation as shown in Schr\"{o}%
dinger's 1st installment on quantum mechanics [1]. Below, in this paper we
shall prove that Schr\"{o}dinger's definition of Huygens' principle is
indeed correct in a rigorous mathematical sense. \ If this is so, then if
electrons and photons\footnote{%
And other heavier particles} produce the same interference patterns in the
double slit experiment, why then formalism developed in optics [8] cannot be
applied to electrons and heavier particles? What makes use of Born's
probabilistic \ interpretation of the wave function in quantum mechanics
superior to that used in optics for description of the two- slit experiment
? \ This issue was carefully investigated by David Bohm in his classical
monograph on quantum mechanics [12], pages 97-98. He found that differences
do exist but just in few places. The updated comparison was recently made by
Sanz and Miret-Artes in their book [13], chapters 4 and 7. From these
chapters it follows that all objections made by Bohm \ in the remaining few
places can be removed. The results presented \ in [13] along with references
on which these results are based are incomplete \ to a some extent. In [14]
we eliminated this deficiency so that it should be read alongside with
reading of this manuscript. Since nowadays the sophisticated quantum
mechanical experiments are mainly done optically [15,16], ref. [14] supplies
helpful additional guidelines for understanding of these experiments. In
view of this, it makes sense to claim that our understanding of all
subtleties of quantum mechanics is contingent upon our understanding of
optics where Huygens' principle (HP)\ is playing a very prominent role. \
Although the essence of HP is summarized in Definition 3.1. \ in section 3,
details are essential. They are presented in sections 4,5 and Appendix B. In
addition, the HP is linked with the conformal invariance of the
Huygens-trivial equivalence class of equations. The notion of Huygens'
triviality was formulated by Hadamard \ (details are presented in sections 4
and 5 and in Appendix B). The equivalence class is made of all 2nd order
partial differential equations (PDE's) which can be transformed in a
prescribed way (described in the text) into standard wave equation obeying
the HP. This equation is conformally invariant. Its conformal group is $%
SO(4,2).$It is the largest symmetry group leaving invariant free Maxwell's
equations or, which is equivalent, the massless Klein-Gordon (or wave)
equation [$14$]. $SO(4,2)$ is the conformal group of Minkowski spacetime [$%
21 $]. It contains\ 4 translations, 6 rotations, 1 dilatation and 4
inversions. Hadamard conjectured that Huygens' -trivial equivalence class
which he defined is the only one possible. Results of Appendix B demonstrate
that this is not the case. Subsequent studies revealed that there are
conformally invariant PDE operators which cannot be transformed into
D'Alembertian using the transformation rules set up by Hadamard. But they do
respect the HP nevertheless! Unexpectedly, their form is determined by the
nature of cosmic gravitational plane-wave background. \ This is explained in
the Appendix B.

Our work is made of 7 sections and 4 appendices. The role of appendices is
not just auxiliary. Each of them serves as a nucleus for some further work.
Therefore, they cannot be omitted upon first reading. In section 2,
following the original but not widely known ideas by Schr\"{o}dinger, we \
discuss the Schr\"{o}dinger-style derivation of the uncertainty relations
based on relativistic arguments. These arguments are such that they allow to
restore the stationary Schr\"{o}dinger equation. In section 3 we take into
account that inclusion of time-dependence, that is replacement of the
hyperbolic (wave) equation from which the stationary Shr\"{o}dinger equation
was derived, by the parabolic (truly Schr\"{o}dinger) equation had occurred
only in the 4th installment of Schr\"{o}dinger's papers on quantum mechanics
[1]. Such an inclusion was associated with a lot of difficulties for Schr%
\"{o}dinger. The biggest of this difficulties was the apparent departure
from ideas by De Broglie on the physical nature of waves of matter. Clearly,
the relativistic considerations leading to \ Schr\"{o}dinger's formulation
of the uncertainty principle also need to be sacrificed (e.g. read \ the
additional comments in the Appendix A). \ As result, the role of HP in the
formalism of quantum mechanics had become obscured resulting in some
erroneous statements. Subsequent theoretical and experimental works cited in
section 3 contain erroneous claims requiring modification of the
superposition principle of quantum mechanics-one of the pillars (e.g. the
double slit experiment) of quantum mechanics. These circumstances caused us
to write a very detailed section 4 describing the mathematical aspects of
the HP. Additional information is contained in the Appendix B. In section 4
and Appendix C we find new solutions of the Schr\"{o}dinger equation-Dupin
cyclides. These are having both micro and macro (cosmological) importance.
The micro importance ultimately originates from the seminal work of Madelung
on hydrodynamical formulation of quantum mechanics which, to our knowledge,
was never reproduced in its entirety in English. The cosmological
significance of these solutions originates from the remarkable work \ by
Roger Penrose discussed in the Appendix B and also briefly in the Appendix
D. Thanks to work by Ward (e.g. read Appendix B), induced by work of
Penrose, instead of cosmic microwave background we may think about the
cosmic gravitational plane-wave background. It is this background which is
the real cause for the Schr\"{o}dinger equation to exist in its known form.
Indeed, Dupin cyclides emerge from analysis of Madelung equations of quantum
mechanics, on one hand and, of Huygens- nontrivial (as compared to
Huygens-trivial) conformally invariant class of \ PDE's\ which should exist
in the gravitational plane-wave background, \ on another. Use of \ conformal
transformations (not to be confused with transformations defined by Hadamard
discussed in section 4) relates Hugens-nontrivial and trivial equations to
each other.

To strengthen \ these conclusions, we also developed alternative paths for
reaching the same goals. These are coming from the detailed study of
group-theoretical properties of the Schr\"{o}dinger equation initiated some
time ago by Niederer [$17$]. By analogy with Newton's mechanics, where
Newton's second law is invariant with respect to Galileo-type
transformations, one would expect the same for the non relativistic Schr\"{o}%
dinger's equation. This happens not to be the case, however. This fact
somewhat complicates the recovery of Newton's equations from quantum
mechanics via Ehrenfest theorem [$18$]. Studies by many authors had firmly
established \ that the group of symmetries of, say, hydrogen atom is $%
SO(4,2) $ [$19$]. This happens to be symmetry of all atoms of periodic
system of elements as well as of molecules (at least diatomic) [$20,21$].
But this symmetry group is the largest symmetry group which leaves invariant
free Maxwell's equations or, which is equivalent, the massless Klein-Gordon
equation [$14$].The de Sitter and anti-de Sitter groups $SO(4,1)$ and $%
SO(3,2)$ are subgroups of $SO(4,2).$ These subgroups are not isometry groups
of the Minkowski spacetime though. Details about these groups and their Lie
algebras are summarized in [$21$]. These \ groups \ are also discussed in
section 7 and Appendix D. It is fundamentally important that static
Einsteinian spcetimes are conformally Minkowskian. Therefore they are
invariant under the action of the conformal group $SO(4,2)$ [$22$]. \ The
group $SO(4,2)$ is a typical representative of the so called dynamical
symmetry groups and their spectrum generated algebras [$19,23$].The
relationship between geometry of the underlying spacetimes and quantum
mechanics was noticed and developed to a some extent already by Schr\"{o}%
dinger. In his book [$24$] he made an attempt to extend his apparatus of
quantum mechanics for flat spacetimes to curved spacetimes. He was
interested in finding those spacetimes which permit quantum mechanics to
exist. Since the group $SO(4,2)$ supports electromagnetic waves, that is
dynamics of massless photons, the same should be true for all massless
particles, e.g. neutrino, graviton, gluon, etc. Penrose's twistor theory was
developed initially for description of dynamics of such types of particles [$%
25$]. However, it was not used in the atomic and molecular physics because
atoms and molecules are massive. This obstacle \ is possible to by pass as
explained in section 6.\ Thus, in view of this, \ in this work we initiate
use of twistors in the atomic physics.\ 

The Huygens principle is linked with the conformal invariance. Details are
provided in section 4 and the Appendix B. The conformal invariance is very
nontrivially associated with the Lie sphere geometry. Details on this
geometry are given in section 7. By design, this geometry transforms circles
into circles and \ spheres into spheres. Since points are spheres of zero
radius and hyperplanes are (hyper)spheres of infinite radius the Dupin
cyclides are covering \ all surfaces made of spheres whose centres are
moving along some prescribed \ curve \textbf{c}(t) parametrized by time $t$
so that the radii of spheres $\mathbf{r}(t)$ also change with time. \ By
design, thus made surfaces are invariants of the Lie sphere geometry. The
simplest possible examples are: spheres, planes, cones and cylinders. Use of
M\"{o}bius transformations (this is the subgroup of the Lie sphere
transformations group) transforms these simple objects into the whole
variety of Dupin cyclides which stay invariant under the Lie sphere
transformations. In the same section 4 we discovered an unusual property of
the stationary Schr\"{o}dinger equation which makes this equation to be
treatable gauge-theoretically by employing \ deep mathematical results of
Andreas Floer. This result came as a by product of our efforts to adopt the
progressive wave solution method by Friedlander\ [70] to the stationary Schr%
\"{o}dinger equation. His method was applied initially to the conformally
invariant D'Alembert wave equation resulting in Dupin cyclides \ as
solutions. In section 4 using \ mathematically formulated \ Huygens'
principle \ we demonstrated that the Dupin cyclides are also valid solutions
of the stationary Schr\"{o}dinger equation. The same result was reobtained
in the Appendix C by different methods. In doing so we were motivated by the
results obtained in sections 5 and 6. Section 5 \ begins with removing the
mass parameter from the relativistic Klein-Gordon (K-G) equation using
Hadamard -type transformations. These are making the K-G equation
Huygens-trivial. Next, we notice that every spinor component originating
from the Dirac equation is obeying the K-G equation. This means that the
massive Dirac equation \ can also be made Huygens-trivial. Next, in this and
in section 6, we discuss all relativistically invariant massive equations
with integer or half integer spins and conclude that they are also
Huygens-trivial. We continue our study of Huygens triviality in section 6
where we demonstrate Huygens triviality of the stationary Schr\"{o}dinger
equation being influenced by the seminal paper by Vladimir Fock published in
1935 [101],[102]. His goal was not to demonstrate Huygens triviality of the
Schr\"{o}dinger equation (say, for the hydrogen atom). Nevertheless, he came
very close to this task. Huygens triviality of the stationary Schr\"{o}%
dinger equation enabled us to use twistor methods for solving hydrogen atom
problem in section 6 while in section 7 \ following ideas of Sophus Lee and
Felix Klein we demonstrate the isomorphism between the twistor methods and
that of the Lie sphere geometry. In the same section we discuss some
physical applications such as the interrelation between the Dupin cyclides
and torus knots. Such unexpected utility of the Lie sphere geometry we
pushed further. In section 7 and the Appendix D, we demonstrated its major
role in establishing the AdS-CFT correspondence and in establishing the
Penrose limits.

\bigskip

\textbf{2. \ Schr\"{o}dinger-style derivation of uncertainty relations
leading to }

\ \ \ \ \ \ \textbf{the Schr\"{o}dinger equation\medskip \medskip }

\bigskip 2.1. \ \textsl{Shr\"{o}dinger-style derivation of uncertainty
relations. Role of relativistic arguments\medskip \medskip }

According to Feynman [6], the formalism of quantum mechanics is lying on two
pillars: a) Heisenberg's uncertainty principle and b) the two-slit
experiments with photons, electrons, etc. Being guided by these ideas, we
found it very illuminating to develop these ideas from scratch based on some
considerably lesser known writings by Schr\"{o}dinger. In an obscure
publication [26$]$ he sketched a derivation of the uncertainty relations not
found in any other textbooks on quantum mechanics. We begin, however, with
equally interesting Schr\"{o}dinger's remarks on page 50 of the same
reference. " Now the special technique by which classical mechanics dodges
the awkward fact of indeterminateness (the fact that equal initial
conditions are followed by different consequences) consists in including
initial velocity within \ the initial conditions....the initial velocity is
taken as forming part$\footnote{%
In mechanics (our observation)}$ of the initial condition at any given
moment. Velocity, after all, is defined as a differential quotient with
respect to time%
\begin{equation}
\frac{dx}{dt}=\lim_{\Delta t\rightarrow 0}\frac{x_{2}-x_{1}}{\Delta t}%
;\Delta t=t_{2}-t_{1}.  \tag{2.1}
\end{equation}%
This definition refers to two moments of time and not to the state at one
moment... \ It may be \ that the mathematical apparatus devised by Newton is
inadequately adapted to nature; and the modern claim that the concept of
velocity becomes meaningless for a precisely defined position in space
points strongly in that direction." These arguments \ by Schr\"{o}dinger
were taken into consideration seriously only recently, e.g. in 2002 in the
book "Quantum Calculus" $[27]$. Obviously, his arguments serve as precursors
for arguments leading to the uncertainty relations. Surprisingly, no such
relations can be found in "Quantum Calculus" [$27$] while in the book by Schr%
\"{o}dinger [$26$], on page 126, we find the following comments : "
...according to the fundamental equation of the Quantum theory:\footnote{%
Here $E$ is energy, $h$ is Planck's constant $\hbar =2\pi \hbar $ and $\nu $
is frequency.} 
\begin{equation}
E=h\nu .  \tag{2.2}
\end{equation}%
To measure frequency we need a certain time. Let us think of the primitive
procedure of counting $n$ vibrations within a definite time $\Delta t.$
Then, 
\begin{equation}
\nu =\frac{n}{\Delta t}  \tag{2.3}
\end{equation}%
but manifestly with a possible error of $\Delta \nu =\frac{1}{\Delta t}$
because the process of counting necessarily results in giving the whole
number, which is subject to an error of $\pm 1/2.$This entails a possible
error with respect to energy of $\Delta E=\frac{h}{\Delta t};$ hence $\Delta
t\cdot \Delta E=h...$ Now, \textsl{relativistically\footnote{%
Emphasis is ours}}, the energy is the fourth component of the
energy-momentum vector....Therefore the uncertainty relation can be
transferred to the other components as well, for example:%
\begin{equation}
\Delta x\cdot \Delta p_{x}=h,  \tag{2.4}
\end{equation}%
where $p_{x}$ is the momentum in the $x$-direction." The same result Schr%
\"{o}dinger obtains differently on page 129. \ There, he writes:

"We need but replace the particle by a wave-group and let the wave -length $%
\lambda $ and the momentum $p$ have a relation\footnote{%
This is clear from the dispersion relation for the light: $\omega =ck$ or $%
\nu =c/\lambda .$ Here $c$ is speed of light. Multiplying both sides of this
dispersion relation by $h$ we obtain: $E=cp,$ where $p$ is the same as in
Eq.(2.5). Multiplying both sides of Eq.(2.5) by $c$ and taking into account
that $\nu =c/\lambda ,$ we obtain: $cp=h\nu =E.$ Therefore, it follows that
the De Broglie relation is\ also valid for light, that is for the massless
particle.} 
\begin{equation}
p=\frac{h}{\lambda }.  \tag{2.5}
\end{equation}%
In order to build such a group a certain $\lambda -$interval is required.
Let $\Delta x$ be the length of the group, then...the ratio must be allowed
to wary by one unit...Thus: 
\begin{equation*}
\Delta x\cdot \Delta (\frac{1}{\lambda })=1.
\end{equation*}%
Multiply by $h$, then $h\cdot \Delta \left( \frac{1}{\lambda }\right) $ is
the uncertainty $\Delta p.$ And so, $\Delta x\cdot \Delta p=h."$

From these extensive quotations from writings by Schr\"{o}dinger it follows
that:

a) the relationship $\Delta t\cdot\Delta E=h$ implies the relationship $%
\Delta x\cdot\Delta p_{x}=h;$

b) the relationship $\Delta t\cdot \Delta E=h$ is obtained with the
assumption that the

\ \ \ \ motion is periodic;

c) the De Broglie wavy relation (2.5) is consistent with the relation (2.4)
obtained

\ \ \ with the help of relativistic arguments.

\bigskip

It happens, that the stationary Schr\"{o}dinger's equation can be obtained
based on these three observations only. This can be seen from reading of Schr%
\"{o}dinger's Physical Review paper.[28$]$ In it, Schr\"{o}dinger
acknowledges that his theory came as result of elaboration on works by De\
Broglie. \ It is helpful to reproduce Schr\"{o}dinger's arguments in order
to emphasize some essential \ elements which were overlooked subsequently in
physics literature. Specifically, the connection between \ the
nonrelativistic Schr\"{o}dinger equation \ and the relativistic mechanics
sketched by Schr\"{o}dinger has not found its \ well deserved place in
physics literature to our knowledge.\bigskip

\bigskip 2.2. \textsl{From uncertainty relations to stationary Schr\"{o}%
dinger's equation }

\bigskip

Relations a) and c) of previous subsection are consistent with the
relativistically invariant scalar wave equation 
\begin{equation}
\square _{4}\text{ }\phi (x,y,z;t)=0,  \tag{2.6a}
\end{equation}%
where the D'Alembertian $\square _{4}$ is given by 
\begin{equation}
\square _{4}=\frac{\partial ^{2}}{\partial t^{2}}\frac{1}{c^{2}}-\frac{%
\partial ^{2}}{\partial x^{2}}-\frac{\partial ^{2}}{\partial y^{2}}-\frac{%
\partial ^{2}}{\partial z^{2}}\text{ }.  \tag{2.6b}
\end{equation}%
Here, as before, $c$ is the speed of light in the \ vacuum. The simplest
solution of (2.6a) is given by the plane wave $\phi (x,y,z;t)=A\exp
\{i(\omega t\pm $ $\mathbf{k}\cdot \mathbf{x)\},x=(}x,y,z\mathbf{),}$ $%
A=const$ \ [$29$]. When it is substituted into Eq.(2.6a), it leads to the
dispersion relation 
\begin{equation}
\omega ^{2}-\mathbf{k}^{2}c^{2}=0  \tag{2.7}
\end{equation}%
implying $\omega =\pm \left\vert \mathbf{k}\right\vert c$. \ However, $%
E=\hbar \omega $ and $p=\hbar k.$ Therefore, we obtain: $E=\pm \left\vert 
\mathbf{p}\right\vert c.$ This result was mentioned in the previous
subsection. Let furthermore $E^{\prime }=E+\Delta E$ and $\left\vert \mathbf{%
p}\right\vert ^{\prime }=\left\vert \mathbf{p}\right\vert +\left\vert \Delta 
\mathbf{p}\right\vert ,$ then $\Delta E=\left\vert \Delta \mathbf{p}%
\right\vert c.$ Since $\Delta E\simeq \frac{h}{\Delta t},$ we obtain$:$ $%
\frac{h}{\Delta t}=\left\vert \Delta \mathbf{p}\right\vert c,$ or $%
\left\vert \Delta \mathbf{p}\right\vert c\Delta t=\left\vert \Delta \mathbf{p%
}\right\vert \left\vert \Delta \mathbf{x}\right\vert \simeq h,$ in accord
with Schr\"{o}dinger [$26]$. \ The question arises: \ Are there solutions of
(2.6a) other that plane wave(s) or their linear combinations$\mathbf{?}$
This issue was studied in detail in $[29].$ Evidently, the plane waves are
valid solutions in the vacuum. But there are other types, e.g. progressing
waves, to be discussed below in section 4.2., etc. If the waves are
propagating in some (innhomogeneous) medium with refractive index $n(x,y,x)$%
, \ the results from optics require us to replace $c$ by $c/n$. In such a
case the simple plane wave solution is no longer suitable and should be
replaced by $\phi (x,y,z;t)=A\exp \{i(\frac{2\pi Et}{h})\}\psi (x,y,z)%
\footnote{%
More on this, please, read in Appendix C.}$. Clearly, $\frac{2\pi E}{h}=%
\frac{2\pi \hbar \omega }{h}=\omega ,$ $\hbar \omega =h\nu .$ Substitution
of this ansatz into Eq.(2.6a) leads to the equation 
\begin{equation}
\left( \frac{\omega ^{2}n^{2}}{c^{2}}+\frac{\partial ^{2}}{\partial x^{2}}+%
\frac{\partial ^{2}}{\partial y^{2}}+\frac{\partial ^{2}}{\partial z^{2}}%
\right) \psi (x,y,z)=0.  \tag{2.8}
\end{equation}%
However, $2\pi \nu =\omega .$ Therefore, since (in view of Eq.(2.5)) $\nu
^{2}=\frac{1}{\lambda ^{2}}\left( \frac{c}{n}\right) ^{2}$, we obtain : $\ 
\frac{\omega ^{2}n^{2}}{c^{2}}=\frac{4\pi ^{2}}{\lambda ^{2}}$\textit{.}
Using the De Broglie relation $p=\frac{h}{\lambda }$ and keeping in mind
that the total energy $E$ of the dynamical system made of a particle moving
in the potential $V$ is

\begin{equation}
E=\frac{\mathbf{p}^{2}}{2m}+V(x,y,z),  \tag{2.9a}
\end{equation}%
we obtain:%
\begin{equation}
p=\sqrt{2m(E-V)}.  \tag{2.9b}
\end{equation}%
With help of this result, Eq.(2.8) acquires the familiar form of \textit{%
stationary} Schr\"{o}dinger's Eq.(1.4). \ In such a form this equation was
used by Sch\"{o}dinger in his first three communications on quantum
mechanics [$1$]\ (not to be confused with the his paper published in
Physical Review [$28]).$ Only in the 4th paper he found a way of introducing
the time dependence correctly. \ This topic is discussed in the next
section.\bigskip

\bigskip

\textbf{3. \ Inclusion of time-dependence into Schr\"{o}dinger's equation }

\ \ \ \ \ \textbf{by Schr\"{o}dinger and its impact on the latest
experimental }

\ \ \ \ \ \textbf{and theoretical works }

\bigskip

Inclusion of time-dependence into Schr\"{o}dinger's equation was made only
in\ the 4th installment of Schr\"{o}dinger's series of papers on quantum
mechanics [$1].$ There is a good reason why this was done so late. It will
be explained at length in the companion paper\footnote{%
In the meantime, please, read [30].}. In this paper we employ physical
arguments, beginning with those by Schr\"{o}dinger. On page 103 of [$1]$ \
we find the following comment :

" Thus, when we designated equation (1) or ($1^{\prime }$)\footnote{%
Our Eq.(1.4).} on various occasions as "the wave equation", \textsl{we were
really wrong}\footnote{%
Emphasis is ours}...". \ To fix this problem (that is to make things right)
Schr\"{o}dinger suggest to call Eq.(1.4) as the "\textsl{amplitude equation}%
." He acknowledges, that such an equation is valid only for the conservative
(that is time-independent) systems. Any process of measurement involves,
however, time-dependent perturbation [$31$]. Accordingly, "we must search 
\textit{for the} \textbf{real}\textit{\ wave equation} " \ since " the
amplitude equation is no longer sufficient". Thus, Eq.(1.4) is no longer
real wave equation for Schr\"{o}dinger. Furthermore, on page 102 he writes:
" Equation (1)$\footnote{%
That is our Eq.(1.4)}$ contains the energy-or frequency-parameter $E$, and
is valid, as expressly emphasized in Part II\footnote{%
That is, read pages 13-40 of [1].}, with a \textit{definite} $E$-value."
Here the italics are Schr\"{o}dinger's. This statement by Schr\"{o}dinger is
very confusing, nevertheless, since in Eq.(1.4) the parameter $E$ is
determined by solving the respective eigenvalue problems. And this is 
\textit{exactly} what Schrodinger did repeatedly, starting with his 1st
installment on quantum mechanics! Next, he suggests \ several ways of
arriving at the "\textbf{real} wave equation" which we do not want to
reproduce here since these results are incorrect (as Schr\"{o}dinger
acknowledges himself). After several wrong attempts, he finally writes: "The
dependence of $\psi $ on the time \ which must exist if ($1^{\prime })%
\footnote{%
That is our Eq.(1.4).}$ to hold, can be expressed by 
\begin{equation}
\frac{\partial \psi }{\partial t}=\pm \frac{2\pi i}{h}E\psi \text{ \ } 
\tag{3.1a}
\end{equation}%
\textsl{as well as by} 
\begin{equation}
\text{\ }\frac{\partial ^{2}\psi }{\partial t^{2}}=-\frac{4\pi ^{2}}{h^{2}}%
E^{2}\psi ."  \tag{3.1b}
\end{equation}%
Obviously, "\textsl{as well as}" only means that: a) either his statement is
erroneous (since equations (3.1a) and (3.1b) are mutually exclusive because
Eq.(3.1a) is leading to the parabolic-type partial differential equation
(PDE) while Eq.(3.1b) is leading to the hyperbolic PDE, b) he had in mind to
use two times for the same equation\footnote{%
Incidentally, two times interpretation of the time-dependent Schr\"{o}dinger
equation was proposed quite recently [$32$], in 2011.}, or c) he suggested
to replace (in his opinion, definite value) the parameter $E$ \ ( present in
the combination $E\psi )$ in his "amplitude" equation (1.4) by $\pm i\hbar 
\frac{\partial \psi }{\partial t}$ in order to arrive at his "\textbf{true} 
\textbf{wave} \textbf{equation.}" \ Finally, the option c) was selected by
Schr\"{o}dinger so that the obtained equation had become a "\textsl{true
wave equation}", known now as Schr\"{o}dinger's equation\footnote{%
Incidentally, selection of the Eq.(3.1a)) leads to the correspondence
between \ the Heisenberg and Schrodinger's version of quantum mechanics.}.
Selection of option c) disconnects this equation \ from \ the wave equation
(2.8) and makes use of the De Broglie particle-wave arguments much more
difficult to implement. On purely mathematical level, selection of the
hyperbolic type equation (based on selection of Eq.(3.1b)) requires
knowledge of the two initial conditions for the Cauchy problem to be set up
correctly, while selection of Eq.(3.1a) requires to use just one initial
condition $[29$] for setting up the Cauchy problem as explained in Appendix
A. This difference was noticed by Feynman. In section 1 we brought the
following Feynman's quotation [4], page 377: "Actually Huygens' principle is
not correct in optics. It is replaced by Kirchhoff's modification which
requires that both the amplitude and its derivative must be known on the
adjacent surface. This is a consequence of the fact that the wave equation
in optics is second order in the time. The wave equation of quantum
mechanics is first order in time; therefore, Huygens' principle\textsl{\ is}
correct for matter waves..."\footnote{%
But not corresct in optics according to statement by Feynman cited in
section 1.} . Unfortunately, this statement by Feynman is not correct as we
shall explain below. \ To begin our explanation, we start with \ some
quotations from the book by Feynman and Hibbs on path integrals [$7$] as
well as from Feynman's lectures on physics [$6$]. In section 1-1 of [$6$],
on the 1st page, we find the following statement: " There is one lucky
break, however-electrons behave just like light. The quantum behavior of
atomic objects (electrons, protons, neutrons, and so on) \textsl{is the same
for all}, they are all "particle waves"... \textsl{\ So, what we learn about
the properties of electrons...we will apply also to all "particles}" \textsl{%
including photons of light\footnote{%
Here italic empasis is ours. Historically, the \ events occured just in
reverse order though, as explained in section 1.}}". The interpretation of
outcomes of the double slit interference experiment is central for
understanding of quantum mechanics as explained in section 1 and in [14].
Because the light is being treated as "particle waves" by Feynman,we shall
begin with the light in the context of the double slit experiment in optics.
This experiment was described in detail long before quantum mechanics was
born [3,8, 33]. It involves use of Huygens' principle. Mathematically
accurate definition of this principle is given in the next section. For now
it is sufficient to think about this principle as follows.\bigskip

\textbf{Definition 3.1}. \textsl{The Huygens principle}. The new wave front
at later time $t^{\prime }$\ is the envelope of secondary waves emanating
from each point of the original wavefront\bigskip .

Stated in such a form it leaves entirely open the problem: Why the envelope
of secondary waves is moving only forward? Later on \ Fresnel added a
superposition principle \ for the amplitudes of the secondary waves to
explain the phenomenon of \textit{diffraction}. Subsequently, the same
problem was looked upon by Kirchhoff, Fraunhofer, Beltrami, Volterra,
Hadamard, Arnol'd, and many others.\ In optics literature the description of
diffraction begins with study of the scalar Helmholtz equation 
\begin{equation}
\left( \nabla ^{2}+k^{2}\right) \Psi =0.  \tag{3.2}
\end{equation}%
If we replace $\frac{c}{n}$ in (2.8) by $\tilde{c}$ \ and introduce $%
k=\omega /\tilde{c}$ then, Eq.(2.8) becomes the Helmholtz equation, provided
that the potential $V=0$. In such a form \ Eq.(3.2) \ is used \ in
scattering theory of quantum mechanics [$34]$. It is appropriate \ to notice
at this point that it is commonly believed that only Schr\"{o}dinger's
ansatz, Eq.(3.1a), makes the wave function complex and, with this, it makes
Born's interpretation of quantum mechanics possible. \ This opinion is
erroneous though since the simplest plane wave solution of Eq.(3.2) is known
to be [34] 
\begin{equation}
\Psi =\exp (\pm i\mathbf{k}\cdot \mathbf{r})  \tag{3.3}
\end{equation}%
leading to the probability current%
\begin{equation}
\mathbf{j}=\frac{\hslash }{2mi}(\Psi ^{\ast }\mathbf{\nabla }\Psi -\Psi 
\mathbf{\nabla }\Psi ^{\ast }).  \tag{3.4}
\end{equation}%
The question arises: Is it possible to find an analog to this current in
optics? From the book by Born and Wolf [$8$], chapter 8, it follows that the
analog of the current in optics can be obtained as result of application of
Green's theorem, also used in quantum mechanics for derivation of the
current, Eq.(3.4). In optics $\Psi $ is an amplitude of the wave.
Experimentally the intensity is measured. It is expressible through the
combination $\Psi ^{\ast }\Psi .$ In optics it is possible to replace the
vector wave equations with the scalar ones when diffraction is discussed.
Eq.(3.2) is the scalar wave equation for $\Psi $. In optics \ the intensity $%
\Psi ^{\ast }\Psi $ is measured experimentally while in quantum mechanics
the very same quantity is the probability. Very detailed analysis of
similarities and differences between optics and quantum mechanics was made
in our work, ref.[14]. It was stimulated by the footnote on page 387 of [8].
In it, the \ reference to the 1959 work by Emil Wolf \ was made. In this
work it is demonstrated "that both the (time averaged) energy density and
the energy flow in unpolarized quasimonochromatic (optical) \ wave field may
always be derived from \textsl{one component complex} \ and time-harmonic
scalar wave function." \ Neither Born nor Wolf \ attempted to develop these
observations subsequently. This was done in [14] where it is demonstrated
that Born probabilistic interpretation of quantum mechanics and the optical
interpretation (especially for the double-slit experiments) can be made
mathematically completely indistinguishable. Initially, the discrepancy of
interpretations \ of quantum mechanics was noticed by De Broglie. On Page
127 of his book [$35$] we find the following statement: "The probability
that the presence of \ the photon will be made \ known by photographic
action in the apparatus is everywhere proportional to the resultant
intensity of the wavetrain.....\textsl{It} \textsl{is almost certain that \
the same considerations are valid for the diffraction of material particles%
\footnote{%
The italics are ours.}}". Thus, results of [$14$] provide rigorous
mathematical support \ to De Broglie's intuition.

From Feynman's lectures [$6$], it follows that in all double slit
experiments the detectors were used (connected with loudspeakers) to produce
"clicks." The number of clicks was counted per unit time as a function of
the position ($x$) of the detector. Clearly, the detectors register particle
hits. These are not\ exactly the probability amplitudes but, traditionally
treated , they are represented through these amplitudes nevertheless! These
amplitudes were experimentally calculated with the help of protocol
described in section 1-1 of [$6$ ] and, indeed, when compared against the
interference patterns for light, the complete agreement was found, e.g. read
page 8 of section 1-1[6] and page 5 of [$7$]. Because of this, further steps
were made. On pages 5 and 6 of [$7$], \textit{without saying it explicitly},
Feynman formulates the Huygens-Fresnel principle in accord with the results
known from optics [$8$], page 371. Specifically, on pages 5-6 of [$7$] we
find the following description of the 2 hole experiment: " Furthermore, $%
\phi (x)$ is the sum of two contributions: $\phi _{1}$ the amplitude of
arrival through hole 1, plus $\phi _{2}$, the amplitude of arrival through
hole 2. In other words, there are complex numbers $\phi _{1}$ and $\phi _{2}$
such that $P=\left\vert \phi \right\vert ^{2},$ $\phi =\phi _{1}+\phi _{2%
\text{ }}$ and $P_{1}=\left\vert \phi _{1}\right\vert ^{2},P_{2}=\left\vert
\phi _{2}\right\vert ^{2}...$ Here we say only that $\phi _{1}$, for
example, \textsl{may be evaluated as a}\textbf{\ \textsl{solution of a wave
equation representing waves spreading from the source to 1, and from 1 to x}.%
\footnote{%
This is exactly Huygens' principle in the form of the "major premise" by
Hadamard described in Appendix B.} \textsl{This reflects the wave properties
of electrons\ (or in the case of light,} }\textsl{photons})\footnote{%
Here the emphasis is ours.}". But the "motion" of photons is described by
Eq.(2.6 a), while electrons-by Schr\"{o}dinger's wave equation Eq.(1.3).
Hence, the mathematical description is visibly different: Two different
equations- the hyperbolic (wave) and the parabolic (Schr\"{o}dinger) cannot
describe the same reality in the same way! The confusion is caused by the
fact that the Helmholtz Eq.(3.2) is used both in optics and in quantum
mechanics, e.g. read survey on atomic optics [$36$].When time-dependence
effects can be ignored, then mathematically these equations are
indistinguishable! With this observation in \ our hands we came all the way
back to Schr\"{o}dinger's dilemma: which one of the equations: Eq.(3.1a) or
(3.1b), should be used? Please, recall that the switch from (3.1b) to (3.1a)
was caused by Schr\"{o}dinger's desire to develop correct description of the
time-dependent \ quantum phenomena. For the time-independent situations,
however, it is completely safe to use the wave Eq.(2.8) producing the
stationary Schr\"{o}dinger Eq.(1.4). In this and the following
(time-dependent) paper we shall resolve Schr\"{o}dinger's dilemma by
treating the time-dependent case in such a way that the time-dependence is
rigorously eliminated. In such a case, even for the time-dependent
situations to be treated in the next paper we can begin with the wave
Eq.(2.6a) and, by doing so, we shall bring into correspondence treatments of
photons and the rest of "particle waves" (in Feynman's terminology).

As an input and guidance for what follows, we need now to comment \ on
several recently published papers-all reflecting the existing confusion
associated with uses of various definitions of Huygens' principle in quantum
mechanics. The confusion is exacerbated by several \ additional sources not
described above. To illustrate our point, we begin with the Helmholtz
Eq.(3.2). The Green's function $G(\mathbf{r},\mathbf{r}^{\prime })$ for this
equation is obtained from the equation 
\begin{equation}
\left( \nabla ^{2}+k^{2}\right) G(\mathbf{r},\mathbf{r}^{\prime })=\delta (%
\mathbf{r}-\mathbf{r}^{\prime }).  \tag{3.5}
\end{equation}%
From quantum \ mechanics [$34$], we obtain (for the outgoing standing wave)%
\begin{equation}
G_{(+)}(\mathbf{r},\mathbf{r}^{\prime })=-\frac{\exp (ik\left\vert \mathbf{r}%
-\mathbf{r}^{\prime }\right\vert )}{4\pi \left\vert \mathbf{r}-\mathbf{r}%
^{\prime }\right\vert }  \tag{3.6}
\end{equation}%
(for the incoming wave $k$ should be replaced by -$k$ in the exponent). Now, 
$G_{(+)}(\mathbf{r},\mathbf{r}^{\prime })$ is not at all \ the Green's
function of the wave equation. Accordingly, one cannot use the
Fresnel-Huygens principle in the form hinted by Feynman and implemented in [$%
37$]. Specifically, what the authors of [37] are calling as the
Huygens-Fresnel principle 
\begin{equation}
G_{(+)}(\mathbf{r}_{x},\mathbf{r}_{0})=\dint d\mathbf{r}_{1}G_{(+)}(\mathbf{r%
}_{x},\mathbf{r}_{1})G_{(+)}(\mathbf{r}_{1},\mathbf{r}_{0})  \tag{3.7}
\end{equation}%
is not all mathematically valid expression! Accordingly, the results of [$37$%
] are not valid. Since these results question the double slit experiment
lying at the heart of quantum mechanics, we now provide the detailed \
explanation of what went wrong with the results of [$37$]. First of all, the
propagator $G_{(+)}(\mathbf{r},\mathbf{r}^{\prime })$ does not admit the
path integral representation and, accordingly, it does not posses the
Markovian property essential for all path integral treatments. This could be
seen by direct calculation of Eq.(3.7) in which Eq.(3.6) is used. For the
record we quote the result from the book by Ito and McKean [$38$], paragraph
7.21, containing a remarkable formula connecting the Brownian motion
propagator with the Newton (or Coulombic) potential%
\begin{equation}
\dint\limits_{0}^{\infty }dt(2\pi t)^{-\frac{d}{2}}\exp \{-\frac{\left\vert 
\mathbf{b}-\mathbf{a}\right\vert ^{2}}{2t}\}=const\left\vert \mathbf{b}-%
\mathbf{a}\right\vert ^{2-d}.  \tag{3.8}
\end{equation}%
Here $d$ is the dimensionality of space, $d\geq 3.$ On the left hand side of
this equality we find \ well known Feynman's propagator for the free
particle of unit mass upon transition $t\rightarrow i\hbar t.$ Accordingly,
for \textbf{this} propagator one can safely use Eq.(3.7). Unfortunately,
Eq.(3.8) cannot be replaced by that involving the Yukawa-type potential,
Eq.(3.6). While the attempts to find an identity analogous to Eq.(3.8) for
the Yukawa-type potentials, e.g. Eq.(3.6), are still ongoing, it is obvious
that, based on current knowledge of this topic, Eq.(3.7) is invalid. Because
of this, the\ main statement of [$37$] questioning correctness of the
superposition principle of quantum mechanics is also incorrect. This is so
because of the following. The quantum mechanical superposition principle for
two slits A and B in its orthodox form reads: $\psi _{AB}=\psi _{A}+\psi
_{B}.$ However, the authors of [$37$], following some earlier works, write
instead $\psi _{AB}=\psi _{A}+\psi _{B}+\psi _{L}$. The third term, \
supposedly, originates from the entanglement of the Brownian paths
originating at the source with two slits, e.g. A and B. That is to say, $%
\psi _{L}$ accounts for the possibility of the Brownian path going through
the slits and coming back to the source/the origin. \ However, according to
Huygens' principle the secondary waves can move only forward! According to
Feynman and the cited experiments, both the light and the particle waves
produce the same interference patterns. In optics the description of such
patterns does involve uses of Huygens' principle. Therefore, the same should
be true for the "particle waves." Accordingly, the equality $\psi _{AB}=\psi
_{A}+\psi _{B}+\psi _{L}$ is incorrect. In many papers attempting to use
Feynman's path integrals for description of the double and triple slit
experiments, e.g. read [$39$] and references therein, the authors used the
Markovian analog of Eq.(3.7) (e.g. see Eq.(2.31) on page 37 of the book by
Feynman and Hibbs [$7]$). Use of this analog is illegitimate though because
the Brownian propagator in Eq.(3.8) when integrated over time time is not
the propagator for the wave equation. The parabolic time-dependent Schr\"{o}%
dinger equation is surely not \ the same thing as the hyperbolic wave
Eq.(2.6a) since in the first case we need just one initial condition while
in the second-two. For the wave equation Huygens's principle rigorously
holds (as explained in the next section and Appendix B). An attempt at
rigorous implementation of Huygens' principle in quantum mechanics
formulated in terms of Feynman path integrals was made by Gutzwiller [5]
whose sketchy and incomplete results were discussed in the 1st section.

Replacement of the hyperbolic (wave) equation by the parabolic (Schr\"{o}%
dinger) could be a source of wrong conclusions about the outcome of
three-slit experiments with photons. These were reported in "Science" in
2010 [$40$] with the purpose of rigorous testing of the superposition
principle. These experiments found no deviations from the superposition
principle of quantum mechanics. The quality of these experimental studies
was further improved in [$41$] with the same outcome thus confirming the
validity of both, Huygens' and superposition principles. Subsequent \
analytical and computer studies of \ the three slit configurations [$39$],
culminating in the latest PRL [$37$] -all criticized \ the experimental
results of Sinha et al [$40$] as well as those by S\"{o}llner et al [41].
This critique \ is invalid however for reasons \ already \ explained. We
noticed in section 2 that the De Broglie relation $p=\frac{h}{\lambda }$ is
valid also for photons. But the dispersion relation for photons is $\omega
=ck$ while that for the Schr\"{o}dinger's equation for the free particle of
mass $m$ is $\omega =\frac{\hbar k^{2}}{2m}$ (e.g. read Appendix A). This
difference in dispersion relations is equivalent to the replacement of
Eq.(3.1b) by (3.1a) as done by Schr\"{o}dinger. It is fundamental since only
under such conditions Heisenberg and Schrodinger pictures of quantum
mechanics are in agreement with each other! The paradoxicality of the
existing situation can be understood by reading some earlier descriptions of
Huygens' principle, the descriptions of diffraction, etc. [$3$]. They all
involved the wave equation and time-dependence. However, this
time-dependence happened to be not essential as could be seen, for example,
by reading the authoritative book on optics by Arnold Sommerfeld [42$]$.
Subsequently, the time-dependence was wiped out from books on optics [8,43].
Once it was wiped out, use of the time-independent optics formalism, e.g.
describing the double slit diffraction, makes it possible to replace the
optical formulas by the quantum mechanical ones. More details are given in [$%
14$]. Inclusion of time dependence leads to dramatic effects. Details will
be provided in future publications. This can be seeing already from \ the \
observation \ that switching from the hyperbolic (wave) Eq. (2.6a) requiring
two functions for the initial conditions (the Cauchy data) to the parabolic
(Schr\"{o}dinger) equation requiring only one function is highly nontrivial
, e.g. read subsection 6.2.1. below. \ Being parabolic, the Schr\"{o}dinger
equation possess the Markovian property while the wave equation is not. To
by pass this difficulty, the semigroup analysis of operators was invented [$%
44$] \footnote{%
Also, read \ the subsection 6.2.1 and \ ref.[14] where the Duffin-Kemmer
formalism is briefly discussed leading to analogous results.} In section 1
we quoted Gutzviller [5] who said that "\textsl{neither Hyugens' principle
nor the path -integral applies to the stationary waves} which are the
solutions of Eq.s (1.4) and (1.5)". \ This statement of Gutzviller happens
to be incorrect as we shall demonstrate below. Furthermore, in [5]
Gutzwiller was using the semigroup analysis [44] for claiming that the
time-dependent Schr\"{o}dinger equation is obtainable from the relativistic
(wave-like, a la Dirac, equation ) in the limit when $c\rightarrow \infty .$
First of all, this limit is unphysical (see, however, Appendix A) and,
second of all, we shall demonstrate that Gutzwiller's claims regarding
applicability of Hyugens' principle to Eq.s (1.4) and (1.5) is also
incorrect. However, his claims about the Markovian property of path
integrals are completely consistent with ours. Eq.(3.7) is surely non
Markovian. There is no way of \ rewriting Eq.(3.7) in the path integral
form. \textsl{Thus, the Markovian property is not coinciding with the
Huygens' principle property! } The Hugens principle\ (and property) in its
refined form permits propagation of wave packets only forward (as explained
in the following section) while Schrodinger's equation allows spreading of
the wave packets both ways [45]. Interestingly enough, the Huygens principle
had been put by Schr\"{o}dinger into center of his developments of wave
mechanics. This can be seen from reading of Part II [$1],$pages 13-40, of
his four foundational installments on quantum mechanics. In view of the
variety of opinions in physics literature about the essence of Huygens'
principle (as compared to the variety of opinions about Huygens' principle
prevailing in mathematics literature), in the next section (and \ in
Appendix B)\ we are presenting some basic facts about Huygens' principle as
it is developed by mathematicians.

\bigskip

\textbf{4. Basic facts about Huygens' principle (mathematicians
perspective\bigskip )}

\textsl{4.1. Wavy (that is partial differential equations ) -type versus
contact-geometric-type aspects of Huygens' principle}

\bigskip

Definition 3.1. provides \ the essence of Huygens' principle. In plain words
Huygens' principle can be concisely formulated as follows. Consider a \
point at the wave front at the moment $t$ as the source of a new (secondary)
wave emanating from this point. The Huygens principle (HP) as formulated by
Huygens himself leaves entirely open the problem: Why the envelope of
secondary waves is moving only forward? \ In the previous section we stated
that to fix this problem subsequently Fresnel added a superposition
principle for amplitudes of the secondary waves. This helped him to explain
the phenomenon of \textit{diffraction}. We had already mentioned that the
same problem was looked upon later by Kirchhoff, Beltrami, Volterra,
Hadamard, Arnol'd, and many others. Different authors used the HP with
different purposes in mind. This caused differences in interpretation and
confusion among users. Basically, the split in interpretations originates
from the split of opinions about what is light. If we take the side of
proponents of the wavy nature of light, then it is instructive to read
carefully the fundamental work by Hadamard [$46].\func{Re}$ad also [29],
chapter 6. If we take the side of proponents of corpuscular nature of light,
then it is instructive to read woks by Arnol'd [47-49] in which methods of
contact geometry and topology are used. To our knowledge, the detailed
connections between these two directions of thought still do not exist.
Works by Maslov [50,51] indicate that the connection problem is solved, in
principle$.$

It is appropriate again to bring the quotation by Schr\"{o}dinger on this
topic [$26].$ On page 154 he writes:\footnote{%
This is conclusion of his Nobel Address delivered on December 12th, 1933.}"
The light ray, or track of the particle, corresponds to the \textit{%
longitudinal} continuity of the propagating process (that is to say, in the
direction of spreading); the wave front, on the other hand, to the \textit{%
transversal} one, that is to say, perpendicular to the direction of
spreading. \textit{Both} continuities are undoubtedly real. The one has been
proved by photographing of particle tracks, and the other by interference
experiments. As yet we have not been able to bring the two together into
uniform scheme. It is only in extreme cases that the transversal -the
spherical-continuity or the longitudinal -the ray -continuity shows itself
so predominantly that \textit{we believe} we can avail ourselves either of
the wave scheme or the particle scheme." \ Here the italics are Schr\"{o}%
dinger's. To complete this chain of quotations we cite Arnold's "Lectures on
Partial Differential Equations" [$48].$\ At the beginning of Lecture 1 he
writes: \ "In this lecture we shall consider a case in which there is a
complete theory, namely the case of one first order equation. From the
physical point of view this case (displays) the duality that occurs in
describing \ a phenomenon using waves\textit{\ or} particles. \ The field
(of waves) satisfies a certain first-order partial differential equation,
the evolution of the particles is described by ordinary differential
equations, and there is a method of reducing the partial differential
equation to a system of ordinary differential equations; \textsl{in that way
one can reduce the study of wave propagation to the study of the evolution
of particles}.\footnote{%
Words put in curly brackets as well as italic emphasis are ours}" \ 

In view of particle-wave duality just described it is appropriate to define%
\textit{\ the dual of Huygens' principle. }In the theory of ordinary
differential equations [52,53] (that is for "rays" in optical terminology)
there is so called \textit{rectification} theorem which, when translated
into language of Hamiltonian dynamics, is known as \textit{symplectic
rectification} theorem [54]$.$ In plain words it is the statement which can
formulated as follows. \bigskip \textbf{\ }

\textbf{Definition 4.1.} \textsl{The} \textsl{dual Huygens' principle}. 
\textit{Any point on the dynamical trajectory}\footnote{%
That is ray trajectory} \textit{can serve as the origin of (new) motion}%
.\bigskip

That is to say, at any point along the trajectory it is possible to find
canonical variables which are making the Hamiltonian \ of the dynamical
system to vanish. At all such points along the trajectory the canonical
Hamiltonian equations \ become trivial. When such dynamical system is
quantized, the unitary operator describing system evolution becomes trivial
too. Since the dual Huygens' principle applies to any point $(p,q)$ along
the trajectory in phase space, this means that Heisenberg's uncertainty
principle fails. The way out of this difficulty was suggested by Maslov [50]
who suggested to use the so called Lagrangian manifolds $\mathcal{L}$. These
manifolds naturally arise in contact geometry [47-49]. An introduction to
this field can be found in [55]. In the context of Hamilton-Jacobi formalism
the Lagrangian manifolds \ naturally originate based \ on familiar equations
of the type $p_{i}=\partial S(\{q_{j}\},t)/\partial q_{i}.$ Here $%
i,j=1,...,N $, $N$ is the dimension of the configurational (tangent) space, $%
S$ is the action of the dynamical system. For the prescribed $p_{i}^{\prime
}s$ just stated set of equations defines the Lagrangian manifold $\mathcal{L}
$. Given that the dimension of the symplectic manifold is $2N$, the
dimension of the Lagrangian manifold is $N$. \ Since the symplectic manifold
is determined locally by the 1-form $\theta =\sum_{i=1}^{N}p_{i}dq_{i}$ ,
the Lagrangian manifold is determined by the requirement: $d\theta
=\sum_{i=1}^{N}dp_{i}\wedge dq_{i}=0.$ At the same time, the contact 1-form $%
\alpha $ is determined by $\alpha =dS-\sum_{i=1}^{N}p_{i}dq$ while the
contact plane is determined by the condition $\alpha =0.$ \ The Bohr
-Sommerfeld quantization conditions for trajectories on $\mathcal{L}$ are
trivial, that is $\doint p_{i}dq_{i}=0,$ since \ by design $\{p,q\}=0$ on $%
\mathcal{L}$ where $\{,\}$ is the Poisson bracket. Existence of the
Lagrangian manifolds does not \ preclude the existence of quantum mechanics,
at least in the asymptotic sense [50], even though the Groenvold- van Hove
theorem [56],[57] causes severe difficulties in systematic development of
quantum mechanics.

Resolution, if any, of the above mentioned difficulties caused us to discuss
only the wavy (that is, the PDE) side of Huygens' principle [$29],[33$]. The
dual (particle) side requires uses of \ methods of contact geometry and
topology as \ we\ just explained. Much more detailed account of this side of
the Huygens principle is given in $[58$], Chr.10. Reference [$59$] contains
some alternative treatment while very readable but rigorous \ ref.[60]
claims along with Schr\"{o}dinger [1], part II, that Huygens' principle and
the Hamilton-Jacobi equation are equivalent concepts. More on this
contact/symplectic aspects of Huygens' principle will be presented in \ our
forthcoming publications. \ 

We begin our discussion of the wavy aspects of the HP by summarizing results
by Hadamard [$46]$ \ In ref.[$46]$ he formulated the following\bigskip

\textbf{Hadamard} \textbf{problem}. \textit{How to describe/to classify all
second order hyperbolic partial differential equations satisfying HP}%
?\bigskip

Hadamard discovered that the HP is working only in spaces in which the
number of \textsl{spatial} dimensions is odd, e.g. in 3+1 spacetimes of
Lorentzian signature. From here, it follows that the behavior of solutions
of wave equations in 2+1 and 3+1 dimensional spaces is quite different. A
short time point-like perturbation in \textbf{R}$^{3}$ will produce sharply
defined moving spherical wave so that when it reaches the observer he/she
will see (or hear) just a short lasting flash/or splash. At the same time,
for waves on the surface of water (this is an example used by Feynman, e.g.
read the discussion pertinent to Fig.1-2 \ in [6] for his preliminary
description of the two-slit experiment), that is in \textbf{R}$^{2},$ the
whole region inside the moving (spreading) circle will be disturbed. They
say, that the 3 dimensional event is the result of the action of Huygens'
principle while the 2 dimensional event, say, on the surface of water, is
caused by the \textit{wave diffusion}. \ The following theorem proven
independently by Mathisson[$61]$ and Asgeirsson[$62]$ is playing the central
role in studying the HP -type of problems\bigskip

\textbf{Theorem\ 4.1}. \textit{If \ the hyperbolic equation in space-time of
Lorentzian signature is satisfying HP, then it is equivalent to
Eq.(2.6a).\bigskip }

Thus, $\square _{4}$ is the Huygens operator and a question emerges: How to
define the equivalence? \ Following [$63,64]$ this equivalence can be
defined as follows. Let L[$\phi ]$ be a Huygens' operator (that is the
operator which is satisfying HP) and let \~{L}[$\phi ]$ be another Huygens
operator. Then they are equivalent if:\medskip

a) \ \~{L}[$\phi ]$ can be obtained from L[$\phi ]$ by non-singular
transformations of the independent variables.

b) \ \~{L}[$\phi ]=\lambda ^{-1}$L[$\lambda \phi ]$ for some positive,
smooth function $\lambda $ in the causal domain $\Omega $ (to be defined
momentarily).

c) \ \~{L}[$\phi ]=\rho $ L[$\phi ]$ for some positive smooth function $\rho 
$ in $\Omega .\medskip $

To define the causal domain $\Omega $ we need to define the distance
function [$63]$ $\Gamma (t,x;\tau ,y)$. It is given by 
\begin{equation}
\Gamma (t,x;\tau ,y)=c^{2}(t-\tau
)^{2}-\dsum\limits_{i=1}^{3}(x_{i}-y_{i})^{2}.  \tag{4.1}
\end{equation}

Each pair $(t,x)\in \mathbf{R}^{1+3}$ is called an \textit{event}. The
Euclidean \ line segment $r(x,y)=\sqrt{\dsum%
\limits_{i=1}^{3}(x_{i}-y_{i})^{2}}$ between two events $(t,x)$ and $(\tau
,y)$ can be used to formally define the velocity $\mathit{v}=r(x,y)/(\tau
-t),\tau >t.$ This definition then allows us to define an open (\textit{%
future}) set D$_{+}(t,x)\in \mathbf{R}^{1+3}$ of those events that can be
reached from $(t,x)$ with the velocity $v\leq c$. Analogously, it is
possible to define \ an open(\textit{past}) set D$_{-}(t,x)\in \mathbf{R}%
^{1+3}$ of events $(\tau ,y)$ from which $(t,x)$ can be reached with the
velocity $v\leq c.$ Thus, $(\tau ,y)\in $D$_{+}(t,x)$ if $(t,x)\in $D$%
_{-}(\tau ,y).$ The boundary of D$_{+}(t,x)$ is called \textit{forward}
(future) (respectively \textit{backward (}past)) characteristic cone C$%
_{+}(t,x)$ (respectively, C$_{-}(t,x)).$ The \textit{lightcone} is defined
by the requirement $\Gamma (t,x;\tau ,y)=0.$ This is \ an equation of a 3d
sphere for a fixed $(\tau -t)^{2}.$ Based on this information, we are now in
the position to define\bigskip

\textbf{\ Definition 4.2. }\textsl{Hadamard criterion}. The operator L[$\phi
]$ is of\textit{\ Huygens-type }if\textit{\ }%
\begin{equation}
\text{L}[\phi ]\mid _{\Gamma (t,x;\tau ,y)=0}=0.  \tag{4.2}
\end{equation}%
Accordingly, the causal domain $\Omega $ is determined by the following
requirement [$63]$ : event $(t,\ x)\in \Omega $ only if $(\tau ,y)\in $ C$%
_{\pm }(t,x).$ \bigskip

\textbf{Definition} \textbf{4.3.} A Huygens operator L[$\phi ]$ that arises
from $\square _{4}$ via operations a), b) and c) is called \textit{trivial}
Huygens operator.\bigskip

\textsl{Hadamard conjecture}\textbf{. }\textit{Every Huygens operator is
trivial.\bigskip }

Evidently, the Hadamard conjecture is fully compatible with Theorem 4.1. The
question arises: \ If at any point of (pseudo)Riemannian 3+1 spacetime the
metric can be brought into the diagonal form of Lorentzian signature, will
the Hadamard conjecture be valid in some open domain of such space? \ Some
studies of this problem were made by Friedlander [$65],$ McLenagan [$66],$
Goldoni [$67],$ Ibragimov [$68]$ , W\"{u}nsch et al [$69$]. To discuss these
works further from physical standpoint, we need to relate them to Schr\"{o}%
dinger's work [1]. Incidentally, Schr\"{o}dinger later on in his life
studied the problem of wave propagation in curved spacetimes [$24].\ $More
details on this topic are given in the Appendix B.

\bigskip

\textsl{4.2. \ Connection of mathematical works on Huygens principle with
Schr\"{o}dinger's foundational papers on quantum mechanics\bigskip \bigskip }

\textsl{4.2.1. General background\bigskip }

Already in section 1 we noticed that for Schr\"{o}dinger (2nd installment in
[1]) HP is synonymous with the H-J equation. This point of view is being
shared by such famous mathematicians as Gelfand and Fomin [60], pages
208-217. \ In our opinion, based on [29,33], Gelfand and Fomin results can
be presented \ in a such a way that their consistency with results \ of
Hadamard and other mathematicians who used the PDE methods for description
of the HP will become obvious. Following [33], we shall assume that at time $%
t_{0}$ a light signal had been originated at the point (x$_{0}$,y$_{0}$,z$%
_{0}$). At the later time $t>t_{0}$ this signal had penetrated into domain
of space enclosed by a surface $V$ given analytically in the form 
\begin{equation}
V(x_{0},y_{0},z_{0};x,y,z)=c(t-t_{0})  \tag{4.3a}
\end{equation}%
The surface $V$ is called \textsl{wavefront surface}. The velocity of the
wavefront \ can be measured with help of the velocity along the trajectories
orthogonal to the wavefront (Appendix A). For points \ P$_{1}$ and P$_{2}$
on such trajectory we obtain:%
\begin{equation}
V(x_{0},y_{0},z_{0};x_{2},y_{2},z_{2})-V(x_{0},y_{0},z_{0};x_{1},y_{1},z_{1})=c(t_{2}-t_{1})
\tag{4.3b}
\end{equation}%
or, in the differential form, 
\begin{equation}
V_{x}dx+V_{y}dy+V_{z}dz=cdt.  \tag{4.3c}
\end{equation}%
Mathematically, this result is identical to the condition $\alpha =0$ for
the contact plane mentioned in the previous subsection. \ From this
observation it follows that it is perfectly reasonable to apply methods of
contact geometry mentioned in the previous subsection for description of the
HP. At the level of PDE this principle can be further elaborated as follows.
Since Eq.(4.3a) is describing a wavefront, that is a two -dimensional
surface $\Gamma $ in three-dimensional space, \ it makes sence to introduce
coordinates $\xi ,\eta $ on this surface so that 
\begin{equation}
x=f(\xi ,\eta ),y=g(\xi ,\eta ),z=h(\xi ,\eta ).  \tag{4.4a}
\end{equation}%
Since the surface $\Gamma $ is moving, the above parametrization can be
extended as follows%
\begin{equation}
x=x(\xi ,\eta ;\tau ),y=y(\xi ,\eta ;\tau ),z=z(\xi ,\eta ;\tau ). 
\tag{4.4b}
\end{equation}%
Evidently, 
\begin{equation}
x(\xi ,\eta ;0)=f(\xi ,\eta )\equiv x_{0},y(\xi ,\eta ;0)=g(\xi ,\eta
)\equiv y_{0},z(\xi ,\eta ;0)=h(\xi ,\eta )\equiv z_{0}  \tag{4.5a}
\end{equation}%
and 
\begin{equation}
\dot{x}(\xi ,\eta ;0)=a(\xi ,\eta ),\dot{y}(\xi ,\eta ;0)=b(\xi ,\eta ),\dot{%
z}(\xi ,\eta ;0)=c(\xi ,\eta ).  \tag{4.5b}
\end{equation}%
To use these equations, we need to make some detour into theory of PDE
following [29]. In particular, equations\textsl{\ }\ like Eq. (4.3a) are
equations of \textsl{characteristics}. These are equations for some
surfaces, e.g. $\Gamma .$ These are encountered not only for the first order
PDE's \ known to standardly trained physics professionals but also to the
higher order \ PDE's too. Suppose, we are having, say, the 2nd order PDE,
e.g like Eq.(2.6a), for which the Cauchy probem (Appendix A) can be set up
and is well posed.\textsl{\ \ }Then, one can think about extending the
Cauchy initial values, e.g. see Eq.s(4.5a),(4.5b), prescribed on $\Gamma $ to%
\textsl{\ }solutions of, say, Eq.(2.6a).\textsl{\ }

For this purpose it is useful \ to consider some auxiliary problems first.
E.g. let us \textsl{\ }consider a cone\textsl{, }Eq.(4.1), defined on C$%
_{+}(t,x).$ Following [29], page 558, \ we introduce the function $\chi
=\left( ct\ \right) ^{2}$\ $-x^{2}-y^{2}$\ $-z^{2}.$ By direct calculation \
we obtain ($c=1$):%
\begin{equation}
\chi _{t}^{2}-\chi _{x}^{2}-\chi _{y}^{2}-\chi _{z}^{2}=4\chi .  \tag{4.6a}
\end{equation}%
\ On the cone\ $C_{+}(t,x)$ we have the condition \ $\chi =0$ leading to the
1st order PDE 
\begin{equation}
\chi _{t}^{2}-\chi _{x}^{2}-\chi _{y}^{2}-\chi _{z}^{2}=0.  \tag{4.6b}
\end{equation}%
Just obtained resullt admits broad generalization. For instance, consider
instead (c=1) $\phi =t-\sqrt{x^{2}+y^{2}+z^{2}\ }=const.$ Then, \ we again
obtain%
\begin{equation}
\phi _{t}^{2}-\phi _{x}^{2}-\phi _{y}^{2}-\phi _{z}^{2}=0.  \tag{4.7}
\end{equation}%
Clearly, \ we can make these results to hold in any number of dimensions.
Furthermore, Eq.s (4.6b) and (4.7) are looking the same. \ Now we
demonstrate that: a) there are many other functions than $\chi ,\phi ,$
satisfying \ the same 1st order PDE, Eq.(4.7), b) solutions $\chi ,\phi ,$
etc. are also solutions of Eq.(2.6a). \ The demonstration can be achieved
with help of the progressive wave solution method developed by Friedlander [$%
70$] in 1946. Since his resuts, to our knowledge, are unknown in physics
literature, we would like to make a detour and to discuss his results. This
will enable us to develop a variety of physical applications.\bigskip

\textsl{4.2.2. Progressive \ wave solution by Friedlander. Introduction}

\bigskip

We have to look for solutions \ of Eq.(2.6a) in the form%
\begin{equation}
\phi (x,y,z;t)=u(x,y,z)F(ct-f(x,y,z))  \tag{4.8a}
\end{equation}%
where $F$ is an arbitrary well behaving function. The ansatz (4.8a) can be
further complicated if, instead, we shall look for solutions in the form%
\begin{equation}
\phi (x,y,z;t)=\dsum\limits_{m=1}^{N}u_{m}(x,y,z)F_{m}(ct-f(x,y,z)), 
\tag{4.8b}
\end{equation}%
where 
\begin{equation}
F_{m}^{^{\prime }}(\xi )=F_{m-1}(\xi ),F_{m}(\xi )=\dint\limits^{\xi }d%
\tilde{\xi}F_{m-1}(\tilde{\xi}).  \tag{4.8c}
\end{equation}%
In this work we shall \ use only the ansatz given by Eq.(4.8a). Substitution
of this result into Eq.(2.6a) results in the following three coupled \ PDE's
: 
\begin{equation}
f_{x}^{2}+f_{y}^{2}+f_{z}^{2}=1,  \tag{4.9a}
\end{equation}%
\begin{equation}
2(u_{x}f_{x}+u_{x}f_{x}+u_{x}f_{x})+u\nabla ^{2}f=0,  \tag{4.9b}
\end{equation}%
\begin{equation}
\nabla ^{2}u=0.  \tag{4.9c}
\end{equation}%
Clearly, the simplest case is obtained when $u=const$. In this case we are
left \ only with Eq.(4.9a). Eq.(4.7) will coincide with Eq.(4.9a) \ if we
choose $\phi (x,y,z;t)=t-f(x,y,z).$ In view of Eq.(2.8) the obtained results
can be extended further. For instance, Eq.(2.6a) can be replaced by%
\begin{equation}
\left( \frac{\partial ^{2}}{\partial t^{2}}\frac{n^{2}(x,y,z)}{c^{2}}-\frac{%
\partial ^{2}}{\partial x^{2}}-\frac{\partial ^{2}}{\partial y^{2}}-\frac{%
\partial ^{2}}{\partial z^{2}}\right) \phi (x,y,z;t)=0,  \tag{4.10}
\end{equation}%
where $n^{2}(x,y,z)$ is some dimensionless function of spatial variables. In
optics it is associated with the index of refraction [$33$]. With such a
replacement, Eq.(4.9a) is changed into 
\begin{equation}
\tilde{f}_{x}^{2}+\tilde{f}_{y}^{2}+\tilde{f}_{z}^{2}=n^{2}(x,y,z). 
\tag{4.11}
\end{equation}%
Next, we return back to Eq.(4.3c) which we rewrite as 
\begin{equation}
V_{x}\dot{x}+V_{y}\dot{y}+V_{z}\dot{z}=c.  \tag{4.12}
\end{equation}%
With a wavefront, Eq.(4.3a), a two parameter ($\xi ,\eta )$ family of rays
orthogonal to the wavefront is associated. \ A current point on a ray
trajectory is described by \{x($\tau ),y(\tau ),z(\tau )\}.$

Let $\mathcal{F}$(x($\tau ),y(\tau ),z(\tau ))$ be some yet arbitrary
function on this trajectory. Then, 
\begin{equation}
\frac{d\mathcal{F}}{d\tau }=\mathcal{F}_{x}\dot{x}+\mathcal{F}_{y}\dot{y}+%
\mathcal{F}_{z}\dot{z}.  \tag{4.13}
\end{equation}%
Without loss of generality, we can let $\mathcal{F}$(x($\tau ),y(\tau
),z(\tau ))=\tilde{f}($x($\tau ),y(\tau ),z(\tau )).$ Furthermore, let%
\begin{equation}
\frac{dx}{dt}=v\frac{\tilde{f}_{x}}{\sqrt{\ \tilde{f}_{x}^{2}+\tilde{f}%
_{y}^{2}+\ \tilde{f}_{z}^{2}\ \ }},\frac{dy}{dt}=v\frac{\tilde{f}_{y}}{\sqrt{%
\ \tilde{f}_{x}^{2}+\tilde{f}_{y}^{2}+\ \tilde{f}_{z}^{2}\ \ }},\frac{dz}{dt}%
=v\frac{\tilde{f}_{z}}{\sqrt{\ \tilde{f}_{x}^{2}+\tilde{f}_{y}^{2}+\ \tilde{f%
}_{z}^{2}\ \ }},  \tag{4.14}
\end{equation}%
where $v$ is the absolute value of the velocity. With such identifications
we now replace $V(x,y,z)$ in Eq.(4.12) by $\tilde{f}(x,y,z)$ and use the
r.h.s. of Eq.s(4.14) in Eq.(4.12) to arrive at 
\begin{equation}
\sqrt{\ \tilde{f}_{x}^{2}+\tilde{f}_{y}^{2}+\ \tilde{f}_{z}^{2}\ \ }=\frac{c%
}{v}.  \tag{4.15}
\end{equation}%
But $\dfrac{c}{v}=n.$ Therefore, by squaring we reobtain back Eq.(4.11). \ \
In addition, however, using Eq.s (4.14) we obtain 
\begin{equation}
\tilde{f}_{x}=\frac{1}{v}\frac{dx}{dt}\frac{c}{v}\equiv \frac{c}{v^{2}}\dot{x%
}.  \tag{4.16a}
\end{equation}%
Accordingly, 
\begin{equation}
\tilde{f}_{x}^{2}+\tilde{f}_{y}^{2}+\tilde{f}_{z}^{2}=\left( \dfrac{c}{v}%
\right) ^{2}=\left( \dfrac{c}{v^{2}}\right) ^{2}[\dot{x}^{2}+\dot{y}^{2}+%
\dot{z}^{2}]\text{ \ or }\dot{x}^{2}+\dot{y}^{2}+\dot{z}^{2}=v^{2} 
\tag{4.16b}
\end{equation}%
By combining Eq.s (4.5b) and (4.16b) we obtain:%
\begin{equation}
a^{2}+b^{2}+c^{2}=v^{2}(x_{0},y_{0},z_{0})  \tag{4.17}
\end{equation}%
This result is holding on $\Gamma $ along with Eq.s(4.5a). Therefore $%
V(x_{0},y_{0},z_{0};x_{0},y_{0},z_{0})=0=\tilde{f}(\xi ,\eta ).$ Let $%
t_{0}=0 $ in Eq.(4.3a) and rewrite Eq.(4.3a) as 
\begin{equation}
\tilde{f}(\xi ,\eta ;x,y,z)-ct=0  \tag{4.3.d}
\end{equation}%
so that at $t=0$ the characteristic surface $\Gamma $ is $\tilde{f}(\xi
,\eta )=0$. During the time evolution each point ($\xi ,\eta )$ of $\Gamma $
\ creates its own wavefront according to Eq.(4.3b). Thus, a \ two-parameter
set of wavefronts is obtained. \textsl{The essense of Huygens' principle
lies in demonstration that the envelope of all these wave fronts \ is again
a wavefront }\ $\psi (x,y,z)-ct=0$ \textsl{such that for} $t=0$ we reobtain $%
\psi (x,y,z)\mid _{t=0}=\tilde{f}(\xi ,\eta ).$ \ The procedure for finding
an envelope is reduced to eliminating the parameters ($\xi ,\eta )$ from
three equations 
\begin{equation}
\tilde{f}_{\xi }(\xi ,\eta ;x,y,z)=0,  \tag{4.18a}
\end{equation}%
\begin{equation}
\tilde{f}_{\eta }(\xi ,\eta ;x,y,z)=0,  \tag{4.18b}
\end{equation}%
\begin{equation}
\tilde{f}(\xi ,\eta ;x,y,z)-ct=0.  \tag{4.18c}
\end{equation}%
Suppose that 
\begin{equation}
\xi =A(x,y,z),\eta =B(x,y,z)  \tag{4.19}
\end{equation}%
are calculated from Eq.s(4.18a) and (4.18b), respectively, then $\tilde{f}%
(A(x,y,z),B(x,y,z);x,y,z)$ is a solution of Eq.(4.11). Indeed, 
\begin{equation}
\tilde{f}_{x}=\tilde{f}_{\xi }A_{x}+\tilde{f}_{\eta }B_{x}+\tilde{f}_{x}%
\text{ }=\tilde{f}_{x}  \tag{4.20}
\end{equation}%
in view of Eq.s(4.18a),(4.18b). Analogously, $\tilde{f}_{y}=\tilde{f}_{y}$
and $\tilde{f}_{z}=\tilde{f}_{z}$ implying that Eq.(4.11) holds. Thus, the
surface (the characteristic) $\Gamma :$ $0=\tilde{f}(\xi ,\eta )$ \ at time $%
t$\ is converted into surface described by Eq.(4.3d). Evidently, since $%
\tilde{f}(\xi ,\eta ;x,y,z)=$ $\tilde{f}(A(x,y,z),B(x,y,z);x,y,z)$ it is
obeying the same H-J equation (4.11) in view of Eq.(4.20) (and those for y
and z components). \ As we know already (e.g. read section 1), \ the essence
of Huygens' principle for Schr\"{o}dinger is the H-J Eq.(4.11). With
explanations just made superimposed with those in [60], it should be clear
to our readers that this is indeed the case.\bigskip

\textsl{4.2.3. \bigskip\ Progressive \ wave solution method by Friedlander.
From \ mechanics of Bohm}

\ \ \ \ \ \ \ \ \ \textsl{\ to gauge-theoretic mechanics of Floer\bigskip }

Results of previous subsection were obtained under the assumption that $%
u(x,y,z)$ in Eq.(4.8a) is a constant. If it is not a constant, situation
becomes \ much more complicated mathematically [$65],[70$]. \ To decide what
to do with these complications physically requires some work. First, we
notice that the H-J equation was used in the 1st paper by Schr\"{o}dinger on
quantum mechanics [1], pages 1-12, as an input for obtaining the stationary
Schr\"{o}dinger equation. \ Although his method of deriving this equation
was endorsed by Courant and Hilbert [2], pages 445-450, to our knowledge, it
was left without attention in physics literature. For the sake of results we
shall develop momentarily, we are going to reproduce the 1st Shr\"{o}dinger
method now. For this purpose, we should notice that: a) the ansatz,
Eq.(4.8a), was made without account of dimensionality arguments; b) when the
dimensionality arguments are taken into account, the H-J equation, e.g.
Eq.(4.11), becomes a simple statement about the classical momentum of the
particle\footnote{%
Following Schr\"{o}dinger, and for the sake of argument, we are discussing
only the one and two-body problems.}%
\begin{equation}
\left( \frac{\partial S}{\partial x}\right) ^{2}+\left( \frac{\partial S}{%
\partial y}\right) ^{2}+\left( \frac{\partial S}{\partial z}\right)
^{2}=2m(E-V).  \tag{4.21}
\end{equation}%
Schr\"{o}dinger makes the following\textsl{\ reversible} substitution $%
S\rightleftarrows $ $\hbar $ln$\psi $\textsl{\ }into Eq.(4.21) resulting in
his Eq.($1^{\prime \prime }$) (ours Eq.(1.4)). For our readers convenience
we rewrite it here again: 
\begin{equation}
\left( \frac{\partial \psi }{\partial x}\right) ^{2}+\left( \frac{\partial
\psi }{\partial y}\right) ^{2}+\left( \frac{\partial \psi }{\partial z}%
\right) ^{2}=\frac{2m}{\hbar ^{2}}(E-V)\psi ^{2}.  \tag{4.22}
\end{equation}%
Because of noticed reversibility, this is still the classical H-J equation,
even though it has $\hbar $ in it. It is \textsl{exactly} the same as
Eq.(4.21). \ Next, using Eq.(4.22), instead of Eq.(4.21), Schr\"{o}dinger \
considers the following optimization problem: \ Find the minimum of the
functional 
\begin{equation}
J[\psi ]=\frac{1}{2}\dint d^{3}x[(\mathbf{\nabla }\psi )^{2}-\frac{2m}{\hbar
^{2}}(E-V)\psi ^{2}]  \tag{4.23}
\end{equation}%
under the subsidiary condition 
\begin{equation}
\dint d^{3}x\psi ^{2}=1.  \tag{4.24}
\end{equation}%
The result of such a minimization is the stationary Schr\"{o}dinger
Eq.(1.4). By design, any solution of \ Eq.(1.4) should be a minimum of $%
J[\psi ].$If \ $\psi $ coming \ as solution of Eq.(1.4) is such that $J[\psi
]=0,$ then it is describing the \textsl{classical trajectory} according to
Eq.(4.22) since the substitution $S\rightleftarrows $ $\hbar $ln$\psi \ $is
reversible$.$ But, in view of results of previous subsection, this is indeed
the case! E.g. solutions of Eq.(4.11) are solutions of Eq.(4.10) if \ in the
ansatz Eq.(4.8a) we replace $c$ by $\omega $ and take into account Eq.s
(2.8), (2.9). This observation allows us to reinterpret variational results
of the 1st Schr\"{o}dinger paper in terms of the gauge-theoretic formalism
developed by Floer. A quick introduction to Floer's theory is provided in
[71] while the detailed account can be found, for instance, in [72]. Even
though Floer theory studies a multitude of closed orbits on symplectic
manifolds classically, the \ extremely sophisticated computational methods
developed by Floer \ exactly parallel those used in nonperturbative
(instanton) treatments of the Yang-Mills theory. These results establish a
connection between \ the non relativistic Schr\"{o}dinger equation (with
time-independent Hamiltonian) and the gauge-theoretic Yang-Mills-type
theory. We shall say more on this subject in subsection 7.4.

If $u(x,y,z)$ in Eq.(4.8a) is not a constant then, at the very least, we end
up with the formalism of quantum mechanics developed by David Bohm [73]. See
also Appendix C. It also can be considered as classical because quantum
mechanical corrections in Bohmian mechanics are\ treated by methods of
classical mechanics (in fact, of hydrodynamics). Bohm formalism is based on
representation given by Eq.(4.8a) with an extra restriction $%
F(ct-f(x,y,z))\rightarrow \exp \{\frac{i}{\hbar }(Et-S(x,y,z))\}.$ It
remains to be investigated \ how results of Bohm formalism \ might change if
Eq.(4.8b) is used instead. This \ task is left for further study. \bigskip

\textsl{4.2.4. } \ \textsl{Progressive \ wave solution method by
Friedlander. Dupin cyclides\bigskip , the Lie}

\ \ \ \ \ \ \ \ \ \ \ \textsl{sphere geometry \bigskip and the conformal
group SO(4,2)}

From what was discussed thus far it follows that both the hyperbolic
(wave-like) and the parabolic (Schr\"{o}dinger-like) equations \ can be
treated with help of the progressive wave solution method. \ If this is so
then, according to Friedlander [70], the most general solution in both cases
should be expressible in terms of cyclides of Dupin. For the hyperbolic
PDE's this was demonstrated by Friedlander [65,70]. In this subsection, and
also in subsection 7.5.3. and Appendix C, we \ shall demonstrate that this
is also true for the stationary Schr\"{o}dinger equation. By doing so, we
shall put the obtained results in a much broader context. \ This context
will allow us to find \ a place for (thus far) very exotic \ Dupin cyclides
in atomic, high energy physics and cosmology.

\ To begin, we notice that 1946 result by Friedlander [70] was reconsidered
in 2005 by Sym [74]. \ Not only he reobtained it in a much shorter \ and
simpler form in ref.s[74,75] but, in addition, he was able to find some (not
serious) mistakes in the original work. \ Both Friedlander and Sym were
solving the system of \ Eq.s(4.9a-c). Sym very cleverly used symmetry to
solve these equations. His solution strategy can be summarized as follows.
First, solve Eq.(4.9a) by cleverly using the symmetry built into the Dupin
cyclides. Second, to take the full advantage of this symmetry, rewrite
Eq.s(4.9a-c) in curvilinear (actually geodesic) coordinates reflecting this
symmetry. The choice of coordinates in Eq.s(4.9b,c) is determined by the
fact that Eq.(4.9a) should \ admit a natural solution \ respecting \ the
symmetry of Dupin cyclides. Third, use this solution in Eq.(4.9b) and insure
that Eq.(4.9c) is solved \ in such coordinates as well.

\ Being armed with these results, we apply them to the Helmholtz Eq.(3.2) \
which is obtainable from both the D'Alembert and the free particle Schr\"{o}%
dinger equations. Therefore, the results of separation of variables in
Eq.(4.9c) can be used for solving the Helmholtz equation as well. The next
step is made by checking \ what \ kinds of potentials \ in the stationary
Schrodinger's \ equation \ allow the full separation of variables with help
of these geodesic coordinates. This problem is non trivial. It was studied,
for example, in [76,77]. We shall not go into full details, however, since
we \ shall develop much broader vision of \ the role of Dupin cyclides in
such kind of problems which, in addition, will make such calculation much
simpler and much more physically appealing.

For this purpose we need to explain: a) What are the Dupin cyclides? b) How
their presence/absence affects the results of Schr\"{o}dinger and Bohm? c)
What happens to the Huygens principle if Dupin cyclides are taken into
account? \ The answer to b) is given in part in the already cited [76,77].
But it is also given in Apendices B and C from the entirely different
standpoint. The positive answer to c) follows in part from works by
Sym[74,75] just mentioned and also follows from results of the Appendix B.
Therefore, we need only to provide the answer to a). The answers to b) and
c) are provided in just mentioned appendices and in section 7. The answer to
a) \ is provided immediately below.

We begin with ref.[78]. It\ is associated with the notion of \textsl{canal
surface. } Such a surface \ can be designed as follows. Choose some sphere $%
S_{c,r}(t)$ \ whose radius $\mathbf{r}(t)$ is changing in time. The \ the
center of $S_{c,r}(t)$ is moving along some curve (the trajectory) $\mathbf{%
c(}t\mathbf{)}$ parametrized by time $t$ so that such a motion is described
analytically as 
\begin{equation}
F(\mathbf{x},t)=\left\Vert \mathbf{x}-\mathbf{c}(t)\right\Vert ^{2}-\mathbf{r%
}^{2}(t)=0.  \tag{4.25}
\end{equation}

\textbf{Definition 4.4.} \textsl{Canal surface. } The canal surface $\Sigma
_{c,r}\subset \mathbf{R}^{3}$ is an envelope of a 1-parameter family $%
S_{c,r}(t)$ of spheres centered at the \textsl{spine curve} $\mathbf{c}(t)$%
.\bigskip

The envelope is being defined as joint solution of two equations%
\begin{eqnarray}
F(\mathbf{x},t) &=&0,  \TCItag{4.26a} \\
\frac{\partial F(\mathbf{x},t)}{\partial t} &=&<\mathbf{x}-\mathbf{c}(t),%
\mathbf{\dot{c}}(t)>+\mathbf{r}(t)\cdot \mathbf{\dot{r}}(t)=0. 
\TCItag{4.26b}
\end{eqnarray}%
Here $<,>$ is the scalar product in $\mathbf{R}^{3}.$ The canal \ surface $%
\Sigma _{c,r}$ is fully determined by a curve in 4 dimensional Minkowski
space \textbf{R}$^{3,1}.$ Clearly, every point $(\mathbf{c}(t),\mathbf{r}%
^{2}(t))$ of \textbf{R}$^{3,1}$ corresponds to a sphere $S_{c,r}(t).$
\bigskip Dupin cyclides are made \ of canal surfaces but not \ another way
around. This follows from the ingenious observation by J.C. Maxwell who
constructed some of the first Dupin cyclides following the orininal work by
Chales Dupin done in 1803. \ \ In 1868 Maxwell noticed that, even though \
every point $(\mathbf{c}(t),\mathbf{r}^{2}(t))$ \ belongs to a sphere $%
S_{c,r}(t)$ , \ there could be \ two different sets: a) \textbf{c}$_{1}$ and 
\textbf{r}$_{1},$ b)$\ $and \textbf{c}$_{2}$ and \textbf{r}$_{2}$ producing
the same canal surface\bigskip .

\textbf{Dedinition 4.5 \ a). }\textsl{Dupin cyclides} are the only canal
surfaces\textbf{\ }which can be designed in two different ways.

\bigskip

Consider the simplest example-the torus

a) the spine curve $c_{1}(t)=a(\frac{1-t^{2}}{1+t^{2}},\frac{2t}{1+t^{2}}%
,0), $ the radius function $\ r_{1}(t)=c;$

b) the spine curve $c_{2}(t)=a(0,0,\frac{2t}{1-t^{2}}),$ \ \ \ \ the radius
function $\ r_{2}(t)=c-a\frac{1+t^{2}}{1-t^{2}}.$

Using of a) and b) in Eq.s(4.26a,b) and eliminating $t$ leads \ to the
following quadric (describing the torus-the canal surface)%
\begin{equation}
(x^{2}+y^{2}+x^{2}+a^{2}-c^{2})^{2}-4a^{2}(x^{2}+y^{2})=0.  \tag{4.27}
\end{equation}%
From this elementary example the following general definition follows\bigskip

\textbf{Definition 4.5 b). }Dupin cyclides are the surfaces \ whose lines of
curvatures are circles\textbf{. }Elementary examples of (degenerate) Dupin
cyclides include \ spheres, planes \ (that is spheres of infinite radius),
toruses and cones of revolution.

\bigskip

The above results admit generalization to higher dimensions [79]. More on
this will be said in section 7. In the meantime, we would like to notice the
following. \ An affine isometry \textbf{f }of \textbf{R}$^{n}$ is described
as follows. Let \textbf{x}$\in $\textbf{R}$^{n}$, then \textbf{f}(\textbf{x}%
)=\textbf{Qx}+\textbf{b }where\textbf{\ Q} is $n\times n$ orthogonal matrix
(that is \textbf{QQ}$^{T}$=\textbf{1}) whose entries belong to \textbf{R}, 
\textbf{b }$\in $\textbf{R}$^{n}.$ \bigskip

\bigskip \textbf{Definition 4.6. }It \textbf{f}(\textbf{c})=\textbf{c }\
(respectively \textbf{f}($\Sigma _{c,r})=\Sigma _{c,r}$ ), then \textbf{f }%
is called \textsl{symmetry} of \textbf{c} (respectively of $\Sigma _{c,r})$

To get a feeling of this symmetry it is sufficient to consider the space 
\textbf{R}$^{2}.$ Since the canal surfaces are made of circles moving along
the spine curve, the symmetry transformations should convert circles into
circles. This is possible with help of the M\"{o}bius (or conformal)
transformations $f(z)=\frac{az+b}{cz+d},$ $z\in \mathbf{C,}$ \textbf{C}=%
\textbf{R}$^{2}\cup \{\infty \}.$ It is clear then, that the transformation 
\textbf{f}($\Sigma _{c,r})=\Sigma _{c,r}$ is a conformal transformation. \
It is possible to extend these ideas to higher dimensions naively. This will
yield the multidimensional M\"{o}bius transformations. These are not quite
yet the transformations having the Dupin cyclides unchanged. They form a
subgroup of the Lie sphere geometry group [80] which by design leaves the
Dupin cyclides unchanged. In two dimensions the M\"{o}bius group is \ the
isometry group of the Poincare$^{\prime }$ hyperbolic upper half plane. The
interrelationship between the hyperbolic and the Minkowski spaces is nicely
described in the book by Ratcliffe [81]. From it we find that the M\"{o}bius
group is also group of isometries of the Minkowski spacetime. In sections
6,7 we shall demonstrate that the group of isometries of the compactified
Minkowski spacetime is the conformal group SO(4,2). It is described, for
example, in [19], pages 345-348. The significance of this group was noticed
already in section 1. In section 7 we shall demonstrate that this group
coincides with the Lie sphere group. This observation \ is of fundamental
physical significance to be explained in section 7 and the Appendix B. In
the meantime we would like to explain the differences between the M\"{o}bius
and the Lie sphere groups. Since the M\"{o}bius group is the subgroup of the
Lie sphere group, the additional symmetry elements originate from the fact
that a) spheres could have an extra label- \textsl{orientation}. This can be
seen already in dimension two. A circle \ can have the radius -vector with
the base sitting on the circle and directed toward the center of circle.
Another option is \ for the same vector to play the role of the outward
normal to the circle. Besides, one should pay attention to the fact that the
circles can touch each other and each circle could have an infinite family
of circles sitting inside or outside of the given one and touching it at the
same point. To these objects one should add points-circles of zero radius
and, also -lines, the circles of infinite radius (in two dimensions) or the
planes in dimensions higher than two. According to Klein's Erlangen program,
the following definition is \ the most appropriate.\bigskip

\textbf{Definition 4.7.} The essence of Lie sphere geometry lies in the
study of properties of transformations mapping oriented spheres (including
points and planes) to oriented spheres \ while preserving the oriented
contact of sphere pairs.\bigskip

\textbf{Corollary 4.8}. Dupin cyclides are invariants of the Lie sphere
geometry.\bigskip

\textbf{Corollary 4.9}. The conformal group SO(4,2) was used in atomic [$19$%
]\ and in high energy physics [$83$] without any reference to the Lie sphere
geometry symmetry group. It was discussed already in section 1. Use of
conformal group caused introduction of the unphysical two-times formalism in
[$83$]. Use of Lie sphere geometry removes this deficiency. In the rest of
this paper we \ demonstrate how the conformal group SO(4,2) is identified
with the Lie sphere geometry group. This identification opens the door for \
new results in conformal dynamics, conformal quantum mechanics, conformal
quantum field theory, conformal gravity, etc. [$84$]. The AdS-CFT
correspondence to be discussed in section 7 and Appendix D is part of this
"conformal program". It is important to remember that the source-free
Maxwellian electrodynamics is invariant \ with respect to SO(4,2) group as
well [$82$] while the results of conformal quantum mechanics recently were
utilized with great success for recovery of the Regge-like spectrum of
hadrons [$85$].

\bigskip

\textbf{5. \ Closer look at the Hadamard conjecture. Recovery of some known}

\ \ \ \ \ \textbf{equations for massive/massless particles with or without
spins \bigskip \medskip }

In this paper we are not going to discuss the developments associated with
counterexamples to the Hadamard conjecture. All these counterexamples \ are
discussing the validity of the Huygens' principle in spacetimes of
dimensionality higher than four [$68]$. In the light of results of section 1
use of spacetimes of dimensionality higher than four for multielectron atoms
and \ for molecules is also \ not necessary, apparently. Fortunately, in
four dimensional spacetimes no counterexample to the Hadamard conjecture was
found. Results of Appendix B should be considered as a cosmologically
inspired ramification\footnote{%
Associated with accounting for Penrose limits of \ physical spacetimes.} of
\ the \ already known results. Because of this, it is convenient to restate
the Theorem 4.1. as \bigskip follows

\textbf{Fundamental Principle 5.1.} \textit{Quantum mechanical behavior of \
all elementary particles (massive and massless, of integer and half integer
spin) is inseparably linked with \ the Lorentzian \ signature \ of ambient
spacetimes.\bigskip }

\textbf{\ \ Conjecture} \textbf{5.2.} \textit{Although not immediately
obvious}, \textit{the} \textit{Fundamental Principle is synonymous with the
central role of the Lie sphere geometry \ acting in \ conformally flat
spacetimes of Lorentzian signature. Its influence on physics is ranging from
conformal mechanics to conformal wave mechanics and conformal gravity.} 
\textit{Group-theoretical} \textit{classification of physically sensible
spacetimes is contingent upon their ability to sustain quantum mechanics%
\footnote{%
Further details are presented in section 7 and Appendix B.}.\bigskip
\bigskip }

\textsl{5.1. Hadamard triviality and mass generation. Panoramic view\bigskip 
}

We continue this section with demonstration of Huygens' triviality of
telegrapher's and the Klein-Gordon equations. Using Eq.(2.6) and the
equivalence condition \ \~{L}[$\phi ]=\lambda ^{-1}$L[$\lambda \phi ]$ with $%
\lambda =e^{at},$ where $a$ is some constant, we obtain ($c=1$):%
\begin{equation}
e^{-at}\{[\frac{\partial ^{2}}{\partial t^{2}}e^{at}\phi ]-[\nabla
^{2}e^{at}\phi ]\}=\frac{\partial ^{2}}{\partial t^{2}}\phi +2a\frac{%
\partial }{\partial t}\phi -\nabla ^{2}\phi +a^{2}\phi =0.  \tag{5.1}
\end{equation}%
Next, let $\lambda _{1}=$e$^{ibx}$ and $\lambda _{2}=$e$^{-icx}.$ Substitute
these factors into Eq.(5.1) \ and apply\ again Hadamard's equivalence rules.
After a short calculation we arrive at 
\begin{equation}
\frac{\partial ^{2}}{\partial t^{2}}\phi +2a\frac{\partial }{\partial t}\phi
-\nabla ^{2}\phi +2ib\frac{\partial }{\partial x}\phi -2ic\frac{\partial }{%
\partial x}\phi \text{ }+\text{ }\left( a^{2}-b^{2}-c^{2}\right) \phi =0. 
\tag{5.2a}
\end{equation}%
If now $b=c$ and $a^{2}=2b^{2}$, we obtain telegrapher's equation%
\begin{equation}
\frac{\partial ^{2}}{\partial t^{2}}\phi +2a\frac{\partial }{\partial t}\phi
-\nabla ^{2}\phi =0.  \tag{5.2b}
\end{equation}%
One dimensional version of this equation is discussed at length in the book [%
$86].$ \ In one and two dimensions this equation admits the path integral
treatment. \ In the meantime, the Klein-Gordon (K-G) equation is obtained
now if we make a replacement $\phi =e^{at}\psi $ in Eq.(5.2b) with
subsequent replacement of $a$ by $ia$. After this, we obtain:%
\begin{equation}
\frac{\partial ^{2}}{\partial t^{2}}\psi -\nabla ^{2}\psi +a^{2}\psi =0. 
\tag{5.3}
\end{equation}%
Thus, we just demonstrated that the K-G equation is Huygens-equivalent to
the D'Alembert Eq.(2.6a). According to $[87$], page 99, every spinor
component of the Dirac equation with nonzero mass is satisfying the K-G
equation. This fact establishes the Huygens equivalence between the Dirac
and D'Alembert equations. Apparently, the particles with higher spin, e.g.
spin-2 gravitons, etc. also belong to the same Huygens equivalence class [$%
87-89$].\bigskip

\textbf{Corollary 5.3. }Huygens\textbf{\ }equivalence between the D'Alembert
and all relativistic equations of integer and half integer spin explains why
the double slit experiments made with photons and massive particles produce
the same fringe patterns. \bigskip More details on this is given in ref.[14].

Thanks to the seminal work by Mark Kac [$90]$ the propagator for
telegrapher's and Dirac's equations can be presented in the path integral
form at least in 1+1 dimensions [$91$].\ In view of Eq.s (5.2b) and (5.3)
the same is true for the K-G equation as explained in detail in [$92$].

It should be noted that in all these cases the associated path integrals\ do
not involve the Gaussian-type random processes. They are designed with help
of the Poissonian- type random processes. Excellent description of \ these
types of processes in conjunction with the telegrapher's equation is given
in [$86$]. \ The noticed connection with random walks is helpful but not
crucial for the tasks to be completed in the rest of this paper.
Furthermore, the above path integrals can be designed rigorously only in 1+1
and 2+1 dimensions. In [$93$] it was rigorously demonstrated that these
results cannot be extended to higher dimensions.

To stay focused, we \ are not going \ to discuss any further the connection
between the random walks of various kinds and PDE's. Instead, we notice the
following

\textbf{First}. \ The diffusion/Schr\"{o}dinger is not the wave-type
equation studied by Hadamard. The diffusion/Schr\"{o}dinger equation
describes the dispersive waves discussed in the Appendix A. Therefore,
contrary to Feynman's claims made in [$4$], it \ apparently does \textsl{not}
obey the Huygens principle. However,\ the results of sections 4 and 6 \
indicate \ that this apparent deficiency of the Schr\"{o}dinger equation can
be repaired, and quite rigorously, for as long as the quantum Hamiltonian of
this equation is manifestly time-independent. What remains to be proven is
hyugens triviality of Schr\"{o}dinger's equation with time-independent
Hamiltonian.

\textbf{Second}. Initially, Hadamard obtained his results in 3+1 dimensions
as explained in previous section. Subsequently, he developed the \textit{%
method of descent} \ allowing use of 3+1 dimensional results as an input for
obtaining 2+1 dimensional solutions. Still later, these results were
extended by others to 1+1 dimensions, \ again with use of the method of
descent,[$94$], pages 315, 316.

\textbf{Third}. Clearly, by going down from 3+1 to 1+1 dimensions some
information is lost. Otherwise the distinction between, say, 2 and 3
dimensional wave propagation \ is going to disappear. But it is not!
Therefore, the attempt to use the method of descent in reverse cannot help
us in extending 2+1 dimensional results obtainable with help of, say, the
Poissonian statistics to 3+1 dimensions. Other methods, e.g. those using
Grassmann variables should be used instead [$95$]. Different method, also
using Grassmann variables but employing differential-geometric
considerations for 3 dimensional paths evolving in 3+1 dimensional spacetime
(leading to 3+1 Dirac propagator) was developed in [$96$]. In connection
with [$96$], the following observation is appropriate.

\textbf{Forth.} \ The Standard Model of particle physics uses\ the widely
accepted Higgs mechanism responsible for the mass generation. At the same
time, the twistor formalism describing all massless particles had been
extended recently by accounting for the rigidity of the world-lines of the
massless particles [$97$]$.$ Such differential-geometric mass generation
method is analogous to that proposed in [$96$]. The latest results \ in this
direction can be found in [$98$].

\textbf{Fifth. }After Eq.(5.3) we stated that the K-G, the Dirac and other
basic equations for massive particles can be made Huygens-trivial using
Hadamard's transformation rules. Since establishing \ of this equivalence is
reversible process, this means that it is possible to avoid use of the Higgs
mechanism. The two-time formalism developed by Itzhak Bars (summarized in [$%
83$]) leads to the same conclusions. Because the \ De Broglie relation is
valid for massless photons as well as for the massive particles as
demonstrated in the footnote 6 and because the stationary Schr\"{o}dinger
equation can be restored from the relativistically obtained uncertainty
relations (as demonstrated by Schrodinger (e.g. read section 2)) the notion
of mass should be linked with the De Broglie relation. Its actual value is
controlled by the rigidity of world lines [97],[98]\footnote{%
The latest results on this topic are mentioned in the "Note added in proof"}.

In view of these remarks, our first task is to address and to solve the
problem formulated in the first remark. In section 3 \ we provided the
following Schr\"{o}dinger's comments : " Thus, when we designated equation
(1) or ($1^{\prime }$)\footnote{%
Our Eq.(4.22).} on various occasions as "the wave equation", we were really
wrong..." Well, actually, Schr\"{o}dinger was not at all wrong! Eq.(1.4) was
used systematically by Schr\"{o}dinger himself in Parts I-III of [$1]$ to
solve the Hydrogen atom, the harmonic oscillator, the rigid and non rigid
rotators and the Stark effect quantum mechanically. He also developed the
perturbation theory based on the exact results which he obtained. All these
results are correct! \ In section 3 we explained what made Schr\"{o}dinger
unhappy with his "amplitude equation", that is with our Eq.(1.4). In section
4 \ using methods of characteristics and progressive waves we demonstrated
that Eq.(1.3) is indeed the correct wave equation in the sense of De Broglie
[$35]$ obeying the Huygens' principle in the sense of Hadamard. Furthermore,
in section 6 \ we demonstrate that for the\ time-independen Hamiltonians
Schrodinger's "\textbf{real} \textbf{wave equation}" \ can be embedded into
the Hadamard scheme of calculations relating a given 2nd order PDE with the
wave Eq.(2.6a).

In conclusion, we would like to mention that by using the two-time formalism
Itzhak Bars established the equivalence (in the sense of gauge
equivalence(duality) defined in his paper) between the D'Alembert Eq.(2.6a),
on one hand, and the Schr\"{o}dinger equations for the hydrogen atom and the
harmonic oscillator, on another [$99$]. In [99] Bars acknowledged that he
was not able yet to provide a complete classification of all
gauge-equivalent(dual) quantum mechanical systems originating from the same
(gauge-invariant) model. The attempt to do so was made later, in [$100$].
However, making choices between the two times in such a formalism still
remained mysterious. Below, we shall obtain the same results using entirely
different methods enabling us to avoid the two-times formalism altogether.

\bigskip

\textbf{6. \ Huygens triviality of the stationary Schr\"{o}dinger
equation\bigskip }

\bigskip \textsl{6.1. \ Some comments about 1935 work by V. Fock on hydrogen
atom\bigskip }

On February 8, 1935, Vladimir Fock presented his seminal lecture entitled
"On theory of the hydrogen atom" \ at the theory seminar of the Leningrad
State University. English translation \ of his talk can be found in [101].
It is based on Fock's article [102] published in German.

To begin our comments on his paper we rewrite Eq.(1.4) in the following
standard form%
\begin{equation}
-\frac{\hbar ^{2}}{2m}\nabla ^{2}\psi +V\psi =E\psi .  \tag{6.1a}
\end{equation}%
Without loss of generality we shall consider only potentials for which the
above equation is exactly solvable. At the classical level use of canonical
transformations makes all such exactly solvable problems equivalent since
they all can be brought into standard action -angle form known for the
harmonic oscillator Accordingly, it is sufficient to consider only the
hydrogen atom problem for which in known system of units the potential $V(%
\mathbf{r})$ is defined as $V(\mathbf{r})=-\frac{Ze^{2}}{r}\equiv -\frac{k}{r%
}.$ For this (Coulombic) potential Fock replaces the partial differential
Eq.(6.1a) by the equivalent integral equation 
\begin{equation}
(\frac{\mathbf{p}^{2}}{2m}-E)\psi (\mathbf{p})=\frac{k}{2\pi ^{2}h}\int 
\frac{d^{3}\mathbf{p}^{\prime }\psi (\mathbf{p}^{\prime })}{\left\vert 
\mathbf{p}-\mathbf{p}^{\prime }\right\vert ^{2}}.  \tag{6.1b}
\end{equation}%
Citing Fock, the rationale for using the integral equation method instead of
solving the partial differential equation is caused by the following
observations.

" It has long been known that the energy levels of the hydrogen atom are
degenerate with respect to the azimutal quantum number \textit{l}... But any
degeneracy of eigenvalues is linked to the transformation group of the
relevant equation: e.g. the degeneracy with respect to the magnetic quantum
number $m$ is attributed to the usual rotational group." Here Fock refers to
the degeneracy of the following type. For a given\textit{\ l} (associated
with the rigid rotator energy \textit{l}($\mathit{l}$+1)) one has 2\textit{l}
+1 wavefunctions labeled by the magnetic number $m:-\mathit{l}\leq m\leq 
\mathit{l}.$ He calls such a degeneracy "\textit{accidental}" and in his
paper he finds the symmetry group causing this accidental degeneracy. He
demonstrated that the group causing accidental degeneracy is
four-dimensional rotational group $SO(4)$.

Instead of copying \ Fock's arguments, we shall arrive at the same results
much more economically. For this purpose we introduce the notations:%
\begin{eqnarray}
\text{H} &=&\frac{\mathbf{p}^{2}}{2m}-\frac{k}{r};  \TCItag{6.2a} \\
\mathbf{L} &=&\mathbf{r}\times \mathbf{p;}  \TCItag{6.2b} \\
\mathbf{A} &=&\mathbf{p\times L-}mk\mathbf{\hat{r},}\text{ \ }\mathbf{\hat{r}%
=}\frac{\mathbf{r}}{r}.  \TCItag{6.2c}
\end{eqnarray}%
It can be easily demonstrated that in addition to the Hamiltonian H which is
constant of motion, both the angular momentum \textbf{L} and the
Laplace-Runge-Lentz vector \textbf{A} are also constants of motion [101].
These objects can be looked upon either at the classical level with
subsequent quantization or at the quantum level. The last route was chosen
initially by Pauli [$103$]. The results in both classical and quantum cases
depend upon the value of the energy constant $\ E=$H$:$ $E>0,$ $\ E<0$ and $%
E=0$. In this work the case $E=0$ will not be considered since it is not
related to the tasks we would like to accomplish. \ Traditional analysis
involving bound orbits/states typically begins with the case $E<0$. \ In
this case it is more convenient to replace \ the vector \textbf{A} with the
rescaled vector \textbf{D} defined as ([104], page 421): 
\begin{equation}
\mathbf{D}=\frac{\mathbf{A}}{\sqrt{-2mE}}\mathbf{.}  \tag{6.3a}
\end{equation}%
In the case of $E>0$, the vector \textbf{D} is defined accordingly as 
\begin{equation}
\mathbf{D}=\frac{\mathbf{A}}{\sqrt{2mE}}.  \tag{6.3b}
\end{equation}%
In terms of such notations the Poisson brackets $\{,\}$ commutation
relations are readily obtained with the result [$104$]%
\begin{eqnarray}
\{L_{i},L_{j}\} &=&\varepsilon _{ijk}L_{k},  \TCItag{6.4} \\
\{D_{i},L_{j}\} &=&\varepsilon _{ijk}D_{k},  \notag \\
\{D_{i},D_{j}\} &=&\varepsilon \varepsilon _{ijk}L_{k}.  \notag
\end{eqnarray}%
Here $\varepsilon =1(E<0),\varepsilon =-1(E>0).$ Using this Poisson algebra,
the respective Lie algebra is obtained via standard Dirac quantization
prescription. As result, the so(4) Lie algebra (for $E<0$) and \ so(3,1) Lie
algebra (for $E>0$) is obtained. The first one is the Lie algebra of the
SO(4) rotation Lie group while the second is the Lie algebra of the SO(3,1)
Lorentz group. In his talk Fock did mention the Lorentz group but provided
no computational details.

\bigskip

\textsl{6.2. \ Huygens' triviality \ of the stationary Schr\"{o}dinger
equation in the light}

\textsl{\ \ \ \ \ \ \ of Fock's work\bigskip }

\textsl{6.2.1. \ General consideration}

\bigskip

Just obtained results, when superimposed with results by\ Fock, are
sufficient for group-theoretical proof of Huygens' triviality of the
stationary Schr\"{o}dinger equation. Nevertheless, it is instructive to
arrive at the final destination via extremely informative detour. This
detour is possible to perform \ by using, for example, fundamental work by
Gelfand, Milnos and Shapiro [$105$]. Alternatively, the same results can be
obtained \ using the Duffin-Kemmer formalism \ [$14$]. The authors of \ [$%
105 $] discussed carefully the problem of classification of all
relativistically invariant equations. That is of all Lorentz-invariant
equations. \ Clearly, the protocol of study of these equations \ is the same
as that for study of rotationally invariant equations. Therefore, we shall
study both problems simultaneously. The study begins with the equation of
the type 
\begin{equation}
\dsum\limits_{i=0}^{n}g_{ij}\frac{\partial ^{2}\psi }{\partial x_{i}\partial
x_{j}}+m^{2}\psi =0,  \tag{6.5a}
\end{equation}%
where $n+1$ is the dimensionality of spacetime, $m$ is the mass parameter,
and the metric tensor $g_{ij}$ is given by 
\begin{eqnarray}
g_{ij} &=&\delta _{ij}\text{ (rotational group); }  \notag \\
g_{ij} &=&1(i,j,=0),g_{ij}=-1(i,j,\neq 0,i=j),g_{ij}=0,i,j\neq 0,i\neq j.%
\text{ (Lorentz group).}  \TCItag{6.5b}
\end{eqnarray}%
Typically, $n=3$ \ but other dimensionalities could be considered as well. \
With such defined metric tensor, the 2nd order PDE, Eq.(6.5a), \ is
convenient to rewrite in the form of the system of the 1st order PDE's.
Incidentally, in such a case, one can think about the method of
characteristics for such PDE's and the H-J equations, etc.[$29$]. We shall
not touch this topic in this section though since the relevant information
was already provided in section 4. Instead, following [$105$], we introduce
an auxiliary functions $\psi _{i}$ via equation%
\begin{equation}
m\psi _{i}=\frac{\partial \psi }{\partial x_{i}},i=0,...,n.  \tag{6.6}
\end{equation}%
A simple calculation produces (for the Lorentzian case, $n=3$)%
\begin{equation}
\dsum\limits_{i=1,2,3}\frac{\partial \psi _{i}}{\partial x_{i}}-\frac{%
\partial \psi _{0}}{\partial x_{0}}-m\psi =0;\frac{\partial \psi }{\partial
x_{i}}=m\psi _{i},  \tag{6.7a}
\end{equation}%
and (for the Euclidean case, $n=3$)%
\begin{equation}
\dsum\limits_{i=1,2,3}\frac{\partial \psi _{i}}{\partial x_{i}}+\frac{%
\partial \psi _{0}}{\partial x_{0}}+m\psi =0;\frac{\partial \psi }{\partial
x_{i}}=m\psi _{i}.  \tag{6,7b}
\end{equation}%
At this point it is convenient to introduce the extended wave vector $\Phi
=\{\psi ,\psi _{0},\psi _{1},\psi _{2},\psi _{3}\}^{T}$ (where $T$ means
"transpose") and to rewrite the system of \ the above equations into the
form 
\begin{equation}
\dsum\limits_{k=0}^{n}\mathcal{L}_{k}\frac{\partial }{\partial x_{k}}\Phi
\pm im\Phi =0.  \tag{6.8}
\end{equation}%
Here $\mathcal{L}_{k}$ are the $5\times 5$ matrices. E.g. in the Lorentzian
case%
\begin{equation}
\mathcal{L}_{0}=\left( 
\begin{array}{ccccc}
0 & i & 0 & 0 & 0 \\ 
-i & 0 & 0 & 0 & 0 \\ 
0 & 0 & 0 & 0 & 0 \\ 
0 & 0 & 0 & 0 & 0 \\ 
0 & 0 & 0 & 0 & 0%
\end{array}%
\right) ,  \tag{6.9}
\end{equation}%
etc. Let $G$ be the matrix of either Euclidean rotations or of Lorentz
transformations acting on coordinates $x_{i}$ as $x^{\prime }=Gx.$ Let $%
T_{G} $ be some unitary operator such that 
\begin{equation}
\Phi ^{\prime }(x^{\prime })=T_{G}\Phi (x)\text{ and }\Phi
(x)=T_{G}^{-1}\Phi ^{\prime }(x^{\prime })  \tag{6.10}
\end{equation}%
Accordingly, Eq.(6.8) can be rewritten \ now as 
\begin{equation}
\dsum\limits_{k,l}\mathcal{L}_{k}T_{G}^{-1}\left( \frac{\partial }{\partial
x_{l}^{\prime }}\Phi ^{\prime }(x^{\prime })\right) G_{lk}\pm
imT_{G}^{-1}\Phi ^{\prime }(x^{\prime })=0  \tag{6.11a}
\end{equation}%
or, equivalently, as 
\begin{equation}
\dsum\limits_{k,l}T_{G}\mathcal{L}_{k}T_{G}^{-1}G_{lk}\left( \frac{\partial 
}{\partial x_{l}^{\prime }}\Phi ^{\prime }(x^{\prime })\right) \pm im\Phi
^{\prime }(x^{\prime })=0.  \tag{6.11b}
\end{equation}%
The condition of invariance follows immediately from Eq.(6.11b) 
\begin{equation}
\dsum\limits_{k}T_{G}\mathcal{L}_{k}T_{G}^{-1}G_{lk}=\mathcal{L}_{l} 
\tag{6.12}
\end{equation}%
\textsl{Since this condition is unchanged in the massless case},\textsl{\ it
is sufficient to consider the massless case only}. Clearly, the simplest
equation invariant with respect to either the Euclidean rotation or Lorentz
group transformations is the four-dimensional Laplacian in Euclidean case or
D'Alembertian \ in Lorentzian case respectively. In his paper [$102$] Fock
obtained indeed the four -dimensional Lapacian equation as an equation whose
solutions are those of the integral Eq.(6.1b). To obtain solutions in the
Lorentzian case requires us only to perform the Wick rotation, that is to
make a replacement: $x_{0}\rightarrow \pm ix_{0}.$ This innocently looking
operation routinely used in physics literature required 126 pages of proof
in mathematics literature [$106-108$]. Thanks to its existence, we are
spared from the necessity to repeat needed proofs. The same results \ could
be obtained group-theoretically via development of the unified treatment of
SO(4) and S(3,1) Lie groups [$109$]. A transparent and motivating example of
such an interrelation is given in the pedagogically written article by John
Milnor [110]. Nevertheless, below we shall provide yet another derivation.
It involves the twistor formalism. Use of twistor formalism enables us to
present Fock's results in a different light. To accomplish this task
requires several steps which we would like to describe now. We begin with
the following.

\bigskip

\textsl{6.2.2. \ Grassmannians and Pl\"{u}cker embedding} \bigskip \bigskip

First of all, we would like to take a careful look at the Hamiltonian H for
the hydrogen atom at the classical level \ 
\begin{equation}
\text{H}=\frac{\mathbf{p}^{2}}{2m}-\frac{k}{r}=E.  \tag{6.13}
\end{equation}%
In particular, let initially $E<0$. This condition connects the momenta and
coordinates. In particular, it should also hold for $r\rightarrow 0.$ In
such a case, to maintain the equality, the momenta should become infinite.
If initially we had $\mathbf{p}=\{p_{x},p_{y},p_{z}\}\in \mathbf{R}^{3}$,
now we must add a point at infinity $\mathbf{p}_{\infty }$ to keep the
relation, Eq(6.13), unchanged. The addition of $\mathbf{p}_{\infty }$ leads
to the compactification of $\mathbf{R}^{3}$ thus converting it to $S^{3}=%
\mathbf{R}^{3}\cup \{\infty \}.$ The $SO(4)$ group is the group of
isometries of $S^{3}[111].$ Next, instead of rescaling the Runge-Lentz
vector \textbf{A} in Eq.(6.3) we can rescale the angular momentum $\mathbf{L}
$. This is effectively done in Fock's paper as we would like to explain now.
For this purpose we follow [55], pages 361-364.

Consider 2 points \textbf{x} and \textbf{y} on the line $\mathcal{L}$ in 3d
space. Since the coordinates $\mathbf{x}=\{x_{1},x_{2},x_{3}\}$ and $\mathbf{%
y=\{}y_{1}\mathbf{,}y_{2}\mathbf{,}y_{3}\mathbf{\}}$\textbf{\ }are \ taken
with respect to some fixed origin\textbf{, }these are vectors. We can
construct from these vectors two other vectors: $\mathbf{Z}=\mathbf{y}-%
\mathbf{x}$ and $\mathbf{L}=\mathbf{x}\times \mathbf{y.}$ The new element is
coming from the following step. We enlarge the embedding into Euclidean
space by increasing its dimensionality, that is by replacing vectors \textbf{%
x} and \textbf{y }by\textbf{\ x}$^{p}$=\{x$_{0}$,\textbf{x}\} and \textbf{y}$%
^{p}$=\{y$_{0}$,\textbf{y}\}. With such an enlargement the vectors \textbf{x}%
$^{p},$\textbf{y}$^{p}$ can be looked upon either as vectors in \textbf{R}$%
^{4}$ or as points in the projective space \textbf{P}$^{3}.$ Now we take
into account that in 3 dimensions $a\times b\rightleftarrows a\wedge b$
[55]. We would like to apply this correspondence to the vectors \textbf{x}$%
^{p}$ and \textbf{y}$^{p}$. Specifically, we consider the exterior product (x%
$_{0}$, x$_{1}$, x$_{2}$, x$_{3}$)$\wedge (y_{0},y_{1},y_{2},y_{3})\equiv
(x_{0},\mathbf{x})\wedge (y_{0},\mathbf{y})$ $\equiv (l_{01}$,$%
l_{02},l_{03},l_{23,},l_{31},l_{12})$. For the sake of illustration, we can
choose x$_{0}$=y$_{0}$=1, then $%
l_{01}=y_{1}-x_{1}=Z_{1},...,l_{23}=x_{2}y_{3}-y_{2}x_{3}=L_{1},...,l_{12}=x_{1}y_{2}-y_{1}x_{2}=L_{3}. 
$ Clearly, $L_{1},L_{2},L_{3}$ can be looked upon as components of the
angular momenta \textbf{L}. In such a case, following Fock [101,102] we can
make an identification \textbf{y}$^{p}$=\{y$_{0}$,\textbf{y}\}$\rightarrow
\{p_{0},\mathbf{p}\}$ with $p_{0}=\sqrt{-2mE},E<0.$ Evidently, in such a
case we no longer can choose $x_{0}=y_{0}=1$. This is not essential,
however. \ In particular, this could be seen if we notice that $%
Z_{1},Z_{1},Z_{3}$ are proportional to the components of the
Laplace-Runge-Lentz vector. If this is so, then we require that 
\begin{equation}
\mathbf{Z}\cdot \mathbf{L}=0.  \tag{6.14a}
\end{equation}%
However,\ for the Kepler problem we have: 
\begin{equation}
(\mathbf{p\times L-}mk\mathbf{\hat{r})\cdot L=}\left( \mathbf{p\times L}%
\right) \cdot \mathbf{L-}mk\mathbf{\hat{r}(r\times p)=}0\mathbf{.} 
\tag{6.14b}
\end{equation}%
At the same time, $\mathbf{p\times L=p\times (r\times p)=r(p\cdot p})-%
\mathbf{p}(\mathbf{p\cdot r}).$ Therefore, $\mathbf{p\times L-}mk\mathbf{%
\hat{r}=}a\mathbf{r}+b\mathbf{p,}$ where $a$ and $b$ are known constants. By
the appropriate rescaling \ this linear combination can be made the same as $%
Z_{1},Z_{1},Z_{3}.$\ There is another meaning of Eq.(6.14a) however. It is
purely mathematical. And, as such, it is totally independent of its
relevance to the Kepler problem. It is associated with the concept of
Grassmannian manifold and of Pl\"{u}cker's embedding of this manifold into
the projective space.

To shorten our discussion, without loss of generality we can always replace
the real space $\mathbf{R}^{n}$ by the complex $\mathbf{C}^{n}.$ Let \{%
\textbf{e}$_{1},...,\mathbf{e}_{n}\}$ be some vector basis in \textbf{C}$%
^{n}.$ Introduction of such a basis requires introduction of the scalar
product. This can be done by analogy with \ the scalar products of quantum
mechanics. \ With help of this basis we introduce the exterior products,
e.g. \textbf{e}$_{1}\wedge \mathbf{e}_{2},...,$\textbf{e}$_{1}\wedge \cdot
\cdot \cdot \wedge \mathbf{e}_{n}$ . \ It is convenient to keep track of
these products by defining the subsets $J_{p}=\{i_{1},...,i_{p}\},p\leq n.$
Let the set of basis vectors \{\textbf{e}$_{i_{1}},...,$\textbf{e}$%
_{i_{p}}\} $ be associated with such a subset$.$ There are exactly $%
n!/p!(n-p)!$ ways to select such a set from the set \{\textbf{e}$_{1},...,$%
\textbf{e}$_{n}\}$ and to make the exterior products \textbf{e}$%
_{i_{1}}\wedge \cdot \cdot \cdot \wedge \mathbf{e}_{i_{p}}$ out of selected
vectors. The Grassmaniann Gr$_{p}$(\textbf{C}$^{n}$) is the manifold made
out of all $p$-dimensional subspaces of \textbf{C}$^{n}$. Suppose that in $%
\mathbf{C}^{n}$ we changed the basis from \{\textbf{e}$_{1},...,\mathbf{e}%
_{n}\}$ to \{$\mathbf{e}_{1}^{\prime },...,\mathbf{e}_{n}^{\prime }\}.$ \
Changes of the basis in \textbf{C}$^{n}$ leads to changes in the basis for
subsets, e.g. 
\begin{equation}
\mathbf{e}_{i_{1}}^{\prime }\wedge ...\wedge \mathbf{e}_{i_{p}}^{\prime
}=A_{i_{1}j_{1}}\mathbf{e}_{j_{1}}\wedge \cdot \cdot \cdot \wedge
A_{i_{1}j_{p}}\mathbf{e}_{j_{p}}\equiv \det (A_{ij})\mathbf{e}_{i_{1}}\wedge
\cdot \cdot \cdot \wedge \mathbf{e}_{i_{p}}.  \tag{6.15}
\end{equation}%
This relation can be looked upon as the\ equivalence relation defining a
point in the complex projective space \textbf{CP}$^{N}$. Thus, we just (Pl%
\"{u}cker) embedded the Grassmannian Gr$_{p}$(\textbf{C}$^{n}$) into the
complex projective space \textbf{CP}$^{N}$ \ of dimension $N$, where $%
N=(n!/p!(n-p)!)-1.$ Since $n!/p!(n-p)!=n!/(n-p)!p!$ we also have Gr$_{p}$(%
\textbf{C}$^{n}$)=Gr$_{n-p}$(\textbf{C}$^{n}$) causing the exterior products 
\textbf{e}$_{i_{1}}\wedge \cdot \cdot \cdot \wedge \mathbf{e}_{i_{p}}$ and 
\textbf{e}$_{i_{p+1}}\wedge \cdot \cdot \cdot \wedge \mathbf{e}_{i_{n-p}}$
to be related to each other in the way known from the Hodge theory of
differential forms. Specifically, if we introduce the notations: \textbf{e}$%
_{i_{1}}\wedge \cdot \cdot \cdot \wedge \mathbf{e}_{i_{p}}\in \wedge ^{p}%
\mathcal{E}$ \ and \textbf{e}$_{i_{p+1}}\wedge \cdot \cdot \cdot \wedge 
\mathbf{e}_{i_{n-p}}\in \wedge ^{n-p}\mathcal{E}$ ,\ this \ definition is
implying that in both cases we use the basis \{\textbf{e}$_{1},...,\mathbf{e}%
_{n}\},$ so that \textbf{e}$_{i}\in \mathcal{E}$ $\mathcal{\forall }i$ . If
this is so, we can then construct the products of the type ($\wedge ^{p}%
\mathcal{E}$ )$\wedge (\wedge ^{n-p}\mathcal{E)}.$ Construction of such \
products is subject to the Hodge-type constraints ($\wedge ^{p}\mathcal{E)}$ 
$\wedge (\wedge ^{n-p}\mathcal{E)\sim (}\wedge ^{p+1}\mathcal{E}$ )$\wedge
(\wedge ^{n-p-1}\mathcal{E)\sim \wedge }^{n}\mathcal{E}.$ \ The sign $\sim $
means (Hodge-type) "equivalence". Evidently, it is permissible to have as
well the following equivalences ($\wedge ^{p}\mathcal{E)}$ $\wedge (\wedge
^{n-p}\mathcal{E)\sim (}\wedge ^{p+2}\mathcal{E}$ )$\wedge (\wedge ^{n-p-2}%
\mathcal{E)\sim \wedge }^{n}\mathcal{E}$, and so on. Just described results
lead us to the following\bigskip \bigskip

\textbf{Definition} \textbf{6.1.} Let \ \{\textbf{e}$_{1},...,$\textbf{e}$%
_{n}\}$ be the basis for \textbf{C}$^{n},$ then we define the set \textbf{e}$%
_{J_{p}}=$\textbf{e}$_{i_{1}}\wedge \cdot \cdot \cdot \wedge \mathbf{e}%
_{i_{p}},$ with $1\leq i_{1}<i_{2}<\cdot \cdot \cdot <i_{p}\leq n.$ If $x\in 
\mathcal{\wedge }^{n}\mathcal{E}$ , then $x$ is \textsl{totally decomposable}
if 
\begin{equation}
x=\dsum\limits_{\substack{ J_{p}  \\ p\in \{1,...,n\}}}a_{J_{p}}\mathbf{e}%
_{J_{p}}\equiv \dsum\limits_{1\leq i_{1}<i_{2}<\cdot \cdot \cdot <i_{p}\leq
n}a_{i_{1}i_{2}\cdot \cdot \cdot i_{p}}(\mathbf{e}_{i_{1}}\wedge \cdot \cdot
\cdot \wedge \mathbf{e}_{i_{p}}).  \tag{6.16}
\end{equation}

The homogenous coordinates $a_{J_{p}}$ are called \textsl{Pl\"{u}cker
coordinates} on \textbf{CP}$^{N}.$ \bigskip

From this definition, it follows at once that%
\begin{equation}
x\wedge x=0.  \tag{6.17}
\end{equation}%
This is the condition for the \textsl{Pl\"{u}cker embedding}. We want to
demonstrate now that Eq.(6.14a) is the condition for the Pl\"{u}cker
embedding. For this purpose, consider now the Grassmannian Gr$_{2}$(\textbf{C%
}$^{4}$). For it, we have \{\textbf{e}$_{0}$,...,\textbf{e}$_{3}$\} as the
basis for $\mathcal{E}.$ Next, we construct $\mathcal{\wedge }^{2}\mathcal{E}
$ \ built as follows:%
\begin{equation}
\mathcal{\wedge }^{2}\mathcal{E\sim \{}\mathbf{e}_{0}\wedge \mathbf{e}_{1},%
\mathbf{e}_{0}\wedge \mathbf{e}_{2},\mathbf{e}_{0}\wedge \mathbf{e}_{3},%
\mathbf{e}_{1}\wedge \mathbf{e}_{2},\mathbf{e}_{1}\wedge \mathbf{e}_{3},%
\mathbf{e}_{2}\wedge \mathbf{e}_{3}\}  \tag{6.18a}
\end{equation}%
Since $\mathbf{x}^{p}=\dsum\nolimits_{J_{p}=0}^{3}x_{J_{p}}\mathbf{e}%
_{J_{p}} $ and $\mathbf{y}^{p}=\dsum\nolimits_{J_{p}=0}^{3}y_{J_{p}}\mathbf{e%
}_{J_{p}},$ the condition for the Pl\"{u}cker embedding acquires the
following form: 
\begin{equation}
\mathbf{x}^{p}\wedge \mathbf{y}^{p}=0=\left(
l_{01}l_{23}+l_{02}l_{31}+l_{03}l_{12}\right) \mathbf{e}_{0}\wedge \mathbf{e}%
_{1}\wedge \mathbf{e}_{2}\wedge \mathbf{e}_{3}.  \tag{6.18b}
\end{equation}%
From here we obtain: 
\begin{equation}
l_{01}l_{23}+l_{02}l_{31}+l_{03}l_{12}=0,  \tag{6.18c}
\end{equation}%
where the determinants $l_{ij}$ are defined by 
\begin{equation}
l_{ij}=\left\vert 
\begin{array}{cc}
x_{i} & x_{j} \\ 
y_{i} & y_{j}%
\end{array}%
\right\vert ,i,j=0,1,2,3.  \tag{6.18d}
\end{equation}%
Since, $x_{0}y_{1}-x_{1}y_{0}=Z_{1}=l_{01}$, while $%
x_{1}y_{2}-y_{1}x_{2}=L_{3}$ and so on, the Pl\"{u}cker embedding condition,
Eq.(6.18), is exactly the same as \ the mechanical condition,
Eq.(6.14a).\bigskip

\textsl{6.2.3. \ From Pl\"{u}cker embedding to conformal compactification of
Minkowski}

\textsl{\ \ \ \ \ \ \ \ \ \ space via twistor formalism}

\textsl{\bigskip }

The results obtained in previous subsection enable us to complete our proof
of \ Huygens triviality. In the previous subsection we discussed \ the
rationale for compactification of the momentum space $\mathbf{p}%
=\{p_{x},p_{y},p_{z}\}\in \mathbf{R}^{3}$. Naturally, we ended up with $%
S^{3}.$ The 3-sphere $S^{3}$ \ is living in $\mathbf{R}^{4}$. There is some
advantage in compactification of $\mathbf{R}^{4}$ as well leading to $S^{4}.$
The argument goes as follows. In the previous subsection we introduced
vectors \textbf{x}$^{p},$ \textbf{y}$^{p}$ living in \textbf{R}$^{4}.$ We
noticed that they can be looked upon either as vectors (in $\mathbf{R}^{4}$
) or as points (in $\mathbf{P}^{3}$). By complexification we end up with
these vectors living either in \textbf{C}$^{4}$ or \textbf{CP}$^{3}.$ This
allows us to develop the compactification of the Minkowski space (whose
isometry group is SO(3,1)) and the Euclidean space (whose isometry group
SO(4) ) using the same formalism. \ Furthermore, using the compactification
procedure leads us directly to the formalism of twistors and twistor spaces.

\bigskip

\textbf{Definition 6.2}. The twistor space \textbf{T} is \textbf{C}$^{4}$
with coordinates {\normalsize Z}=\{Z$^{0},$ Z$^{1},$ Z$^{2},$ Z$^{3}\}.$The
projective twistor space is \textbf{PT}=\textbf{CP}$^{3}$ with homogenous
coordinates \{Z$^{0}:$ Z$^{1}:$ Z$^{2}:$ Z$^{3}\}$

\bigskip

Next, we notice that if for the description of the complex plane $\mathbf{C}$
we have to use the complex numbers, e.g. $z=x+iy,$ then for description of 
\textbf{C}$^{2}$ we have to use the quaternions, e.g. $q=t+ix+jy+kz$. Both
complex numbers and quaternions admit matrix presentation [$55$]. For
instance, 
\begin{equation}
z=x+iy\in \mathbf{C}=\mathbf{R}^{2}\rightleftarrows A=\left( 
\begin{array}{cc}
x & -y \\ 
y & x%
\end{array}%
\right) .  \tag{6.19a}
\end{equation}%
The scalar product $\mathbf{x}\cdot \mathbf{y}$ in $\mathbf{R}^{2}$ can be
presented either as $\mathbf{x}\cdot \mathbf{y=}x_{1}x_{2}\mathbf{+}%
y_{1}y_{2}$ or as $\mathbf{x}\cdot \mathbf{y=}\frac{1}{2}(z_{1}\bar{z}_{2}+%
\bar{z}_{1}z_{2}).$ The last presentation can be equivalently rewritten as%
\begin{equation}
\mathbf{x}\cdot \mathbf{y=}\frac{1}{2}tr(A_{1}A_{2}^{T}).  \tag{6.19b}
\end{equation}%
The squared norm $\left\vert z\right\vert ^{2}$=$x^{2}+y^{2}$ of the complex
number can be alternatively represented as 
\begin{equation}
\left\vert z\right\vert ^{2}=\det A.  \tag{6.19c}
\end{equation}%
To extend these results to $\mathbf{C}^{2}$ we notice that $\mathbf{C}%
^{2}\rightleftarrows \mathbf{R}^{4}.$ The quaternion $q=t+ix+jy+kz$ is
encoded by $\ $the $\{t,x,y,z\}\in \mathbf{R}^{4}$. In view of the
correspondence \ $\mathbf{C}^{2}\rightleftarrows \mathbf{R}^{4}$ we
introduce 2 complex numbers $z_{1}=t+ix$ and $z_{2}=y+iz$ and then, by
analogy with Eq.(6.19a), we can write%
\begin{equation}
q=t+ix+jy+kz\in \mathbf{H}=\mathbf{R}^{4}\rightleftarrows A=\left( 
\begin{array}{cc}
z_{1} & -\bar{z}_{2} \\ 
z_{2} & \bar{z}_{1}%
\end{array}%
\right) .  \tag{6.20a}
\end{equation}%
Accordingly, the squared norm $\left\vert q\right\vert ^{2}$=$%
t^{2}+x^{2}+y^{2}+z^{2}$ is given by 
\begin{equation}
\left\vert q\right\vert ^{2}=\det A.  \tag{6.20b}
\end{equation}%
Consider now a spacetime of Euclidean signature $(+,+,+,+)$ in which the
vector $\mathbf{x}$ is given componentwise as \ 
\begin{equation}
(x^{0},x^{1},x^{2},x^{3})\in \mathbf{R}^{4}\rightleftarrows (x^{AB})=\left( 
\begin{array}{cc}
x^{0}+ix^{1} & x^{2}+ix^{3} \\ 
-x^{2}+ix^{3} & x^{0}-ix^{1}%
\end{array}%
\right) ,\text{ \ }A,B=0,1.  \tag{6.21}
\end{equation}%
The square of the Euclidean norm for $\mathbf{x}$ is given by $\det
(x^{AB})=\left\vert z_{1}\right\vert ^{2}+\left\vert z_{2}\right\vert ^{2}.$
Here $z_{1}=x^{0}+ix^{1},z_{2}=-x^{2}+ix^{3}.$ Because of this, we also have 
$(x^{0},x^{1},x^{2},x^{3})\rightleftarrows (z_{1},z_{2})\in \mathbf{C}^{2}.$
Furthermore, we also obtain: $\mathbf{C}^{2}\rightleftarrows \mathbf{R}^{4}=%
\mathbf{H.}$ Accordingly, $\mathbf{C}^{4}\rightleftarrows \mathbf{H}^{2}.$
Next, we recall that the 3-sphere $S^{3}$ can be analytically represented as 
$\left\vert z_{1}\right\vert ^{2}+\left\vert z_{2}\right\vert ^{2}=1.$ The
Hopf map $S^{3}\rightarrow S^{2}$ can be constructed now as follows. Begin
with $\{z_{1},z_{2}\}\in \mathbf{C}^{2}$ and consider the ratio $z_{1}/z_{2}$
(or $z_{2}/z_{1}).$ Since $z=\left\vert z\right\vert \exp (i\varphi )$, it
is clear that $z_{1}/z_{2}\sim z\sim \mathbf{C\sim }S^{2}.$ Alternatively, $%
\{z_{1}:z_{2}\}=\mathbf{CP}^{1}=S^{2}.$ Consider now two quaternions $q_{1}=$%
Z$^{0}+$Z$^{1}j$ and $q_{2}=$Z$^{2}+$Z$^{3}j$ and consider the quaternionic
projective space \ with homogenous coordinates \{$q_{1}:q_{2}\}\in \mathbf{HP%
}^{1}.$ If in the case of complex numbers we had the correspondence $\
z_{1}/z_{2}\sim S^{2}\sim \mathbf{C}$ equivalent to the Hopf map $%
S^{3}\rightarrow S^{2},$ in the quaternionic case we analogously have
another Hopf map: $\mathbf{CP}^{3}\rightarrow S^{4}=\mathbf{HP}^{1}.$To
demonstrate that this is indeed the case, we recall (from the Definition
6.2.) that \textbf{PT}=\textbf{CP}$^{3}.$ At the same time, \{Z$^{0},$ Z$%
^{1},$ Z$^{2},$ Z$^{3}\}\in \mathbf{T}$ and, \ using this fact, we construct
a quaternion $q$ of the type $q=\frac{Z^{0}+Z^{1}j}{Z^{2}+Z^{3}j}.$ In
complete analogy with the complex numbers, where we have: $%
z_{1}/z_{2}=z\rightarrow \frac{q_{1}}{q_{2}}=q,$ now we have: 
\begin{equation}
S^{4}=\mathbf{H}\cup \{\infty \}\ni q=z_{1}+z_{2}j=\frac{Z^{0}+Z^{1}j}{%
Z^{2}+Z^{3}j}.  \tag{6.22}
\end{equation}%
Recall [$55$] that the projective space \textbf{CP}$^{n}$ can be defined as
the quotient%
\begin{equation}
\left\{ \dsum\limits_{j=1}^{n+1}\left\vert \xi _{i}\right\vert
^{2}=1\right\} /(\xi _{i}\rightarrow e^{i\varphi }\xi _{i}),\text{ }\xi
_{i}\in \mathbf{C}.  \tag{6.23}
\end{equation}%
Therefore, each point in \textbf{CP}$^{n}$ should be identified with the
circle $S^{1}$ on $S^{2n+1}.$ From here, we obtain the familiar Hopf
fibration (for $n=1$): $S^{3}/S^{1}\simeq S^{2}$ . This logic fails for for $%
n=3$ where by analogy we formally should expect to have $S^{7}/S^{1}.$ This
quotient is not the Hopf map though. For $n=3$ the Hopf map is given by the
quotient $S^{7}/S^{3}\simeq S^{4}.$ From here we are obtaining the already
mentioned correspondence:%
\begin{equation}
S^{3}\rightarrow z_{1}/z_{2}\sim z\sim \mathbf{CP}^{1}\mathbf{\sim }%
S^{2}\sim \mathbf{R}^{2}\cup \{\infty \}\iff S^{7}\rightarrow
q_{1}/q_{2}\sim \mathbf{HP}^{1}\mathbf{\sim }S^{4}\sim \mathbf{R}^{4}\cup
\{\infty \}.  \tag{6.24}
\end{equation}%
In view of these results and, taking into account that $%
q=z_{1}+z_{2}j;z_{1},z_{2}\in \mathbf{C},$ we can formally write: $q=f(z,w)$%
, $z,w\in \mathbf{C}.$ In addition, we have as well: $\{f+gj:h+kj\}\in 
\mathbf{HP}^{1}$ and think about the correspondence $f\rightleftarrows
Z^{0},g\rightleftarrows Z^{1},h\rightleftarrows Z^{2},k\rightleftarrows
Z^{3} $ as defining some analytic functions of $z$ and $w$. \ In terms of
such notations Eq.(6.22) can be rewritten as 
\begin{equation}
f+gj-(h+kj)(z_{1}+z_{2}j)=0  \tag{6.25}
\end{equation}%
Using the multiplication table for quaternions [$112$], p.39, it is possible
to rewrite this result in more suggestive form as 
\begin{equation}
(h,k)\left( 
\begin{array}{cc}
z_{1} & z_{2} \\ 
-\bar{z}_{2} & \bar{z}_{1}%
\end{array}%
\right) =(f,g).  \tag{6.26}
\end{equation}%
By comparing Eq.s(6.21) and (6.26) we identify $z_{1}$ with $x^{0}+ix^{1}$
and $z_{2}$ with $x^{2}+ix^{3}.$ This identification can be extended by
identifying $Z^{2}$ and $Z^{3}$ with $h$ and $k$ and $f$ and $g$ with $Z^{0}$
and $Z^{1}.$ In literature on twistors [113] the spinor language is used. To
comply with the standard twistor notations we should make an identification
: $Z^{0}=\omega ^{0},Z^{1}=\omega ^{1}$ on the one hand and, $Z^{2}=\pi
_{0},Z^{3}=\pi _{1}$ on the other$.$ In terms of such notations Eq.(6.26)
acquires the following form [114]: 
\begin{equation}
\omega ^{A}=x^{AB}\pi _{B},\text{ \ }A,B=0,1.  \tag{6.27}
\end{equation}%
Here $\omega $ and $\pi $ are the two-component spinors.\bigskip

\textbf{Definition 6.3. }Eq.(6.27) is known in twistor literature as \textit{%
incidence relation. }It connects points in\textit{\ }\textbf{R}$^{4}$ (or
Euclidean) space with points in \textbf{PT}.\bigskip

By changing the matrix $x^{AB}$ describing the Euclidean space to that
describing the Minkowski space%
\begin{equation}
(x^{0},x^{1},x^{2},x^{3})\rightleftarrows (x^{AB})=\left( 
\begin{array}{cc}
-i\left( x^{0}-x^{1}\right) & x^{2}+ix^{3} \\ 
-x^{2}+ix^{3} & -i\left( x^{0}+x^{1}\right)%
\end{array}%
\right) ,\text{ }A,B=0,1,  \tag{6.28}
\end{equation}%
leads to the appropriately changed determinant: $\det (x^{AB})=-\left(
x^{0}\right) ^{2}+(x^{1})^{2}+(x^{2})^{2}+(x^{2})^{2}$ and to the incidence
relation for the Minkowski space $\mathfrak{M}$.

From previous discussion it is clear that the incidence relation, Eq.(6.27),
could be used for both the Euclidean and the Minkowski spaces if we use the
complexification: $(x^{0},x^{1},x^{2},x^{3})\rightleftarrows
(z_{1},z_{2})\in \mathbf{C}^{2}$ leading to the matrix $x^{AB}$ defined by 
\begin{equation}
x^{AB}=\left( 
\begin{array}{cc}
\tilde{z} & w \\ 
\tilde{w} & z%
\end{array}%
\right)  \tag{6.29}
\end{equation}%
in which $\tilde{z}$ and $\tilde{w}$ should not be treated as complex
conjugates of $z$ and $w$ (unless otherwise specified). In terms of such
notations the incidence relation, Eq.(6.27), can be rewritten as 
\begin{eqnarray}
\tilde{z}Z^{2}+wZ^{3} &=&Z^{0},  \notag \\
\tilde{w}Z^{2}+zZ^{3} &=&Z^{1}.  \TCItag{6.30a}
\end{eqnarray}%
Geometrically, this is \ the system of equations for two hyperplanes. In the
language of algebraic/projective geometry they can be conveniently rewritten
as 
\begin{equation}
Z^{\alpha }A_{\alpha }=0\text{ and }Z^{\alpha }B_{\alpha }=0.  \tag{6.30b}
\end{equation}%
The question of \ interest is: Under what conditions these two planes
intersect? Why should we be interested in this question? Because we would
like to connect just described twistor formalism with that presented in
previous subsection. To do so, following [$55$] we consider $2\times 4$
matrix \textbf{M} of the type 
\begin{equation}
\mathbf{M}=\left( 
\begin{array}{cccc}
x_{0} & x_{1} & x_{2} & x_{3} \\ 
y_{0} & y_{1} & y_{2} & y_{3}%
\end{array}%
\right)  \tag{6.31}
\end{equation}%
Its 1st and 2nd rows are made of vectors $\mathbf{x}^{p}$ and $\mathbf{y}%
^{p} $ respectively. All Pl\"{u}cker coordinates are obtainable now as
determinants of columns $i$ and $j$ of \textbf{M}. In view of Eq.(6.16), the
matrix \textbf{M }can be used in the following matrix equation 
\begin{equation}
x_{i}^{p}=\dsum\limits_{j=0}^{3}\text{M}_{ij}\mathbf{e}_{j}\text{ , }i=1,2%
\text{.}  \tag{6.32}
\end{equation}%
In the system of linear equations, Eq(6.30a), one can identify Z$^{0},$ Z$%
^{1},$ Z$^{2},$ Z$^{3}$ with the vectors $\mathbf{e}_{j}$, $j=0\div 3$, in
Eq.(6.32). Then, the matrix \textbf{M} acquires the following form%
\begin{equation}
\mathbf{M}=\left( 
\begin{array}{cccc}
\tilde{z} & w & 0 & 1 \\ 
\tilde{w} & z & 1 & 0%
\end{array}%
\right) .  \tag{6.33}
\end{equation}%
Since the Pl\"{u}cker coordinates for this matrix are: $l_{01}=\tilde{z}z-%
\tilde{w}w,l_{02}=\tilde{z},l_{03}=-\tilde{w},l_{12}=w,l_{13}=-z,l_{23}=-1,$
the Pl\"{u}cker embedding condition, Eq.(6.18), acquires the following form 
\begin{equation}
(-1)(\tilde{z}z-\tilde{w}w)+\tilde{z}z-\tilde{w}w=0  \tag{6.34}
\end{equation}%
and, is trivially satisfied.

Next, we would like to explain how just obtained result is connected with
the intersection of two hyperplanes, Eq.s(6.30b), in \textbf{PT=CP}$^{3}$%
\textbf{.} Following [115], page 83, consider \ a hyperplane through the
origin O in \textbf{CP}$^{3}.$ It is described by the system of two
equations 
\begin{eqnarray}
l_{0}z_{0}+l_{1}z_{1}+l_{2}z_{2}+l_{3}z_{3} &=&0,  \notag \\
m_{0}z_{0}+m_{1}z_{1}+m_{2}z_{2}+m_{3}z_{3} &=&0.  \TCItag{6.35}
\end{eqnarray}%
The description of this plane remains unchanged if instead of $%
l_{0},l_{1},l_{2},l_{3}$ we would use $l_{0}+\lambda m_{0},l_{1}+\lambda
m_{1},etc.,$where $\lambda $ is an arbitrary parameter. This observation
makes the system of Eq.s(6.30a) equivalent to the system of Eq.s(6.35). A
symmetrical set of coordinates \ is obtained \ by defining the six
expressions $l_{i}m_{j}-l_{j}m_{i}=l_{ij},i,j.=0,1,2,3.$ Use of these
expressions in Eq.s(6.35) allows us to eliminate successively $%
z_{0},z_{1},z_{2},z_{3}$ resulting in the following system of equations%
\begin{eqnarray}
l_{01}z_{1}+l_{02}z_{2}+l_{03}z_{3} &=&0,  \notag \\
l_{10}z_{0}+l_{12}z_{2}+l_{13}z_{3} &=&0,  \notag \\
l_{20}z_{0}+l_{21}z_{1}+l_{23}z_{3} &=&0,  \notag \\
l_{30}z_{0}+l_{31}z_{1}+l_{32}z_{2} &=&0.  \TCItag{6.36}
\end{eqnarray}%
Elimination of $z_{0}$, $z_{1}$ and $z_{2}$ from these equations results in
the Pl\"{u}cker relation, Eq.(6.18c), and the reminder equation is the
equation for a complex projective line in \textbf{CP}$^{3}$. So that,
indeed, two planes, Eq.s(6.30b), are intersecting in a line. More details
can be found in [116], pages 141-144. In view of Eq.s(6.15)-(6.17) six Pl%
\"{u}cker coordinates $l_{01}$,$l_{02},l_{03},l_{23,},l_{31},l_{12}$
represent a point in \textbf{CP}$^{5}$ so that Eq.(6.18c) represents a
complex quadric $Q_{4}$ in this projective space \textbf{CP}$^{5}$[116]$.$
Following [$115$] it is useful to rewrite this quadric in terms of new
variables 
\begin{eqnarray}
l_{03} &=&u_{0}+u_{3},l_{13}=u_{1}+u_{4},l_{23}=u_{2}+u_{5},  \TCItag{6.37}
\\
l_{12} &=&u_{0}-u_{3},l_{20}=u_{1}-u_{4},l_{01}=u_{2}-u_{5}.  \notag
\end{eqnarray}%
Use of these variables converts Eq(6.18c) into\footnote{%
In the context of the Lie sphere geometry this equation is discussed \ in
connection with Eq.(7.15) of section 7.} 
\begin{equation}
u_{0}^{2}+u_{1}^{2}+u_{2}^{2}-u_{3}^{2}-u_{4}^{2}-u_{5}^{2}=0.  \tag{6.38}
\end{equation}%
Since $u_{i}^{\prime }s$ are complex we can adopt them for physically
relevant situations. We begin with \bigskip

\textbf{\ Definition 6.4.} The $null$ $cone$ $\Gamma $ of the origin is
defined (according to Eq.(4.1)) as%
\begin{equation}
0=\Gamma (t,x;\tau ,y)=c^{2}(t-\tau
)^{2}-\dsum\limits_{i=1}^{3}(x_{i}-y_{i})^{2}\equiv T^{2}-X^{2}-Y^{2}-Z^{2}.
\tag{6.39a}
\end{equation}

The $generators$ of $\Gamma $- the $null$ $rays-$ are subject to the
constraint%
\begin{equation}
T:X:Y:Z=const.  \tag{6.39b}
\end{equation}%
The Lorentz transformation $\mathcal{L}$ \ sends the set of generators of $%
\Gamma $ into another set of generators of $\Gamma \ $\ preserving $\Gamma .$
Following Penrose [117] we define the group $C(2)$ as the group of all
conformal maps of the compactified complex plane, that is of $S^{2},$ to
itself. The connected component of the identity in $C(2)$ consists \ of the
orientation -preserving \ conformal maps $\varsigma \rightarrow f(\varsigma
) $ of $S^{2},$ given by 
\begin{equation}
\varsigma \rightarrow f(\varsigma )=\frac{\alpha \varsigma +\beta }{\gamma
\varsigma +\delta }.  \tag{6.40}
\end{equation}%
Details are given in [$105$]. From this reference we find that the
homomorphism of the Lorentz group $O(3,1)$ into $C(2)$ is $2\div 1$. The
quadratic form $\Gamma $ defined by Eq.(6.39a) is obtainable from the
quadric $Q_{4}$ defined in Eq.(6.38) if we require $u_{0}^{2}+u_{1}^{2}=0.$
The geometrical and topological meaning of these two extra terms is
associated with conformal symmetry typical for the Lie sphere geometry to \
be described below. In the meantime, at this moment, we restrict ourselves
by the concepts which were already in use in mathematical physics
literature. For this purpose we recall some results from section 1. In it we
mentioned that locally the static Einsteinian space-times are Minkowskian.
Let $\mathcal{M}^{3}$ be some 3-manifold so that topologically all static
Einsteinian spacetimes are of the form $\mathcal{M}^{3}\times \mathbf{R}$
[118], where \textbf{R} represents time. The positivity of mass theorem in
general relativity, used in [$119$], superimposed with use of the Yamabe
theorem [$120$], allows us to replace $\mathcal{M}^{3}\times \mathbf{R}$ by $%
S^{3}\times \mathbf{R}.$ \ The compactification requirement causes us to
replace this manifold by $S^{3}\times S^{1}.$ But analytically $S^{1}$ is $%
u_{0}^{2}+u_{1}^{2}=const.$ Accordingly, Eq.(6.38) does contain information
about compactification. \ We already know that the twistor space is $\mathbf{%
C}^{4}$ which is the complexification of $\mathbf{R}^{4}$ whose
compactification is $S^{4}.$ Thus, we end up with the compactified Minkowski
space $\mathfrak{M\simeq }$ $S^{3}\times S^{1}$ and the compactified
Euclidean $\mathfrak{E}$ space $S^{4}.$ Both are described by the Kleinian
quadric $Q_{4}$ [$116$] defined by Eq.(6.38). The sphere $S^{4}$ is
described by%
\begin{equation}
\mathfrak{E}:u_{0}^{2}+u_{1}^{2}+u_{2}^{2}+u_{3}^{2}+u_{4}^{2}-u_{5}^{2}=0 
\tag{6.41a}
\end{equation}%
while the (compactified) Minkowski space by\footnote{%
Eq.(6.41b) is discussed in terms of the formalism of the Lie sphere geometry
in section 7, in connection with Eq.s (7.10) and (7.20).} 
\begin{equation}
\mathfrak{M}:u_{0}^{2}+u_{1}^{2}+u_{2}^{2}+u_{3}^{2}-u_{4}^{2}-u_{5}^{2}=0 
\tag{6.41b}
\end{equation}%
\bigskip

\textbf{Definition 6.5. }The \textit{null cone} of the origin of the
compactified Minkowski space $\mathfrak{M}$ is described by Eq.(6.41b). The
symmetry group leaving this null cone invariant is the conformal group $%
SO(4,2).$ \bigskip This is the largest symmetry group of the hydrogen atom.

We have reached \ these conclusions using arguments entirely different from
those developed in the group-theoretic [$\mathit{19}$] and high energy
physics[83] literature. Use of the concepts of Lie sphere geometry to be
discussed \ in section 7 provides us with solid theoretical framework for
dealing with just discussed problems. At the same time, the obtained results
are sufficient for finishing our study of Huygens triviality of the
stationary Schr\"{o}dinger equatioin.\bigskip

\textsl{6.3. \ Harmonic analysis and Huygens' triviality}

\textsl{\bigskip }

In Fock's paper [102] the stereographic projection was used to relate $%
\mathbf{p}=\{p_{x},p_{y},p_{z}\}\in \mathbf{R}^{3}$ and $\mathbf{p}%
=\{p_{0},p_{x},p_{y},p_{z}\}\in \mathbf{R}^{4}$. \ The momentum space\ $%
\mathbf{p}$ was one point ($\mathbf{p}_{\infty }$)compactified to 3-sphere $%
S^{3}$ living in $\mathbf{R}^{4}.$ If N is the North pole of the 3-sphere,
N:=(0,0,0,1), we define U$_{N}:=S^{3}\setminus N.$ \bigskip

\textbf{Definition 6.6}. The \textit{stereographic projection} $\pi $ is the
map $\pi :$ U$_{N}\rightarrow \mathbf{R}^{3}$ defined by 
\begin{equation}
\pi :x_{k}=\frac{\xi _{k}}{1-\xi _{4}},\text{ }k=1,2,3.  \tag{6.42a}
\end{equation}

Its inverse $\pi ^{-1}$ is defined by 
\begin{equation}
\pi ^{-1}:\xi _{k}=\frac{2x_{k}}{1+\left\Vert x\right\Vert ^{2}},\xi _{4}=%
\frac{\left\Vert x\right\Vert ^{2}-1}{\left\Vert x\right\Vert ^{2}+1}%
,k=1,2,3.  \tag{6.42b}
\end{equation}%
\bigskip\ The map $\pi $ can be extended to $\bar{\pi}$ defined as follows.
By relating the North pole N for the 3- sphere to the compactification point 
$\mathbf{p}_{\infty }=\{\infty \}$ all $S^{3}$ is covered. Since $\xi
_{4}\neq 1,$ just defined extended stereographic projection $\bar{\pi}$
provides a bijective correspondence between points on $S^{3}$ and points of
the compactified momentum space $\mathbf{p}\cup \{\infty \}$. \ Using such
stereographic projection Fock transformed the integral equation Eq.(6.1b),
defined on $\mathbf{R}^{3},$ into the integral equation on $S^{3}.$ As
result, he demonstrated that solutions of the integral equation \textit{%
inside} $S^{3}$ are \textsl{harmonic} \textsl{functions}. That is they are
solutions of the 4-dimensional Laplacian. The D'Alembertian, Eq.(2.6b), is
obtainable from the Laplacian via formal replacement $x_{0}\rightarrow \pm
ix_{0}$ discussed in subsecton 6.2.1. \ Although Fock did not discuss the $%
E>0$ case in his paper, he did mention about the relevance of the
Lobachevsky (that is hyperbolic) space for the description of $E>0$ case. \
At present, it is possible to find in literature solutions describing $E>0$
case$.$ The detailed calculations are given, for example, in [$121$].
Although these calculations formally solve the $E>0$ problem, they are not
revealing its mathematical essence. This essence was described already in
Fock's paper [$102$] but was left undeveloped. Only recently Fock's remarks
were put into plausible mathematical form. In 2008 in the paper by Frenkel
and Libine [122] in Advances in Mathematics entitled "Quaternionic analysis,
representation theory and physics" Fock's results for both $E<0$ and $E>0$
were rederived. As result, \ it is sufficient to use only results by Frenkel
and Libine for proving Huygens' triviality of the Schr\"{o}dinger equation
(with time-independent Hamiltonian).

\ At this point it is logical to present a condensed summary of Fock's
results along with their subsequent improvements. Fock noticed that
Eq.(6.1b) when stereographically lifted \ to the 3-sphere is looking very
much the same as the Poisson formula for the harmonic functions inside the
circle. It should be noticed, though, that Fock was not referring to the
Poisson formula explicitly. \ The description of this formula employs the
standard complex analysis [123]. Its \ derivation begins with the use of the
Cauchy-Riemann equations. Recall, that these are defined as follows. Let $%
u(x,y)$ and $v(x,y)$ be some analytic functions satisfying the Cauchy-Rieman
equations%
\begin{equation*}
\frac{\partial v}{\partial x}=\frac{-\partial u}{\partial y};\frac{\partial v%
}{\partial y}=\frac{\partial u}{\partial x}.
\end{equation*}%
Since 
\begin{equation*}
\frac{\partial ^{2}u}{\partial x\partial y}=\frac{\partial ^{2}u}{\partial
y\partial x}\text{ and }\frac{\partial ^{2}v}{\partial x\partial y}=\frac{%
\partial ^{2}v}{\partial y\partial x}
\end{equation*}%
both $u$ and $v$ are harmonic functions. That is they satisfy the Laplace
equation: $\triangle u=\triangle v=0,$ $\triangle =\frac{\partial }{\partial
x^{2}}+\frac{\partial }{\partial y^{2}}.$ The crucial theorem leading to the
Poisson formula exploits this fact. \bigskip

\textbf{Theorem 6.7.}\textit{\ Let }$u(x,y)$\textit{\ be the harmonic
function in some simply connected domain }$G$\textit{\ of the complex plane 
\textbf{C}. Then, there is a regular in }$G$\textit{\ function }$f(z)$%
\textit{\ such that }$u(x,y)=\func{Re}f(x+iy).$\textit{\ Function }$f(z)$%
\textit{\ is determined \ with help of }$u(x,y)$\textit{\ with accuracy up
to purely imaginary constant.\bigskip }

The Poisson formula can then be defined as solution of the \textit{Dirichlet
problem}: \ Find a function harmonic in $G$ and coinciding with some
prescribed function $g(z)$ at the boundary of $G$. Since any connected
domain can be converted into a disc of, say, radius R, the solution of the
Dirichlet problem can be represented via the Poisson integral formula as
follows 
\begin{equation}
u(re^{i\psi })=\frac{1}{2\pi }\dint\limits_{0}^{2\pi }u(\func{Re}^{i\varphi
})\frac{\text{R}^{2}-r^{2}}{\text{R}^{2}-2\text{R}r\cos (\varphi -\psi
)+r^{2}}d\varphi .  \tag{6.43a}
\end{equation}%
Let now R$=1$ in Eq.(6.43), then this formula acquires look very similar to
formula Eq.(9.19) of Fock's paper [$\mathit{102}$]. Nevertheless, Fock's
Eq.(9.19) cannot be qualified as the Poisson formula. The Poisson formula is
just a modification of the Cauchy integral formula [$123$]%
\begin{equation}
f(z)=\frac{1}{2\pi }\doint \frac{f(w)dw}{w-z}.  \tag{6.43b}
\end{equation}%
In previous subsections we demonstrated how quaternions replace complex
numbers when $\mathbf{C}\rightarrow \mathbf{C}^{2}$ $\sim \mathbf{R}^{4}.$
The question arises: Can the rest of results of complex analysis \ be
rewritten in the quaternionic language in \textbf{R}$^{4}?$ The development
of the theory of quaternionic functions was initiated by Rudolf Fueter. In
1935, the year of publication of Fock's paper, he produced an exact
quaternionic counterpart of the Cauchy integral formula [$122$],[$124$].
However, it took another 73 years before this formula was converted into
quaternionic analog of the Poisson formula. This happened only in 2008.
Frenkel and Libine [$122$] used the quaternionic version of the Poisson
formula for demonstration that it correctly reproduces both the discrete and
continuous spectrum of the hydrogen atom. The way of derivation of these
spectra is opposite to that used by Fock. Specifically, Fock used the
stereographic projection to bring the 3 dimensional integral equation,
Eq.(6.1b), into 4-dimensional Poisson -like form. Furthermore, the
wavefunction satisfying the Laplace Eq.(9.12) of Fock's paper is \textsl{not}
the same as that obtainable from his 4-dimensional Poisson-like integral
equation. For $E<0$ the integral Eq.(6.1b), when converted into the
4-dimensional form, acquires the following look [125], page 83,%
\begin{equation}
\frac{1}{2\pi ^{2}}\dint\limits_{S^{3}}\frac{\Psi (X^{\prime })}{\left\Vert
X-X^{\prime }\right\Vert ^{2}}d\Omega ^{\prime }=p_{0}\Psi (X)  \tag{6.44}
\end{equation}%
to be compared with Eq.(6.43b). Here $d\Omega ^{\prime }$ is known volume
element of the 3-sphere. The X coordinates are X=\{$\xi _{1},...,\xi _{4}\}.$
These are given by Eq.s(6.42b). The relation between $\Psi (X)$ in Eq.(6.44)
and $\psi (\mathbf{p})$ in Eq.(6.1b) is given by [$125$], page 83, 
\begin{eqnarray}
\Psi (X) &=&\left[ \frac{p^{2}+p_{0}^{2}}{2p_{0}^{2}}\right] ^{2}\psi (%
\mathbf{p}),\text{ }  \TCItag{6.45a} \\
\text{or \ \ \ \ }\Psi (X) &=&\frac{p_{k}^{2}}{\text{X}_{k}^{2}}\psi (%
\mathbf{p}).  \TCItag{6.45b}
\end{eqnarray}%
It would appear that the analogous relation for $E>0$ formally would solve
the Hyugens triviality problem in view of the transformation rules defined
in section 4. This is not the case, however. To explain the existing problem
it is useful to briefly discuss the case $E<0$ first$,$ since it is easier.
In his paper Fock \ introduces the stereographic projection by defining
coordinates $\xi ,\eta ,\zeta $ and $\chi $ such that $\xi ^{2}+\eta
^{2}+\zeta ^{2}$+ $\chi ^{2}=1,\xi =\frac{2p_{0}p_{x}}{p_{0}^{2}+p^{2}},$etc$%
.$ In addition, he introduces in a rather arbitrary fashion still another
set of coordinates 
\begin{equation}
x_{1}=r\xi ;x_{2}=r\eta ;x_{3}=r\zeta ;x_{4}=r\chi  \tag{6.46a}
\end{equation}%
along with the 4-dimensional Laplace equation%
\begin{equation}
\frac{\partial ^{2}u}{\partial x_{1}^{2}}+\frac{\partial ^{2}u}{\partial
x_{2}^{2}}+\frac{\partial ^{2}u}{\partial x_{3}^{2}}+\frac{\partial ^{2}u}{%
\partial x_{4}^{2}}=0.  \tag{6.46b}
\end{equation}%
Since the harmonic function $u(x_{1},x_{2},x_{3},x_{4})$ satisfying this
equation is \textsl{not} a solution of the integral Eq.(6.44)
(surprisingly), clearly Eq.(6.44) cannot be qualified as the Poisson formula
adopted to four dimensions. \ At the same time, if $r$ in Eq.(7.46a) is the
radius of the 3-sphere (eventually $r=1$) then, following Fock, it is
possible to represent the harmonic function as 
\begin{equation*}
u=r^{n-1}\Psi _{n}(X).
\end{equation*}%
Following Cordani [$125$], page 90, we would like to rewrite Fock's results
a bit differently. Thus, let h$_{l}$(X) be a homogenous harmonic polynomial
of degree $\mathit{l}$ in \textbf{R}$^{n+1}.$ That is $\triangle _{\mathbf{R}%
^{n+1}}h_{l}(X)=0$ and $\dsum\limits_{i=1}^{n+1}X_{i}\frac{\partial }{%
\partial X_{i}}h_{l}(X)=lh_{l}(X).$ Then, if $\varsigma \in S^{n},$%
\begin{equation}
\triangle _{\mathbf{R}^{n+1}}h_{l}(X)=\triangle _{\mathbf{R}%
^{n+1}}(r^{l}Y_{l}(\varsigma ))=(\frac{\partial ^{2}}{\partial r^{2}}+\frac{n%
}{r}\frac{\partial }{\partial r}+\frac{1}{r^{2}}\triangle
_{S^{n}})(r^{l}Y_{l}(\varsigma ))=0  \tag{6.47a}
\end{equation}%
leading to 
\begin{equation}
\lbrack (l(l+n-1)+\triangle _{S^{n}}]Y_{l}(\varsigma )=0.  \tag{6.47b}
\end{equation}%
Let $\vec{X}=r\mathbf{n},$ where $\mathbf{n}$ is \ the variable unit vector
on \textbf{R}$^{n+1}.$ In 4 dimensions \textbf{n}=$\{\xi ,\eta ,\zeta ,\chi
\},$e.g$.$see Eq$.$(6.46a)$.$ The normal derivative of $h_{l}(X)$ on the
unit sphere $S^{n}$ is given by $\frac{\partial }{\partial r}%
(r^{l}Y_{l}(\varsigma ))=lY_{l}(\varsigma ).$ Application of \ Green's third
identity [121], page 333, and [$125$], page 87, permits us then to rewrite
the l.h.s.of Eq.(6.44) with help of just defined results in the following
form ($n=3$) 
\begin{equation}
\frac{1}{2\pi ^{2}}\dint\limits_{S^{3}}\frac{Y_{l}(X^{\prime })}{\left\Vert
X-X^{\prime }\right\Vert ^{2}}d\Omega ^{\prime }=\frac{1}{1+l}%
Y_{l}(X),l=0,1,2,...  \tag{6.48}
\end{equation}%
Comparison with the r.h.s. of Eq.(6.44) yields the discrete portion of the
spectrum, $p_{0}=\sqrt{-2mE}=(1+l)^{-1},$ for the hydrogen atom. To get the
continuum portion of the hydrogen atom spectrum requires us to make some
redefinitions of the already obtained results. It is convenient to represent
the continuum and discrete results side-by-side. Specifically, 
\begin{eqnarray}
\text{Discrete spectrum} &\text{:}&\text{ }  \notag \\
\text{unit sphere} &:&S^{n}=\{x\in \mathbf{R}^{n+1}\mid \left\Vert
x\right\Vert _{E}^{2}=1\};  \TCItag{6.49a} \\
\text{ \ \ \ \ \ Continuous spectrum } &\text{:}&\text{ }  \notag \\
\text{unit hyperboloid of two sheets: } &&F^{n}=\{x\in \mathbf{R}^{n+1}\mid
\left\Vert x\right\Vert _{L}^{2}=-1\}.  \TCItag{6.49b}
\end{eqnarray}%
Here $E$ stands for scalar product in space of Euclidean signature while $L$
stands for scalar product in the space of Lorentzian signature. In
particular, in the last case we have%
\begin{equation*}
<x,y>_{L}=x_{1}y_{1}+\cdot \cdot \cdot +x_{n}y_{n}-x_{n+1}y_{n+1}\text{. }%
<x,x>_{L}=\left\Vert x\right\Vert _{L}^{2}
\end{equation*}%
so that the result $\left\Vert x\right\Vert _{L}^{2}=-1$ should be looked
upon either as an option $H_{+}^{n}=\{x\in \mathbf{R}^{n+1}\mid \left\Vert
x\right\Vert _{L}^{2}=-1,x_{n+1}>0\}$ or as an option $H_{-}^{n}=\{x\in 
\mathbf{R}^{n+1}\mid \left\Vert x\right\Vert _{L}^{2}=-1,x_{n+1}<0\}.$%
Therefore$,$ $F^{n}=H_{+}^{n}\cup H_{-}^{n}.$ \ \ 

Use of stereographic projection is defined by [$81$] by analogy with Eq.s
(6.42a),(6.42b) leads to the following identifications:%
\begin{eqnarray*}
\text{Unit hyperboloid of two sheets} &\text{:}&\text{ }F^{n},\pi :\text{ }%
x_{k}=\frac{\xi _{k}}{1+\xi _{n+1}},\text{ }k=1,2,...n; \\
\pi ^{-1} &:&\xi _{k}=\frac{2x_{k}}{1-\left\Vert x\right\Vert ^{2}},\xi
_{n+1}=\frac{\left\Vert x\right\Vert ^{2}+1}{\left\Vert x\right\Vert ^{2}-1}%
,k=1,...,n.
\end{eqnarray*}%
Accordingly, for Eq.(6.44) we obtain:%
\begin{eqnarray}
\frac{1}{2\pi ^{2}}\dint\limits_{S^{3}}\frac{\Psi (X^{\prime })}{\left\Vert
X-X^{\prime }\right\Vert ^{2}}d\Omega ^{\prime } &=&p_{0}\Psi (X)\text{
(3-sphere);}  \TCItag{6.50a} \\
\text{ }\frac{-\varepsilon (\xi _{4})}{2\pi ^{2}}\dint\limits_{F^{3}}\frac{%
\Psi (X^{\prime })}{\left\Vert X-X^{\prime }\right\Vert _{L}^{2}}d\Omega
^{\prime } &=&p_{0}\Psi (X)\text{ (hyperboloid of two sheets).} 
\TCItag{6.50b}
\end{eqnarray}%
Here $\varepsilon (\xi _{4})=1$ for $\xi _{4}\geq 1$ and $\varepsilon (\xi
_{4})=-1$ for $\xi _{4}<1.$

For $E>0$ the Laplace Eq.(6.46b) is replaced by the D'Alembert Eq.(2.6a).
Accordingly, Eq.(6.47a) \ is now being replaced by [121] 
\begin{equation}
\square _{\mathbf{R}^{n+1}}[r^{\lambda }H_{N,\alpha ,\beta }(\varsigma )]=0 
\tag{6.51}
\end{equation}%
with $\lambda =-\frac{1}{2}(n-1)+iN.$ N is being a real number. Since the
solution of Eq.(6.51) presented in [121] contains many gaps in logic it is
not going to be discussed any further in this paper. However one should keep
in mind that it is this solution which is being used in solving the integral
Eq.(6.50b). Thus, the harmonic function $u=r^{n-1}\Psi _{n}(X)$ solving the
Laplace Eq.(6.46b) \ for $E<0$ is being replaced by $u=$ $r^{\lambda
}H_{N,\alpha ,\beta }(\varsigma ),$ where the function $u$ is solution of
the D'Alembert Eq.(2.6a) (or Eq.(6.51)) with the boundary conditions on $%
F^{3}$. As in the case of E$<0,$ the spherical harmonic $\Psi
(X)=H_{N,\alpha ,\beta }(\varsigma )$ solves the integral Eq.(6.50b) for $%
E>0 $. However, the solution protocol for $E>0$ is very different. For us it
is essential that the function $u$ is the solution of D'Alembert equation in
the Minkowski spacetime having the hyperboloid of two sheets $F^{3}$ as
boundary. Eq.s (6.45a) and (6.45b) are being replaced now by 
\begin{eqnarray}
\Psi (X) &=&\left[ \frac{p_{0}^{2}-\mathbf{p}^{2}}{2p_{0}^{2}}\right]
^{2}\psi (\mathbf{p}),  \TCItag{6.52a} \\
\text{or }\Psi (X) &=&\frac{p_{k}^{2}}{\text{X}_{k}^{2}}\psi (\mathbf{p}). 
\TCItag{6.52b}
\end{eqnarray}%
Suppose that solution $H_{N,\alpha ,\beta }(\varsigma )$ of Eq$.(6.51)$ of
the integral Eq.(6.50b) is found, then the solution of the D'Alembert
Eq.(6.51) is given by $r^{\lambda }H_{N,\alpha ,\beta }(\varsigma ).$ In the
case $E<0$ \ the stereographic coordinates X=\{$\xi _{1},...,\xi _{4}\}=%
\mathbf{n}=\{\xi ,\eta ,\zeta ,\chi \}$ had been artificially extended to $%
\vec{X}=r\mathbf{n}$ so that if $\Psi (X)$ defined on the 3-sphere is the
solution of the integral equation, Eq.(6.50), $\Psi (\vec{X})$ is the
solution of the Laplace Eq.(6.46b). Analogous replacement in $E>0$ case
produces solution of the D'Alembert equation from solution of the integral
Eq.(6.50b). When just described extension of stereographic coordinates is
combined with Eq.s (6.52) such hyperbolic analog of $\Psi (\vec{X})$ \
solves the problem of Hadamard triviality. This follows in view of the fact
that the Schr\"{o}dinger operator \~{L}[$\phi ],$ Eq$.(6.1b),$ is obtainable
from the D'Alembert operator L[$\phi ]$ via sequence of non-singular
transformations of independent variables (stereographic projection and
extension). Thus, the Schr\"{o}dinger operator for the time-independent Schr%
\"{o}dinger equation is Hyugens' trivial. At the physical level of rigor the
same result was obtained by I.Bars [99] using two time formalism,

Frenkel and Libine solution of hydrogen atom model resulting in\ spectrum
for bound $E<0$ and scattering $E>0$ states [$122$] improve and considerably
simplify just described results since the quaternionic Poisson formula is
relating the harmonic functions inside $S^{3}$ and $F^{3}$ with solutions of
the stationary Schr\"{o}dinger Eq.(6.1) for both $E<0$ and $E>0$ \ as
required.\bigskip \bigskip

\textbf{7. \ \ Physical uses of Lie sphere geometry. \bigskip }

\textsl{7.1. \ Bird's view of the two-times formalism of \ I.Bars \bigskip }

In [83] the program of two -times formalism was outlined while in [99]
detailed calculations illustrating general principles are presented. In
short, the main idea is to replace 3+1 Minkowski spacetime by more general
(fundamental) 4+2 spacetime having 4 space and 2 time variables so that the
signature of this a spacetime is (+,+,+,+,-,-). Evidently, the symmetry of
such spacetime is either O(4,2) or SO(4,2). In such an extended spacetime
there could possibly exist one fundamental model casting countable number of
"shadows"(projections) \ into \ much more familiar Minkowski spacetime. Each
"shadow" is perceived as known distinct particle physics model. Under an
umbrella of 2 -times formalism, it appears, that apparently different
physical models actually are having the same origin in 4+2 spacetime.
Mathematically speaking, the idea is to distribute all particles(and their
bound states) of high energy physics into equivalence classes in accordance
with the basic (fundamental) models living in 4+2 dimensional spacetime.
Since the already made calculations connect massless relativistic particles
with the massive \ ones (even accounting for the spin), the concept of a
mass of particle and significance of the Higgs mechanism for mass generation
loose their central importance in this formalism. Furthermore, the same
formalism connects, for example, the relativistic particles (massive or not)
with the non relativistic massive particles, the extended (bound) systems,
such as hydrogen atom, with the massless relativistic particles, etc.

In section 6 we demonstrated how the nonrelativistic Schr\"{o}dinger
equation for hydrogen atom is connected with the D'Alembert (in 3+1
dimensions) or the Laplace (in 4 dimensions), that is with equations
describing the massless and spinless relativistic particle. In section 5 we
demonstrated how relativistic equations for massive particles can be reduced
to those whose masses are zero. We were guided by the Hadamard ideas of
Huygens triviality casting all Huygens-trivial equations (e.g. read section
4) into \ the same equivalence class. In another paper [126] Bars connected
the hydrogen atom with the massless relativistic free particle and with the
massive 3-dimensional harmonic oscillator. Thus, he effectively demonstrated
Huygens triviality of the hydrogen-atom and harmonic oscillator. His work
develops results of [127] published previously in which Bars (with
collaborators) \ raised a question: Which of 2 times in the 2 time formalism
is the familiar time coordinate? Unfortunately for his project, no clear-cut
answer to this question was given either in [127] or in the basic reference
[83] summarizing all accomplishments of the 2-times formalism. Accordingly,
the 2-times formalism did not gain much popularity among the high energy
physicists. Nevertheless, in our opinion, upon development, e.g. based on
results of this work, there could be a chance to make formalism developed by
Bars more viable.

We initiate this process of further development with the discussion of the
relationship between the dynamics of hydrogen atom and harmonic oscillator.
Being driven by the same problem of regularization of singularities \ (e.g.
read again comments to Eq.(6.13) for the Kepler problem), Kustaanhemo and
Stiefel (K-S) using the Hopf mapping (section 6) transformed the 3
dimensional classical dynamics of the Kepler problem into the\ 4-dimensional
problem of classical dynamics of the harmonic oscillator on $S^{3}$ $[128].$
Their work was extended by many authors both at the classical and quantum
levels. At the quantum level readable account of uses of the K-S
transformation converting hydrogen atom problem into harmonic oscillator
problem is given in [$129$]. \ An independent and much simpler treatment of
the same conversion problem was given in [130] by Chen. By utilizing results
of [130], Chen demonstrated in [$131$] the isomorphism between the largest
symmetry group SO(4,2) of the hydrogen atom and the group SU(2,2) used in
physics of twistors [25]. Later on he connected group-theoretically
hydrogenic, oscillator and free-particle massless relativistic systems in [$%
132$]. \ In his derivations Chen used only one-time formalism. His results
do not use twistors or compactification of the Minkowski space. Thus, they
cannot be immediately linked \ with discussions involving the Lie sphere
geometry-a tool essential for generalization of these results. At the same
time, the formalism by Bars, perhaps if further developed, can accommodate
the concepts of Lie sphere geometry.\bigskip \bigskip

\textsl{7.2. \ Lie sphere geometry. Fundamentals\bigskip }

\bigskip\ 

The examples discussed in previous subsection indicate that 2-times
formalism invented and developed by Bars \ in some instances can be entirely
replaced by more familiar 1 time formalism. The examples of such a
replacement can hardly be generalized though. At the same time, the
diversity of results obtained with help of the 2 times formalism is
appealing. Thus, in this subsection we would like to suggest a reliable
direction enabling us to replace questionable 2-times formalism by
mathematically well developed formalism of the Lie sphere and the M\"{o}bius
geometries. In physics literature the formalism based of utilizing results
of M\"{o}bius \ geometry\ is known as "geometric algebra".

We begin with the description of Lie geometry since the M\"{o}bius geometry
(and, therefore, geometric algebra) is more restrictive. The Lie geometry is
the geometry of generalized oriented hyperspheres \ living in the
compactified Euclidean space $S^{n}$ =$\mathbf{R}^{n}\cup \{\infty \}.$The
elements of this geometry are:

a) \textsl{Oriented hyperspheres. }These are familiar from section 4 spheres 
$S_{c,r}$ of finite radius $r>0$ with center $\mathbf{c}\in \mathbf{R}^{n}$
so that 
\begin{equation}
S_{c,r}=\{\mathbf{x}\in \mathbf{R}^{n}\mid \left\Vert \mathbf{x}-\mathbf{c}%
\right\Vert ^{2}=r\}.  \tag{7.1}
\end{equation}%
The hypersphere $S_{c,r}$ divides $\mathbf{R}^{n}$ into 2 parts. If we
denote one of these parts as "positive", then another \ part is "negative".
By introducing such a distinction we are introducing the oriented
hyperspheres. In such a case a given $S_{c,r}$ is replaced by $S_{c,r}^{\pm
}.$These results generalize 3 dimensional result, Eq.(4.25), needed for
definition of canal surfaces and Dupin cyclides.

In higher dimensions the cyclides of Dupin are being replaced by the Dupin
hypersurfaces. Such hypersurfaces can live in Euclidean and hyperbolic
spaces [$133$].\bigskip

\textbf{Definition 7.1. \ }Hypersurfaces with constant principal curvatures
are \textsl{Dupin hypersurfaces}.\bigskip

To distinguish \ the oriented hyperspheres analytically it is convenient to
introduce \textsl{signed radius} which is telling us whether we should
consider the inward or outward \ field of unit normals. According to
convention, \ the positive radii $r>0$ are assigned to hyperspheres with the 
\textsl{inward} field of unit normals \ while \ the negative radii $r<0$ are
assigned to hypersheres \ with the \textsl{outward} field of unit normals. \
By doing so we just had introduced a bijection between the hypersurfaces of
non-vanishing radius and tuples 
\begin{equation}
(\mathbf{c},r),\mathbf{c}\in \mathbf{R}^{n},\text{ }r\in \mathbf{R}^{\ast }.
\tag{7.2}
\end{equation}%
Here $\mathbf{R}^{\ast }=\mathbf{R\backslash }0\mathbf{.}$

b) \textsl{Oriented hyperplanes}. A hyperplane \ $P$ in $\mathbf{R}^{n}$ is
characterized by the equation 
\begin{equation}
P=\{\mathbf{x}\in \mathbf{R}^{n}\mid <\mathbf{n},\mathbf{x}>=d\}  \tag{7.3}
\end{equation}%
with a unit normal $\mathbf{n}\in S^{n-1}$ and $d=\mathbf{R.}$ Evidently,
the tuples $(\mathbf{n},d)$ and $(-\mathbf{n},-d)$ represent the same
hyperplane. As a hypersphere, it divides $\mathbf{R}^{n}$ into two half
spaces. \ By declaring one of the two half spaces \ to be positive, we are
getting the notion of oriented hyperplane. Thus, by analogy with
hyperspheres, we obtain a splitting: $P\rightarrow P^{\pm }$ for any $P$.

c) \textsl{Points}. Points are hyperspheres of zero radius.

d) \textsl{Infinity. }Infinity\textsl{\ }point is making $\mathbf{R}^{n}$
compactified, that is $S^{n}$ =$\mathbf{R}^{n}\cup \{\infty \}.$

e) \textsl{Contact elements.} A set of \ all hyperspheres through $\mathbf{x}%
\in \mathbf{R}^{n}$ which are in oriented contact with $P$ and with one
another thus all sharing normal vector \textbf{n} at \textbf{x}.\bigskip

\textbf{Remark 7.2.} The connection/replacement with/of 2 times formalism by
Bars follows now from the observation \ that all just described elements are
modelled as \textsl{points}, respectively \textsl{lines}, in the projective
space \textbf{P}(\textbf{R}$^{n+1,2})$ with the space of homogenous
coordinates \textbf{R}$^{n+1,2}.$ Details follow below.\bigskip

From now on, for simplicity, we shall only discuss the case: $n=3$.
Following Suris [$80$] we equip the space \textbf{R}$^{4,2}$ with 6 linearly
-independent unit vectors $\mathbf{e}_{1},...,\mathbf{e}_{6}$ whose scalar
product is defined as%
\begin{equation*}
<\mathbf{e}_{i},\mathbf{e}_{j}>=\left\{ 
\begin{array}{c}
1\text{ if i=j}\in \{1,...,4\}, \\ 
-1\text{ if i=j}\in \{5,6\}, \\ 
0\text{ if i}\neq j.%
\end{array}%
\right.
\end{equation*}%
To exemplify the properties of Lie sphere geometry in the most efficient
way, it is convenient to make the following redefinitions:%
\begin{eqnarray}
\mathbf{e}_{0} &=&\frac{1}{2}(\mathbf{e}_{5}-\mathbf{e}_{4}),\mathbf{e}%
_{\infty }=\frac{1}{2}(\mathbf{e}_{5}+\mathbf{e}_{4})\text{ \ \ \ \ implying 
}  \TCItag{7.4a} \\
&<&\mathbf{e}_{0},\mathbf{e}_{0}>=<\mathbf{e}_{\infty },\mathbf{e}_{\infty
}>=0;<\mathbf{e}_{0},\mathbf{e}_{\infty }>=-\frac{1}{2}.  \TCItag{7.4b}
\end{eqnarray}%
In terms of just made definitions \ we \ redefine the objects of Lie sphere
geometry as follows

\textsl{Oriented sphere} $\hat{s}$ with center $\mathbf{c}\in \mathbf{R}^{3}$
and signed radius $r\in \mathbf{R:}$%
\begin{equation}
\hat{s}=\mathbf{c+e}_{0}+(\left\vert \mathbf{c}\right\vert ^{2}-r^{2})%
\mathbf{e}_{\infty }+r\mathbf{e}_{6}  \tag{7.5}
\end{equation}

\textsl{Oriented plane \ \ }$\hat{P}$\textsl{\ \ }defined by\textsl{\ }$<%
\mathbf{n},\mathbf{x}>=d$ with $\mathbf{n}\in S^{2}$ and $d\in \mathbf{R:}$%
\begin{equation}
\hat{P}=\mathbf{n+0\cdot e}_{0}+2d\mathbf{e}_{\infty }+0\cdot \mathbf{e}_{6}.
\tag{7.6}
\end{equation}

\textsl{Point} $\mathbf{\hat{x}}\in \mathbf{R}^{3}:$%
\begin{equation}
\mathbf{\hat{x}=x+e}_{0}+\left\vert \mathbf{x}\right\vert ^{2}\mathbf{e}%
_{\infty }+0\cdot \mathbf{e}_{6}.  \tag{7.7}
\end{equation}

\textsl{Infinity:}%
\begin{equation}
\hat{\infty}=\mathbf{e}_{\infty }.  \tag{7.8}
\end{equation}

\textsl{Contact element}:%
\begin{equation}
span(\mathbf{\hat{x}},\hat{P}).  \tag{7.9}
\end{equation}

\textbf{Remark 7.3. }In the projective space \textbf{P}(\textbf{R}$^{n+1,2})$
the first four types of elements are represented by the \textsl{points. }%
These are the equivalence classes of the above vectors with respect to the
equivalence relation $\xi \sim \eta \iff \xi =\lambda \eta .$ $\lambda =%
\mathbf{R}^{\ast },\xi ,\eta \in $\textbf{R}$^{n+1,2}.$

\textbf{Remark 7.4. \ }A contact element in \textbf{R}$^{n}$ is an \textsl{%
isotropic line} in \textbf{P}(\textbf{R}$^{n+1,2}).$This line is defined in
Eq.(7.11) below.

To be specific, we need to introduce the \textsl{Lie quadric}. This is
accomplished as follows. First, we define the set of \textsl{isotropic
vectors} in \textbf{R}$^{n+1,2}$ via 
\begin{equation}
\mathbf{L}^{n+1,2}=\{\mathbf{\hat{v}\in R}^{n+1,2}\mid <\mathbf{\hat{v},\hat{%
v}>=}0\mathbf{\},}  \tag{7.10}
\end{equation}%
then the Lie quadric \textbf{Q}$^{n+1,2}$ is defined as $\mathbf{Q}^{n+1,2}=%
\mathbf{P(L}^{n+1,2}).\bigskip $

\textbf{Remark 7.5.} \bigskip For $n=3$ the Lie quadric just defined
coincides with that defined by Eq.(6.41b) representing properties of the
compactified Minkowski space $\mathfrak{M.\ }$

It was obtained via twistor formalism \ in previous section thus hinting at
the conection between the twistor and the Lie sphere formalisms. That this
is indeed the case will be explained below, in subsection 7.4.\bigskip

Going back to contact elements, choose now two hyperspheres (or spheres if $%
n=3$) $\hat{s}_{1},\hat{s}_{2}\in \mathbf{\hat{v}}\subset $\textbf{R}$%
^{n+1,2}$ such that $<\hat{s}_{1}\mathbf{,}\hat{s}_{2}\mathbf{>=}0.$ Then,
these spheres are in \textsl{oriented contact} with each other. An
elementary proof can be found \ in [$134$], page 15. Using this fact, we can
define a line $\mathcal{L}$ in projective space $\mathbf{P(L}^{n+1,2})$. \
For this purpose \ let $\alpha _{1}$ and $\alpha _{2}$ be some real numbers.
Using these numbers we define a linear combination $\hat{s}=\alpha _{1}\hat{s%
}_{1}+\alpha _{2}\hat{s}_{2},$ a projective line. Using it, we obtain: 
\begin{equation}
<\hat{s},\hat{s}>=\alpha _{1}^{2}<\hat{s}_{1}\mathbf{,}\hat{s}_{1}\mathbf{>+}%
\alpha _{2}^{2}<\hat{s}_{2}\mathbf{,}\hat{s}_{2}\mathbf{>+}2\alpha
_{1}\alpha _{2}<\hat{s}_{1}\mathbf{,}\hat{s}_{2}\mathbf{>=}0  \tag{7.11}
\end{equation}%
since both $\hat{s}_{1}$ and $\hat{s}_{2}$ are isotropic vectors. If $\hat{s}%
_{1}$ and $\hat{s}_{2}$ in \textbf{R}$^{4,2}$ represent two spheres in
oriented contact, then the line $\mathcal{L}$ in $\mathbf{P(L}^{4,2})$
trough the corresponding points is isotropic as just demonstrated. It lies
entirely on \ the Lie quadric $\mathbf{P(L}^{4,2}).$There are no projective
subspaces of higher dimensions completely contained in $\mathbf{P(L}^{4,2}).$
This is proven in[134], page 17. In just described formalism planes are
spheres of infinite radii and points are spheres of zero radii. The
Definition 4.7. can be restated \ now as follows\bigskip

\textbf{Definition 7.6. \ }The essence of Lie geometry lies in study of \
projective transformations of \ $\mathbf{P(L}^{4,2})$ \ (for $n=3$) leaving\
the Lie quadric $\mathbf{Q}^{4,2}$ invariant. The group of such
transformations is factor group $O(n+1,2)/\{1,-1\}=PO(n+1,2)$. \bigskip

This quotient is just the higher dimensional extension \ of the earlier
discussed quotient $O(3,1)/\{1,-1\}=PO(3,1)$ \ describing a homomorphism \
of embedding of the Lorentz\ group $O(3,1$) into $C(2)$ group introduced in
connection with Eq.(6.40). These transformations are preserving the isotropy
property described \ by Eq.(7.10). \ Furthermore, the (non) vanishing of $%
\mathbf{e}_{0}$ or of $\mathbf{e}_{6}$ component of a point in $\mathbf{P(L}%
^{4,2})$ is not invariant under a general Lie sphere transformation leading
to absence of distinction in this geometry \ between the oriented spheres,
oriented planes and points.\bigskip\ 

Presented background is sufficient for reading of [135], chapter 15, where
many additional facts about the Lie sphere geometry are nicely
explained.\bigskip

\textsl{7.3. M\"{o}bius geometry and geometric algebra. Fundamentals\bigskip 
}

In this subsection we connect the Lie sphere geometry with the M\"{o}bius
geometry known in physics literature as "geometric algebra" [136]. To
demonstrate interconnection between M\"{o}bius geometry and geometric
algebra we have to provide some basics on M\"{o}bius geometry first. This is
easy to do since M\"{o}bius geometry is just part of the Lie sphere
geometry. M\"{o}bius geometry studies subgroups of Lie sphere geometry
preserving subsets of $\mathbf{P(L}^{4,2})$ with vanishing $\mathbf{e}_{6}.$
Thus, it deals with non-oriented spheres, non oriented planes, points $%
\mathbf{x}\in \mathbf{R}^{3},$ the infinity point $\infty $ compactifying $%
\mathbf{R}^{3}$ to $S^{3}.$ The elements $\mathbf{x}\in $ $S^{3}$ are in
one-to-one correspondence with the points on the projectivized light cone $%
\mathbf{P(L}^{4,1})$ where 
\begin{equation}
\mathbf{L}^{n+1,1}=\{\mathbf{\hat{v}\in R}^{n+1,1}\mid <\mathbf{\hat{v},\hat{%
v}>=}0\mathbf{\}}  \tag{7.12}
\end{equation}%
to be compared with Eq.(7.10). Points \ $\mathbf{x}\in \mathbf{R}^{3}$
correspond to points of $\mathbf{P(L}^{4,1})$ with non-vanishing $\mathbf{e}%
_{0}$ while the point $\infty $ corresponds to the only one point of $%
\mathbf{P(L}^{4,1})$ with vanishing $\mathbf{e}_{0}$ component.\bigskip

\textbf{Remark 7.7. }In view of the Remark 7.2\textbf{., }by comparing
Eq.s(7.10) and (7.12) it is clear that results of Bars can be redone only
with help of the Lie sphere geometry. Nevertheless, M\"{o}bius geometry is
used in geometric algebra which found its way into physics for some time
[136]. This fact deserves some further discussion. In particular, we begin
with the following definition\bigskip

\textbf{Definition 7.8.} The essence of M\"{o}bius geometry lies in the
study of properties of nonoriented (hyper)spheres invariant with respect to
projective transformations $\mathbf{P(L}^{4,1})$ mapping points to points,
The group of such transformations is isomorphic to $PO(n+1,1)=O(n+1,1)/%
\{1,-1\}\simeq O^{+}(n+1,1)$. It is the group of Lorentz transformations of 
\textbf{R}$^{n+1,1}$ preserving the time-like direction. Every conformal
diffeomorphism of $S^{n}=\mathbf{R}^{n}\cup \{\infty \}$ is induced by the
restriction of M\"{o}bius transformation to $\mathbf{P(L}^{n+1,1})$. Further
details can be found in [$137$].\bigskip

\textbf{Remark 7.9. \ }Although the Dupin\textbf{\ }cyclides\textbf{\ }are
described by the Lie sphere geometry in 3 dimensions [$138$], additional
studies, e.g. see for example [$139$], demonstrated that in 3 dimensions use
of the M\"{o}bius geometry and, hence, the geometric algebra is sufficient
in the sense described in the next subsection. This algebra describes all
conformal motions. In [$140$] it is \ demonstrated how the Dupin cyclides
emerge \ as sets of orbits of conformal motions. \bigskip

\textsl{7.4. \ Crown achievement of Sophus Lie- discovery of the isomorphism
between the Lie sphere and Pl\"{u}cker line geometries. \bigskip }

The essence of 2-times formalism \ invented by I. Bars and described in his
book [$83$] is summarized in Fig.7.7. of this book. In it, the 2-times
formalism \ is placed at the center of this figure while the twistor
formalism is presented \ in the upper left corner. It is depicted as \ some
kind of corollary of the 2-times formalism. We have already explained that
the 2-times formalism is nothing else but the \ Lie sphere geometry
formalism. Now, following ideas of Sophus Lie [$141$], we are going to
demonstrate that the Lie sphere formalism is \textsl{isomorphic} to the Pl%
\"{u}cker line formalism. Pl\"{u}cker formalism was discussed in section 6
in connection with twistors. \ That such an isomorphism is possible, is
hinted in the Remark 7.5. \ Now we provide the details. By describing the
Lie sphere-Pl\"{u}cker line correspondence we are establishing the
isomorphism between \ the Penrose twistors and the 2-times formalisms. Since
the twistor formalism is associated with \ the\ most of the exactly
integrable systems originating from all kinds of reductions of the Abelian
and non Abelian-Yang Mills gauge fields [$142,143$], \ it is hoped, that \
in view of this isomorphism \ it is sufficient to study properties \ of
these \ gauge fields \ in order\ to address all problems \ of high energy
physics, including those pertinent to the Standard Model and gravity.
Furthermore, the unexpected connection between the gauge-theoretic (Floer)
and Schr\"{o}dinger's formalisms noticed in subsection 4.2.3. now acquires
an independent support.

We develop Pl\"{u}cker's line geometry by analogy with the Mobius and Lie
sphere geometries. For this purpose, we recall first Eq.s(6.16)-(6.17)
describing representative elements of the exterior algebra as well as Pl\"{u}%
cker embedding \ of this algebra into complex projective space. We are
adopting general formalism to \ four dimensional complex (twistor) space 
\textbf{C}$^{4}.$ By treating this space as vector space we have \{\textbf{e}%
$_{0}$,...,\textbf{e}$_{3}$\} as the basis for $\mathcal{E\in }$\textbf{C}$%
^{4}.$ Accordingly, we also can construct a complex six dimensional space $%
\wedge ^{2}\mathcal{E}$ of bivectors: \ 
\begin{equation}
\mathcal{\wedge }^{2}\mathcal{E\sim \{}\mathbf{e}_{0}\wedge \mathbf{e}_{1},%
\mathbf{e}_{0}\wedge \mathbf{e}_{2},\mathbf{e}_{0}\wedge \mathbf{e}_{3},%
\mathbf{e}_{1}\wedge \mathbf{e}_{2},\mathbf{e}_{1}\wedge \mathbf{e}_{3},%
\mathbf{e}_{2}\wedge \mathbf{e}_{3}\}  \tag{7.13}
\end{equation}%
Notice, that in the case of the Lie sphere geometry we used also six
dimensional vector space \{$\mathbf{e}_{1},...,\mathbf{e}_{6}\}.$Therefore,
by analogy with the Lie sphere case, we define the scalar product according
to the rules [$80$]:%
\begin{equation}
<\mathbf{e}_{0}\wedge \mathbf{e}_{1},\mathbf{e}_{2}\wedge \mathbf{e}_{3}>=-<%
\mathbf{e}_{0}\wedge \mathbf{e}_{2},\mathbf{e}_{1}\wedge \mathbf{e}_{3}>=<%
\mathbf{e}_{0}\wedge \mathbf{e}_{3},\mathbf{e}_{1}\wedge \mathbf{e}_{2}>=1 
\tag{7.14}
\end{equation}%
Notice that thus far we nowhere used the fact that the underlying vector
space is complex. Thus, following [$80$] we shall initially treat it as
real. Then, based on the multiplication table given by Eq.(7.14), \ we
conclude that the signature of such a space is (3,3). Accordingly, $\mathcal{%
\wedge }^{2}\mathcal{E\simeq }\mathbf{R}^{3,3}.$ In accord with Eq.(7.10) we
introduce the set of isotropic vectors%
\begin{equation}
\mathbf{L}^{3,3}=\{\mathbf{\hat{g}}\in \mathcal{\wedge }^{2}\mathcal{E\simeq 
}\mathbf{R}^{3,3}\mid <\mathbf{\hat{g}},\mathbf{\hat{g}}>=0\}.  \tag{7.15}
\end{equation}%
The Pl\"{u}cker quadric \textbf{Q}$^{3,3}$ is now defined by \textbf{P}($%
\mathbf{L}^{3,3}).$ It was obtained in section 6, Eq.(6.38). Use of
transformations given by Eq.(6.37) brings us back to the well known Pl\"{u}%
cker embedding result, Eq.(6.18c). Notice that the\ standard Pl\"{u}cker
relation, Eq.(6.18c), does not require use of complex variables. This
happens to be essential as we shall explain momentarily. For this we recall
that Lie discovered line-spherical geometry correspondence in 1870 following
ideas of Poncelet, Pl\"{u}cker and Darboux. All these ideas, including
contributions of Sophus Lie, are presented in the book by Klein [144] while
the detailed account of the Lie Sphere geometry, Dupin cyclides, etc. is
presented in the book by Blaschke [$145$]. In particular, in addition to the
sketchy presentation of the line-sphere geometry correspondence given in
[141], Felix Klein provides accessible \ and comprehensive discussion on the
line-sphere geometry correspondence in his book [144], pages 262-274. Since
books by Blaschke and Klein are both in German, for readers convenience, we
provide only summary \ of what was accomplished which we are seamlessly
connecting with the previous material. We refer to just mentioned books for
further details.

Consider a single sphere $S^{2}$ in \textbf{R}$^{3}.$ Analytically it can be
described as follows%
\begin{equation}
x^{2}+y^{2}+z^{2}-2ax-2by-2cz+D=0.  \tag{7.16}
\end{equation}%
In addition to parameters $a,b,c,D,$ \ it is necessary to introduce the
radius $r$ of the sphere via 
\begin{equation}
r^{2}=a^{2}+b^{2}+c^{2}-D.  \tag{7.17}
\end{equation}%
These parameters are to be considered as \textsl{coordinates in the space of
spheres}. In addition, it is convenient to introduce the homogenous
coordinates 
\begin{equation}
a=\frac{\xi }{\nu },b=\frac{\eta }{\nu },c=\frac{\zeta }{\nu },r=\frac{%
\lambda }{\nu },D=\frac{\mu }{\nu }  \tag{7.18}
\end{equation}%
enabling us to insert the sphere $S^{2}$ into projective space. To do so, we
insert Eq.(7.18) into Eq.(7.17) with the result:%
\begin{equation}
\xi ^{2}+\eta ^{2}+\zeta ^{2}-\lambda ^{2}-\nu \mu =0.  \tag{7.19}
\end{equation}%
Planes (spheres of infinite radius) and points (spheres of zero radius) are
included in this parametrization. For points we have to put $\lambda =0$ in
Eq.(7.19) while for planes we require $\nu =0.$To represent Eq.(7.19) in
terms of the Lie sphere quadric \textbf{Q}$^{4,2}$ (defined after
Eq.(7.10)), it is sufficient to represent \ the combination $\nu \mu $ as $%
\nu \mu =\alpha ^{2}-\beta ^{2}.$ Then, Eq.(7.19) is replaced by 
\begin{equation}
\xi ^{2}+\eta ^{2}+\zeta ^{2}+\beta ^{2}-\lambda ^{2}-\alpha ^{2}=0 
\tag{7.20}
\end{equation}%
which is the same as previously obtained Eq.(6.41b). Questions remains: a)
Is it possible to convert the quadric \textbf{Q}$^{4,2}$ into \textbf{Q}$%
^{3,3}$ by some kind of nonsigular and \textsl{real} transformation? b) If
this is not possible, can the same result be achieved by using complex
transformations? \ In other words : Is it possible to prove that the
projective groups $PO(4,2)$ and $PO(3,3)$ are isomorphic? The answer is "NO"
in the domain of real numbers and "YES" in the domain of complex numbers.
Explicitly, consider Pl\"{u}cker's result, Eq.(6.18c), and perform the
following transformation (attributed to Lie)%
\begin{eqnarray}
l_{01} &=&\xi +i\eta ,l_{23}=\xi -i\eta ,  \notag \\
l_{02} &=&\zeta +\lambda ,l_{31}=\zeta -\lambda ,  \notag \\
l_{03} &=&\mu ,\text{ \ \ \ \ \ \ }l_{12}=-\nu .  \TCItag{7.21}
\end{eqnarray}%
Substitution of this result into Eq.(6.18c) brings us back to Eq.(7.19).
Although algebraically this substitution looks very simple, geometrically,
it is at the heart of the line-spherical geometry correspondence. All
missing details \ of this correspondence are given in the book by Klein
[144]. Thus, all results of 2-times formalism developed by Bars should be
obtainable from the twistor formalism.\bigskip \bigskip

\textsl{7.5. Some physical implications of the Lie sphere geometry \bigskip }

\textsl{7.5.1. Lie sphere vs M\"{o}bius geometry. Important facts}\bigskip\
\ \ \ \ \ \ \ \ \ \ \ \ \ \ \ \ \ \ \ \ \ \ \ \ \ \ \ \ \ \ \ \ \ \ \ \ 

To study physical implications, we need to extend results describing the
line- Lie sphere geometry correspondence presented in Klein's book [144].
This is needed because the Dupin cyclides \ were not discussed by Klein in
the context of this correspondence even though they were discussed in his
book. This is excusable in view of Definitions 4.4 and 4.5. However, in
describing the Lie sphere geometry we introduced isotropic vectors via
Eq.(7.10) and described the oriented contact between two spheres in
Eq.(7.11). A simple calculation done in [$134$], page 15, indicates that the
oriented contact condition $<\hat{s}_{1}\mathbf{,}\hat{s}_{2}\mathbf{>=}0$
is equivalent to 
\begin{equation}
\left\vert \mathbf{c}_{1}(t)-\mathbf{c}_{2}(t)\right\vert
^{2}=(r_{1}(t)-r_{2}(t))^{2}.  \tag{7.22}
\end{equation}%
We would like to test this result using the simplest Dupin cyclide described
after Definition 4.5. Using \ these results, we obtain: $\left(
r_{1}(t)-r_{2}(t)\right) ^{2}=\left( a\frac{1+t^{2}}{1-t^{2}}\right) ^{2}$.
At the same time $\left\vert \mathbf{c}_{1}(t)-\mathbf{c}_{2}(t)\right\vert
^{2}=a^{2}[\left( \frac{1-t^{2}}{1+t^{2}}\right) ^{2}+\left( \frac{2t}{%
1+t^{2}}\right) ^{2}+\left( \frac{2t}{1-t^{2}}\right) ^{2}]=\left( a\frac{%
1+t^{2}}{1-t^{2}}\right) ^{2}$. Thus, we just demonstrated that whenever
there is an oriented contact between two spheres, there is a \ Dupin cyclide
associated with such a contact. The contact between two spheres takes place
along some curve on the surface of the Dupin cyclide. Definition 4.5
indicates that the oriented contact between spheres is the result of
coincidence of envelopes originating from two different canal surfaces. \
According to Maxwell, only Dupin cyclides are being formed in this way.
Evidently, the Lie sphere geometry permits both $<\hat{s}_{1}\mathbf{,}\hat{s%
}_{1}\mathbf{>=}0$ (or $<\hat{s}_{2}\mathbf{,}\hat{s}_{2}\mathbf{>=}0\mathbf{%
)}$ and $<\hat{s}_{1}\mathbf{,}\hat{s}_{2}\mathbf{>=}0$ implying that this
geometry permits the existence of both the canal surfaces and the Dupin
cyclides. \ At the same time, the spine curve \ \textbf{c}(t) could be
closed or not. Accordingly, the most primitive canal surfaces are cylinders,
cones of revolution \ and tori. It happens that use of M\"{o}bius geometry
and, in particular, the inversion transformation, is sufficient for
generation of all known Dupin cyclides [$139,140,146$] starting with
cylinders, cones of revolution and tori. For the case of tori of revolution,
a very accessible proof is given in [147], Proposition 20.36. This fact
allows trivially reobtain the results by Friedlander [65,70] and Sym [74]
starting with cylinders, cones of revolution \ and tori as solutions of the
D'Alembert Eq.(2.6)\footnote{%
More details are given below.}.

The Dupin cyclides as \ much as D'Alembertian, Eq.(2.6), are invariants of
the Lie sphere transformations, e.g. read again the discussion around
Eq.(7.11), while the cyclides \ deform \ under M\"{o}bius transformations.
Complicated Dupin cyclides are formed from the simplest ones via application
of succession of M\"{o}bius transformations. Since the line-Lie sphere
geometry correspondence does not survive under use of M\"{o}bius
transformations, it appears, that M\"{o}bius transformations are playing an
auxiliary role. This conclusion is reinforced by the observation that in M%
\"{o}bius geometry the condition $<\hat{s}_{1}\mathbf{,}\hat{s}_{2}\mathbf{>=%
}0$ is not the contact condition given Eq.(7.22). In M\"{o}bius geometry we
have instead [80]:%
\begin{equation}
\left\vert \mathbf{c}_{1}-\mathbf{c}_{2}\right\vert ^{2}=r_{1}^{2}+r_{2}^{2}
\tag{7.23}
\end{equation}%
which is the condition of two spheres to intersect orthogonally. Notice that
the above condition does not require use of the $t$ parameter. Surprisingly,
these conclusions are wrong! To explain why, following [140], we need to
describe the conformal motion first.

\bigskip

\textsl{7.5.2. \ Fundamentals of conformal kinematics\bigskip\ }

As in physics, everything begins with mechanics and mechanics begins with
kinematics. The Euclidean kinematics is composed of rotations around some
axis followed by translation along the same axis. It is a special \ kind \
of conformal kinematics. \ The kinematics of M\"{o}bius geometry is based on
the known fact that it is an isometry of hyperbolic space [81]. This means
that in such geometry circles are geodesics. They are analogs of straight
lines in Euclidean geometry. Using this observation, \ we cover a plane 
\textbf{R}$^{2}$ (or sphere $S^{2})$ with \ a network of mutually orthogonal
circles (instead of Euclidean plane being covered \ by a lattice made of
mutually orthogonal straight lines). On thus formed \ Euclidean (or non
Euclidean) lattice we can move along geodesics, that is along the segments
of straight lines or along the segments of circles. In both cases the result
of elementary motion along the segment followed by motion along the
orthogonal segment can be interpreted as rotation. In going from two to
three dimensions, in Euclidean geometry, the planar square lattice is
replaced by three dimensional lattice. In hyperbolic space we have to cover%
\textbf{\ R}$^{3}$ by the network of mutually orthogonal spheres. This
explains at once the significance of Eq.(7.22) in M\"{o}bius geometry. By
avoiding technicalities clearly explained in [140] we only notice that any
three dimensional conformal motion can me made of two commuting elementary
motions. These are analogs of Euclidean translation along the axis and
rotation around the same axis (screw motion). Consider now the most
elementary Dupin cyclide-the torus. For the torus the meridians and
parallels are mutually orthogonal. Both are circles. But circles are
crossections of spheres! Accordingly, if we embed such a torus into \textbf{R%
}$^{3}$ (or, better, in $S^{3})$ foliated by mutually orthogonal \ spheres
in such a way that meridians and parallels of the torus are the crossections
of the associated spheres, then we can think about the conformal
trajectories originating on the surface of such a torus. In view of results
of previous subsection, any nontrivial cyclide of Dupin is obtainable from
torus via operation of M\"{o}bius inversion. This means the following:

a) The network of mutually orthogonal circles on the torus will transfer to
the network of mutually orthogonal circles on the cyclide.

b) The set of all orbits originating at the torus will form two dimensional
surface -Dupin cyclide. If we look only at some orbits, they are developing
at the surface of Dupin cyclide.

c) When an initial point \ \textbf{x}(0) is given, it determines both the
orbit and the type of cyclide.\bigskip

\textbf{Remark 7.10.} The above statement a) can be formulated as a theorem.
Originally it was known as \textsl{conjecture} (attributed to Ulrich
Pincall): Cyclides of Dupin are the only surfaces \ in Euclidean space on
which two families of orthogonal circles lie. Although \ we explained why
this is so above, the full proof was given by Thomas Ivey \ [148] \ in 1995.
This proof explains why many researchers, e.g. read [140] or [149], use M%
\"{o}bius geometry for generating Dupin cyclides. Surely, the same results
are achievable with the help of Lie sphere geometric methods as explained in
[150],[151].\bigskip

Just obtained results can be used for reobtaining in the most economical
(physical) way both -the result by by Friedlander [70] discussed in section
4, and \ that for knots made of null fields discussed both in
electrodynamics [152] and in quantum mechanics [14]. Details are given in
the next subsection and Appendices B and C.\bigskip

\textsl{7.5.3. Knots and Dupin cyclides in quantum mechanics and
electrodynamics}

\bigskip

We provided an evidence for existence of Dupin cyclides in quantum mechanics
and in electrodynamics in section 4 and, from different perspective, in
Appendix C. \ We would like \ now to \ reobtain results by Friedlander using
physical arguments. \ Following [14], we begin with the observation that the
set of source-free Maxwell equations in the vacuum can be\ compactly
rewritten as 
\begin{equation}
i\frac{\partial }{\partial t}\mathbf{F}=c\mathbf{\nabla }\times \mathbf{%
F,\nabla \cdot F=0.}  \tag{7.24}
\end{equation}%
Here $c$ is the speed of light, $\mathbf{F}=\mathbf{E}+i\mathbf{H}$ is the
Riemann-Silberstein vector involving both the electric \textbf{E} and
magnetic \textbf{H} fields. By representing this vector as 
\begin{equation}
\mathbf{F}(r,t)=\mathbf{F}_{+}+\mathbf{F}_{-},\mathbf{F}_{\pm
}(r,t)=\dint\limits_{0}^{\infty }d\omega e^{\pm i\omega t}\mathbf{F}_{\pm
\omega }(r)  \tag{7.25}
\end{equation}%
the above set of Maxwell's equations is converted into 
\begin{eqnarray}
\mathbf{\nabla }\times \mathbf{F}_{\omega } &=&k\mathbf{F}_{\omega }, 
\TCItag{7.26a} \\
\mathbf{\nabla \cdot F}_{\omega } &\mathbf{=}&\mathbf{0,}  \TCItag{7.26b}
\end{eqnarray}%
where $\omega $ is +$\omega $ or -$\omega $ and $k=\omega /c$ . In \ plasma
physics Eq.(7.26a) is known as "force-free equation" while in hydrodynamics
it is known as "Beltrami equation". Eq.(7.26a) was discussed in detail in
our book [55] in connection with problems emerging in contact geometry. By
applying the curl operator to both sides of Eq.(7.26a) and taking into
account Eq.(7.26b) we obtain: 
\begin{equation}
\nabla ^{2}\mathbf{F}_{\omega }+k^{2}\mathbf{F}_{\omega }=0.  \tag{7.27}
\end{equation}%
This \ vector version of the Helmholtz equation should be compared with its
more familiar version, Eq.(3.2), discussed in section 3. This (scalar)
version is used both in quantum mechanics and in electrodynamics. The way to
\ relate these two equations to each other is nontrivial [14]. In part, it
is based on the useful identity%
\begin{equation}
(\nabla ^{2}+k^{2})(\mathbf{r}\cdot \mathbf{F}_{\omega })=2\nabla \mathbf{F}%
_{\omega }+\mathbf{r\cdot }(\nabla ^{2}+k^{2})\mathbf{F}_{\omega }. 
\tag{7.28}
\end{equation}%
In [14] it is shown how the vector, Eq.(7.27), can be restored from the
scalar Eq.(3.2) in which $\Psi =(\mathbf{r}\cdot \mathbf{F}_{\omega }).$ By
applying operation div to both sides of Eq.(7.26a) \ and by assuming that $%
k=const=$ $\kappa (x,y,z)$ \ we obtain, div($\kappa \mathbf{F}_{\omega })=%
\mathbf{F}_{\omega }\cdot \nabla \kappa =0.$ Let $\mathbf{r}%
(t)=\{x(t),y(t),z(t)\}$ be some trajectory on the surface $const=$ $\kappa
(x,y,z).$ Since $\frac{d}{dt}\kappa (x(t),y(t),z(t))=v_{x}\kappa
_{x}+v_{y}\kappa _{y}+v_{z}\kappa _{z}=\mathbf{v}\cdot \nabla \kappa =0,$ by
identifying $\mathbf{v}\rightleftarrows \mathbf{F}_{\omega }$ we conclude
that the "velocity" is always tangential \ to the surface $const=$ $\kappa
(x,y,z).$ Since the vector field \textbf{v} is assumed to be nowhere
vanishing, the surface $const=$ $\kappa (x,y,z)$ can only be a torus T$^{2}$%
. In the case if $\kappa $ is rational number the field lines \textbf{v} on T%
$^{2}$ are closed implying that these lines are forming torus knots. \ Using
results of previous subsection and applying M\"{o}bius transformation we end
up with torus-like knots wound around cyclides of Dupin (that is around the
distorted tori). Depending on the type of M\"{o}bius transformation,
different "wrapped" Dupin cyclides will be formed. But this is exactly the
result of the paper by Friedlander [70]!\bigskip

\textbf{Remark 7.11. }Use of conformal transformation for generation of
knots/links was suggested initially by Bateman and Cunningham in 1910. Their
results were \ recently rediscovered and utilized \ for knot/link generation
in [152]. See also [153]. No mention of Dupin cyclides, etc. was made in
these references.\bigskip

\textbf{Remark 7.12 }Results of ref.s [152] and [153] demonstrate how the
torus knots/links can be generated from null and not null electromagnetic
fields (discussed in some detail in in the Appendix B). Being technically
permissible, such torus knot generation is not allowed for the non null
fields on physical grounds.

\bigskip

\textsl{7.6. \ \ Place of Lie sphere \ geometry \ in development of
foundations of AdS-CFT correspondence }

\bigskip

We conclude our paper with a brief discussion of the foundations of AdS-CFT
correspondence. An excellent physical introduction to this subject is given
in the recent book by Nastase [$154$]. Some mathematical aspects of this
correspondence are discussed in our work [$155$]. \ Use of these references
allow us \ to reduce \ our discussion to the absolute minimum. In \ physical
and mathematical literature on AdS-CFT correspondence \ no mention of
relevance of the Lie sphere geometry exist to our knowledge. Thus, what
follows below is the fist attempt at elimination of this deficiency.

We begin with familiar examples, e.g. discussion of the simplest model of
hyperbolic space \textbf{H}$^{2}.$ It is known in literature \ as the
Poincare$^{\prime }$ disc model \textbf{D}$^{2}$ \ Geodesics in this model
are made of the \textsl{horocycles. }These are circular segments whose both
ends lie at the boundary $S_{\infty }^{1}$ of the disc $\mathbf{D}^{2}$. The
boundary $S_{\infty }^{1}$ is considered to be "the spatial infinity"$.$ By
the appropriate choice of constants $a,b,c,d$ in the M\"{o}bius
transformation $f(z)$ given by $f(z)=\frac{az+b}{cz+d},z\in \mathbf{C},%
\mathbf{C}=\mathbf{R}^{2}\cup \{\infty \},$ the Poincare$^{\prime }$ disc
model \ can be transformed into Poincare$^{\prime }$ half plane model of the
hyperbolic space\ \textbf{H}$^{2}$. Clearly, use of the inverse conformal
transformation converts the half plane model back into the disc model. \ Let 
$z$ be some point inside such disc model. Successive applications of the M%
\"{o}bius transformation will propel this point to the spatial infinity $%
S_{\infty }^{1}$ after infinite number of iterations. For detailed examples,
please, read \ [$156$]. This two-dimensional model is generalizable to
higher dimensions where it is known as the \textsl{hyperbolic ball model}
[81]. Thus, when interested in higher dimensions, we need to replace a
combination (\textbf{H}$^{2},S_{\infty }^{1})$ by (\textbf{H}$%
^{n+1},S_{\infty }^{n}).$ \ In \ all described models of hyperbolic geometry
the boundary at infinity is playing an important role. This role can be seen
already in two dimensional model (\textbf{H}$^{2},S_{\infty }^{1}).$ In it,
the deformations of $S_{\infty }^{1}$ lead to the Virasoro algebra -central
object of the conformal field theory [$155,157$], the Teichm\"{u}ller
spaces, etc. Thus, already at the level of two dimensions we are dealing
with a kind of AdS-CFT correspondence. \ By analogy with two dimensions, it
is expected \ that the hyperbolic-like interior- actually, the anti--de
Sitter spacetime (AdS) (in higher dimensions)- is affected by \ (linked
with) \ the conformal field theory (CFT) residing at the boundary. The
higher dimensional AdS-CFT correspondence \ in higher dimensions cannot be
rigorously developed based on the hyperbolic ball model though. This is
because of the Mostow rigidity theorem. The Mostow rigidity makes
deformations of conformal 3-sphere $S_{\infty }^{3}$ impossible to perform.
Nevertheless, the mathematics of this model is serving as a gudeline for
more realistic models as demonstrated by Frances [$158$]. The bottom line is
the following. The hyperbolic space \textbf{H}$^{n}$ is replaced by the
Anti-de Sitter space (AdS)-the Einstein space of constant negative
curvature. This space is obtainable from the vacuum Einstein equations with
added (negative) cosmological constant [155]. The hyperbolic ball at
infinity $S_{\infty }^{n}$ is now replaced by the space of conformally flat
solutions Ein$^{n,1}$ of Einstein's equations. Being conformally flat these
are conformally equivalent to Minkowski spacetime, that is to say their \
conformal symmetry group is SO(4.2). In the Appendix D we rigorously
demonstrate that 
\begin{equation}
Ein^{n,1}=\partial ^{\infty }AdS^{n+2}=\mathbf{Q}^{n+1,2}.  \tag{7.29}
\end{equation}%
Here $\partial ^{\infty }AdS^{n+2}$ is the boundary of the Anti-de Sitter
space while $\mathbf{Q}^{n+1,2}$ is the Lie sphere quadric introduced after
Eq.(7.10).

\bigskip

\textbf{Acknowledgement }\ Authors are deeply grateful to Professor (Dr.)
David Delphenich for supplying us with the original of Madelung's paper,
ref. [177], as well as with the English-German translation of this paper.
Authors are also very thankful to the unknown referee for his thoughtful
remarks which lead to considerable improvement of our presentation.

\textbf{Note added in proof. \ }The\textbf{\ }interplay beween the
differential and the Lie Sphere geometries, the rigidity and quantization
(and hence mass generation) in the context of the transversal knots
(discussed in our earlier work, Ann.Phys.371 (2016) 77) is developed in
great detail in two latest works by E.Musso and L.Nicolodi (Comm.Anal.Geom.
25 (2017) 209 and Nonlin.Anal. 143 (2016) 224). The connection between knots
and the Regge-mass spectrum is discussed in our earlier work ( Int. J.Mod.
Phys. A 30 (2015), id. 1550189-212).

\textbf{\bigskip }

\bigskip

\bigskip

\bigskip \textbf{Appendix A}

\textbf{Comparison between the Cauchy problems for parabolic}

\textbf{and hyperbolic equations}

\textbf{\bigskip }

1. \textsl{Phase velocity}. Without loss of generality, we consider partial
differential equations (PDE) involving just two variables $t$ and $x$.
Suppose that solution $u(t,x)$ can be represented in the traveling wave form
[45] as follows: $\ u(t,x)=f(x-ct).$Here $c$ is the velocity of the profile $%
f$. More generally, in the case when $\mathbf{x}=\{x_{1},...,x_{n}\}$ we get 
$u(t,\mathbf{x})=f(\mathbf{k}\cdot \mathbf{x}-\omega t).$It is called 
\textsl{plane wave} with \textsl{wavefront} normal to \textbf{k, }\textsl{%
speed}\textbf{\ \ }$c\mathbf{=}\frac{\omega }{\left\vert \mathbf{k}%
\right\vert }$ ($\QTR{sl}{phase}$ $\QTR{sl}{velocity}$), and profile $f$.

2. \textsl{Exponential solutions for simplest PDE's and group velocity}%
\textbf{. }When studying linear partial differential equations, it is
convenient to specify the profile function $f$ \ in order to study the
complex-valued plane wave solutions of the form%
\begin{equation}
u(t,\mathbf{x})=\exp \{i(\mathbf{k}\cdot \mathbf{x}-\omega t)\}.  \tag{A.1}
\end{equation}%
Using such an ansatz in the diffusion equation leads to 
\begin{equation}
u_{t}-\triangle u=(-i\omega +\left\vert \mathbf{k}\right\vert ^{2})=0 
\tag{A.2}
\end{equation}%
resulting in the dispersion relation $i\left\vert \mathbf{k}\right\vert
^{2}=\omega .$ Anticipating generalizations, this result can be rewritten as 
$\omega =\omega (\left\vert \mathbf{k}\right\vert ^{2}).$Therefore Eq.(A.1)
can be presented in the form 
\begin{equation}
u(t,\mathbf{x})=\exp \{i(\mathbf{k}\cdot \mathbf{x}-\omega (\left\vert 
\mathbf{k}\right\vert ^{2})t)\}.  \tag{A.3}
\end{equation}%
This representation allows us to introduce the group velocity \textbf{c}$%
_{g} $ as 
\begin{equation}
\mathbf{c}_{g}=\nabla _{\mathbf{k}}\omega (\left\vert \mathbf{k}\right\vert
^{2}).  \tag{A.4}
\end{equation}%
In the case of simplest Schr\"{o}dinger's equation ($m=1,\hbar =1)$ $%
iu_{t}+\triangle u=0,$ the dispersion relation is given by 
\begin{equation}
\left\vert \mathbf{k}\right\vert ^{2}=\omega .  \tag{A.5}
\end{equation}%
Therefore, the group velocity is $\mathbf{c}_{g}=2\mathbf{k.}$ Upon
restoring the usual system of units, we obtain: $E=\hbar \omega =\frac{%
\left( \hbar \mathbf{k}\right) ^{2}}{2m}$ implying $\omega (\mathbf{k}^{2})=%
\frac{\hbar \mathbf{k}^{2}}{2m}.$ Accordingly, $\nabla _{\mathbf{k}}\omega
(\left\vert \mathbf{k}\right\vert ^{2})=\mathbf{c}_{g}=\frac{\hbar \mathbf{k}%
}{m}.$ Using the De\ Broglie relation: $p=\hbar k,$ we conclude that the
group velocity\textbf{\ c}$_{g}$ coincides with the particle velocity in
quantum mechanics.

Consider now the wave equation using the ansatz (A.1). We obtain instead:%
\begin{equation}
u_{tt}-\triangle u=(-\omega ^{2}+\left\vert \mathbf{k}\right\vert ^{2})=0, 
\tag{A.6}
\end{equation}%
resulting in 
\begin{equation}
\left\vert \mathbf{k}\right\vert ^{2}=\omega ^{2}.  \tag{A.7}
\end{equation}%
From here the phase velocity c is obtained as $c=\frac{\omega }{\left\vert 
\mathbf{k}\right\vert }=const$ and the modulus of group velocity \textbf{c}$%
_{g}$ coincides with the phase velocity.

3. \textsl{Cauchy problems}\textbf{. }\ The Cauchy problem for the 1
dimensional diffusion equation is formulated as follows. For the diffusion
equation 
\begin{equation}
u_{t}-Du_{xx}=0  \tag{A.8}
\end{equation}%
defined on the whole line -$\infty <x<\infty ,$ whose solution at the time $%
t=0$ is $u(0,x)=\phi (x),$ find a solution $u(t,x)$ for $t>0$ \ and $-\infty
<x<\infty .$The solution of this problem is given by 
\begin{equation}
u(t,x)=\dint\limits_{-\infty }^{\infty }dyS(x-y,t)\phi
(y)=\dint\limits_{-\infty }^{\infty }dzS(z,t)\phi (x-z)  \tag{A.9a}
\end{equation}%
Here $z=x-y$, \ and%
\begin{equation}
S(z,t)=\frac{1}{\sqrt{4\pi Dt}}\exp (-z^{2}/4Dt).  \tag{A.9b}
\end{equation}%
The last two equations can be combined into 
\begin{equation}
u(t,x)=\dint\limits_{-\infty }^{\infty }dp\exp \{-\frac{p^{2}}{4}\}\phi (x-p%
\sqrt{Dt}).  \tag{A.10.}
\end{equation}%
The analogous problem for the Schr\"{o}dinger's equation is obtained now by
replacing $t$ by $it$ in the above result. This fact is helpful for the
following reason. Suppose that the initial condition $u(0,x)=\phi (x)$ is
zero everywhere, except on some small interval $(a,b)$. Because the
exponential function is never zero, the integral, Eq.(A.9a), is not going to
be zero, even for $t\rightarrow 0^{+}$and $x$ arbitrary far from ($a,b$).
This can be restated as

\medskip

\textbf{Infinite propagation speed} [45]. The initial condition $u(0,x)=\phi
(x)$ affects the solution $u(t,x)$ for all $x$ no matter how small $t$ is.
Thus, heat propagates with infinite speed.\medskip

\bigskip

Evidently, the replacement of $t$ by $it$ in the above results is leading to
the Schr\"{o}dinger equation. But use of the same arguments as in the
diffusion case are not changing this conclusion. This fact is one of the
sources of \ Einstein's spooky action at the distance.

\bigskip

Diffraction of light-main evidence of quantum mechanical behavior- uses
time-independent Helmholtz Eq.(3.2) of the main text. This equation emerges
in both quantum mechanics and in optics. \ However, if one is willing to
study the Cauchy problem for the wave equation $u_{tt}-u_{xx}=0$ defined on
\ -$\infty <x<\infty $, it should \ be formulated as follows. In addition to
the initial data $u(0,x)=\phi _{1}(x)$ one has to supply the initial
velocity $u_{t}(0,x)=\phi _{2}(x).$Thus, while for the well posedness of the
Cauchy problem for the diffusion equation one needs just one initial
condition, one needs two initial conditions for the wave equation. At the
mathematical level no more comments are required. At the physical level,
more comments are needed. These are originating from the fact that, say,
optical waves are generated by atoms [$16$]. The two -level atomic systems -
primary sources of photons-require for their description the time-dependent
Sch\"{o}dinger equation which, is identical in its structure with the Pauli
equation for spin 1/2 particle in varying magnetic field [$16,31,55$]. For
such two-level system two initial conditions are required. In addition,
technologically it is important to generate single polarized photons and to
use the 50:50 beam splitter to cause such a single photon to be
self-entangled \ [$160$]. Thus, even though a dispersion relation Eq.(A.7)
originating from the wave Eq.(A.6) can be used for description of a single
photon \footnote{%
E.g. via the de Broglie relation}, in quantum optics polarization of photons
is exploited essentially [16] causing use of the Pauli-type Schr\"{o}dinger
equation for their description. In such a case quantum phenomenon of
entanglement [$160$] coexist with the classical fact of finite speed
propagation of wave signals.\bigskip

\textbf{Appendix B\bigskip }

\textbf{Unimaginable universality of \ Hadamard premises \bigskip \bigskip }

B.1. \textsl{Brief review of Hadamard's premises\bigskip }

There are 3 "premises" formulated by Hadamard. E.g. read [162], pages
445-450. These are:\bigskip

a) \textsl{Major premise}. \ Suppose we are observing events within the time
interval $0<t<t_{0}.$ In order to find a state at the moment $t=t_{0},$ we
need to know a state at the time $t=0$, then find a state at the time $%
t=t^{\prime }$ and, using this information, find a state at $t=t_{0};$

$b)$ \textsl{Minor premise}. Suppose \ within a short period of time $%
\varepsilon \geq t\geq 0$ there emerges some light disturbance localized in
the vicinity of point $O$ then, at time $t=t^{\prime }$ this disturbance \
will be concentrated in the very\ thin spherical layer \textsl{\ enclosing}
a sphere of radius $\omega t^{\prime }$ centered at $O$ ;

c) \textsl{Corollary}. In order to evaluate the action of the initial light
perturbation, located at the point O at $t=0$, it is permissible to replace
this perturbation by a set of perturbations emerged at time $t=t^{\prime }$
and distributed on the surface of the sphere centered at O and having radius 
$\omega t^{\prime }.\bigskip $

According to Hadamard, different writers identified Huygens' principle
either with a) (e.g. Feynman, as described in section 1) or with b) (e.g.
read our section 4) or with c). In his early work \ on hyperbolic equations
and Huygens' principle Hadamard considered only such wavefronts \ which did
not contain any singularities known as caustics. \ The major premise does
not exclude existence of caustics. Therefore, what is considered \ by
mathematicians as "Huygens' principle", corresponds to Hadamard's minor
premise and to corollary. In such a form it is discussed in the main text,
e.g. read Definition 4.2. Alternatively, read [68], page 138. In short,
Hadamard restricted himself by studying of the wake-free waves. Situations
when wakes are present (leading to caustics) is considered in [50]. Much
more comprehensive and detailed treatment (but not mathematically well
supported) is given in [163].The same but mathematically supported is given
in [59]. \textsl{Huygens triviality} is defined in Definition 4.3. Following
this definition, the \textsl{Hadamard conjecture }is formulated as follows:
\medskip

\textsl{Hadamard conjecture}. \textit{Every Huygens operator is trivial}.
\medskip \medskip

For purposes of quantum mechanics it is almost always sufficient to use the
results of Theorem 4.1. In this appendix we would like to explain that in
the broader (but physically \ also very relevant) context Theorem 4.1.
sometimes fails. This is so if, following [68], we reformulate Theorem 4.1.
as \bigskip

\textbf{Theorem 4.1.1. \ }\textit{If equation of hyperbolic type in
conformally flat Minkowski (3+1) spacetime satisfies Huygens principle, then
it is equivalent of the wave Eq.(2.6a).\bigskip }

The new element here is mention of conformal flatness. The wave Eq.(2.6a) is
surely conformally invariant, e.g. read [82]. Based on this fact, a broader
question can be posed\bigskip :

\textsl{Hadamard-like conjecture}\textbf{.} \textit{Is it true \ that the
Huygens principle holds \ if the underlying equation is conformally
invariant?\bigskip\ That is to say:Is the conformal invariance and the
Huygens principle are equivalent statements (or interdepenent
concepts)?\bigskip }

Stated \ still a bit differently, the above conjecture can be formulated as
follows.\bigskip

\textbf{\ }\textsl{Hadamard-like conjecture. Second version}.\textit{\ Is it
always true that the Huygens principle is obeyed whenever it is possible to
relate the conformally invariant equation of the 2nd order to the wave
Eq.(2.6a)? \bigskip \bigskip }

B.2. \textsl{Conformal invariance versus the Huygens principle\bigskip }

In any given dimension $d$, say, in $d=3+1$, all conformal groups are
classified in flat and curved spaces [164]. \ Many important additional
details are given in [165] and [166]. Accordingly, all conformally invariant
equations can be explicitly written for all Einstein spaces of general
relativity. If this is so, will equations with different conformal groups
lead to the same Eq.(2.6a)?

It is instructive to provide some details to answer this question. Following
[166] we introduce the second order differential (Beltrami) operator $\Delta
_{2}$ as follows%
\begin{equation}
\Delta _{2}u=g^{ij}(\frac{\partial ^{2}u}{\partial x^{i}\partial x^{j}}%
-\Gamma _{ij}^{k}\frac{\partial u}{\partial x^{k}}).  \tag{B.1}
\end{equation}%
With help of this operator any 2nd order linear differential equation can be
written as 
\begin{equation}
\Delta _{2}u+a^{i}(x)\frac{\partial u}{\partial x^{i}}+c(x)u=0.  \tag{B.2}
\end{equation}%
In the same metric conformally invariant 2nd order equation can be written
(in 3+1 dimensions) as 
\begin{equation}
\Delta _{2}u+\frac{1}{6}Ru=0,  \tag{B.3}
\end{equation}%
where $R$ is the scalar curvature. At the same time, using [165],[166] it is
possible to prove the following: \bigskip

\textbf{Theorem B.1.} \textit{Any spacetime }$\mathcal{M}_{n+1},n\geq 3,$%
\textit{\ with a given metric g possesses \textbf{nontrivial} group of
conformal motions if and only if this spacetime \ is conformally equivalent
to the Lorentzian spacetime.\bigskip\ The most general Lorentzian spacetime V%
}$_{3+1}$\textit{\ is described by the metric}%
\begin{equation}
ds^{2}=(dt)^{2}-(dx_{1})^{2}-\dsum%
\limits_{i,j=2}^{4}a_{ij}(x_{1}-t)dx^{i}dx^{j}.  \tag{B.4}
\end{equation}%
Here $a_{ij}$ is the positively definite matrix. \bigskip

\textbf{Corollary B.2. }\textit{Any Lorentzian spacetime with nontrivial
conformal group is conformally equivalent to the space \ for which the Ricci
tensor }$R_{ij}=0.\bigskip $

\textbf{Corollary B.3.} \textit{Every 3+1 Lorentzian spacetime V}$_{3+1}$%
\textit{with nontrivial conformal group admits the 2nd order partial
differential equation } 
\begin{equation}
u_{tt}-u_{xx}-f(x-t)u_{yy}-2\varphi (x-t)u_{yz}-u_{zz}=0  \tag{B.5}
\end{equation}%
\textit{satisfying Huygens' principle}.\bigskip

\textbf{Remark B.4. }\ It is clear that for a special choice of functions $f$
and $\varphi $ Eq.(B.5) will coincide with the wave Eq.(2.6a). However for
other choices these two equations are not always conformally equivalent
since they \ might involve different groups of conformal motions. This will
be further explained below.

\textbf{Remark B.5}. We brought to our reader's attention Eq.(B.5) because
of the remarkable paper by Ward [167] allowing us to connect the results of
this appendix with those in the main text.\bigskip

In section 4 we introduced and discussed the progressive wave solutions of
the wave equation as well as the Dupin cyclides. In section 7 we reobtained
Friedlander's results [70] in much simpler way \ by using methods of
conformal and contact geometries. Now, using these methods, we want to
rederive Ward's results having additional purposes in mind.\ 

In his paper Ward was not discussing the Huygens principle or its connection
with the conformal invariance. Ward's purpose was to demonstrate that if
instead of the wave Eq.(2.6a) one considers the 2nd order wave-like equation
\ with metric taken from the plane-wave gravitational background, one still
can apply the progressive wave solution method by Friedlander to this
equation and to reobtain all types of Dupin cyclides discussed by
Friedlander. Based on results of this appendix and those in sections 4 and 7
of the main text, Ward's goals can be restated as follows:\bigskip

\textbf{Question B.6.} \textit{If the wave Eq.(2.6a) admits progressing
waves solutions resulting in Dupin cyclides, is the same is true for the
Eq.(B.5)}? \bigskip

In our discussion of Eq.(B.5) we have not touched the subject \ of plane
gravitational waves while Ward was not looking any further at the origins of
Eq.(B.5) which he interpreted in terms of \ the plane gravitational pp waves
background. \ We would like now to complement Ward's results with missing
details. By doing so we simplify Ward's results as well and explain why \
plane gravitational waves are fundamentally relevant for our paper.\bigskip

B.3. \ \textsl{Conformal groups and Lie sphere groups, Rainich, Misner and
Wheeler}

\ \ \ \ \ \ \ \ \textsl{geometrodynamics of coupled Einstein-Maxwell fields,
the AdS-CFT }

\ \ \ \ \ \ \ \ \textsl{correspondence and Penrose limit\bigskip }

We begin \ our discussion with results summarized in [168]. From chapter 3
of this book it follows that the flat Minkowski metric%
\begin{equation}
ds^{2}=-(dt)^{2}+(dx)^{2}+(dy)^{2}+(dz)^{2}  \tag{B.6a}
\end{equation}%
can be converted into%
\begin{equation}
ds^{2}=-2dudv+2d\zeta d\bar{\zeta}  \tag{B.6b}
\end{equation}%
with help of substitutions $u=\frac{1}{\sqrt{2}}(t-z),v=\frac{1}{\sqrt{2}}%
(t+z),\zeta =\frac{1}{\sqrt{2}}(x+iy).$ From chapter 17 of the same book we
obtain the metric for the pp-gavitational waves (obtained for the first time
by Brinkman in 1925 and interpreted \ in terms of gravitational waves by
Peres in 1959):%
\begin{equation}
ds^{2}=-2dudv-2H(\zeta ,\bar{\zeta},u)du^{2}+2d\zeta d\bar{\zeta}.  \tag{B.7}
\end{equation}%
The physical meaning of the perturbational term $H(\zeta ,\bar{\zeta},u)$ is
explained in detail in [169],[169]. From these references it follows that
solutions of Einstein's field equations leading to the metric of the type
given by Eq.(B.7) originate from the exact solution of the coupled
Einsten-Maxwell fields. E.g.read pp 324-325 of [168] and ch.r 24 of [169].
Analysis of solutions for these fields made in [170]-[171] indicates that
coupled Einstein-Maxwell fields can be of two types: a) non null and b)
null. The null electromagnetic fields are such for which \textbf{H}$^{2}-%
\mathbf{E}^{2}=0$ and \textbf{E}$\cdot \mathbf{H}=0.$ We used null
electromagnetic fields in [71,96] in connection with formation of torus-like
knots while we explained their relevance to the Schr\"{o}dinger fields in
[14] and in section 7.5.3. Although the content of just cited references
reproduces many results of Ward's paper [167], we still have to add several
important pieces of information. This is so, because we have not made yet a
connection between metrics Eq.(B.7) and (B.4).

Toward this goal, we begin with the observation that the exact solution of
the coupled Einstein-Maxwell fields was obtained by Rainich [$172$] in 1925
and rediscovereed by Misner and Wheeler at the end of 1950ies [173]. They
reformulated results by Rainich in the context of \textit{geometrodynamics}.
In short, this means that the exact solution of Einstein's equations for the
coupled gravity-Maxwell fields is obtainable in terms of the metric from
which it is possible to restore both the gravitational and electromagnetic
fields which are non null. The solution for the null fields was obtained
much later [170],[171].

We are iterested in reinterpreting these null field results in terms of the
Lie sphere geometry discussed in section 7 and Appendix D. To do so,
initially we are following the paper by Kuiper [174]. From this paper it
follows that the (pseudo) Riemannian space \ $\mathcal{M}_{d}$ of
dimensionality $d$ is \textsl{flat }if some region about any point $x\in 
\mathcal{M}_{d}$ \ if it can be covered by a metric -preferred coordinate
system such that 
\begin{equation}
ds^{2}=g_{ij}(x)dx^{i}dx^{j}.  \tag{B.8}
\end{equation}%
Above, $g_{ij}=e_{i}\delta _{ij}$, $e_{i}=+1$ or$-1$ for all points of $%
\mathcal{M}_{d}.\medskip $

\textbf{Definition B.7.}\textit{\ A (pseudo) Riemannian space }$\mathcal{M}%
_{d}$\textit{\ is called conformally flat if \ some region about any point
any point }$x\in M_{d}$\textit{\ can be mapped conformally into a flat
space. A \textsl{conformally -flat} space is obtainable from the \textsl{flat%
} space \ via change of metric,that is}%
\begin{equation}
\tilde{g}_{ij}(x)=\omega ^{2}(x)g_{ij}(x),  \tag{B.9}
\end{equation}%
\textit{where }$g_{ij}(x)$\textit{\ is the metric tensor defined in Eq.(B.8)
while }$\omega ^{2}(x)$\textit{\ \ is some positive function of }$x$\textit{.%
}

\medskip

When $\omega ^{2}(x)$ is constant, then the above transformation is called
"similarity" or "homothety". When $\omega ^{2}(x)=1,$then the $\hom $othety
is an "isometry". Obviously, these definitions are not restricted to the
diagonal metric tensor defined by Eq.(B.8).\medskip

\textbf{Theorem B.8.(}\textsl{Liouville theorem}\textbf{) }\textit{For
spaces of dimensionality \TEXTsymbol{>}2 under the conformal transformations
described by Eq.(B.9) the (hyper) spheres are carried to (hyper)
spheres.\medskip }

\textbf{Corollary B.9.} From the same reference [174] and from section 7,
Eq.(7.10), it follows that the conformal transformations are induced by the
projective transformations leaving the Lie quadric \textbf{Q}$^{n+1,2}$
invariant. From here the connection with the Lie sphere geometry follows. \
More on Liouville theorem can be found in [175].\medskip

\textbf{Definition B.10.} Killing vectors \ \textbf{X} are defined as
solutions of the Killing equations 
\begin{equation}
\mathcal{L}_{\mathbf{X}}g_{ij}=0,  \tag{B.10}
\end{equation}%
where $\mathcal{L}_{\mathbf{X}}$ is the Lie derivative in the direction of 
\textbf{X}. \ These equations are describing the isometric-type of motion on 
$\mathcal{M}_{d}.$ Accordingly, \textsl{\ }the\textsl{\ conformal Killing
vectors} \ \textbf{X} are defined as solutions of the \textsl{conformal
Killing equations } 
\begin{equation}
\mathcal{L}_{\mathbf{X}}g_{ij}(x)=2\psi (x)g_{ij}(x).  \tag{B.11}
\end{equation}%
These are describing the conformal-type of motion on $\mathcal{M}_{d}$.
Examples of solutions of Eq.s(B.11) are demonstrated in [166].\medskip

\textbf{Definition B.11. }The\textbf{\ }\textsl{conformal Lie group }\^{C}($%
\mathcal{M}_{d})$ of a connected $d-$dimensional (pseudo) Riemannian
manifold $\mathcal{M}_{d}$ \ is \textsl{conformally trivial} if by using
transformations defined by Eq.(B.9) it is possible to find \ a
conformally-equivalent space $\mathcal{\tilde{M}}_{d}$ in which Eq.(B.11) is
replaced by Eq.(B.10).\medskip

\textbf{Definition B.12. }If use of Eq.(B.9) cannot bring Eq.(B.11) into
Eq.(B.10), such conformal group is \textsl{nontrivial}.\medskip

In [166] the following remarkable theorem is proved\medskip

\textbf{Theorem B.13.}\textit{\ (pseudo) Riemannian space }$M_{d}$\textit{\
has nontrivial conformal group only if it is conformally equivalent to the
conformally flat space\medskip }

\textbf{Remark B.14.} Following [174] for conformally flat spaces about any
point $x\in \mathcal{M}_{d}$ \ there is a region \ in which (for $d\geq 3)$\
the $\frac{1}{2}(d+2)(d+1)-$parameter group of infinitesimal conformal
transformations may exist. For $d=4$ we obtain 15 parameters group of
conformal motions. This is SO(4,2) group \ describing conformal symmetries
of hydrogen atom described in section 1. Based on Theorem \ B.8., \ this is
the Lie sphere geometry group of motions.\medskip

\textbf{Remark B.15. }Following\textbf{\ [}176\textbf{] }the result\textbf{\ 
}$\frac{1}{2}(d+2)(d+1)$ provides maximal number of conformal generators for
conformally flat spacetimes. However \ further studies \ demonstrated that
Theorem B.13. can be extended. For \textit{(pseudo) Riemannian spaces }which
are not conformally flat the number of conformal generators is strictly less
than the maximal number. It can be demonstrated [177]$\mathbf{,[}178\mathbf{]%
}$ that for the metric of the type given by Eq.(B.7) the maximal number is 7.

\textbf{Remark B.16. \ }Such\textbf{\ }not conformally flat metrics are of
no physical interest, however, for variety of reasons. First, we immediately
loose the connection with the Lie sphere geometry in view of Theorem B.8. \
Second, once this connection is lost, we also loose the connection with the
AdS-CFT correspondence (section 7 and Appendix D).

\textbf{Remark B.17.} In [166] it was proven that every Lorentzian spacetime
with nontrivial conformal group is conformally equivalent to the Ricci flat
space, e.g. read \ the Corollary B.2. In [1 $\mathbf{7}$] it is argued (by
invoking \ the content of Brinkmann's theorem) that the only Ricci flat but
nonflat 4-manifolds (that is 4-manifolds whose scalar curvature is not
identically zero) admitting nonhomothetic conformal vector fields are \
manifolds described in terms of the pp waves metric given by Eq.(B.7). In
the same reference it is being argued that for nonconformally flat
spacetimes the dimension of the conformal group is at most 7.

Based on the last two remarks, it is physically meaningful to consider the
conformally flat pp waves. In such a case we obtain the pp wave metric with
circle-preserving 15 components conformal group \ which is Ricci flat. The
condition of Ricci flatness \ is known to be ([169], page 386): 
\begin{equation}
\frac{\partial ^{2}}{\partial \bar{\zeta}\partial \zeta }H(\zeta ,\bar{\zeta}%
,u)=0.  \tag{B.12}
\end{equation}%
The pp metric, Eq.(B.7), with $H(\zeta ,\bar{\zeta},u)$ determined from
Eq.(B.12) is describing the coupled Einstein-Maxwell null fields. Eq.(24.43)
of [169], page 384, explains why the condition, Eq.(B.12), is equivalent to
the condition of Ricci flatness. The $u-$dependence of $H(\zeta ,\bar{\zeta}%
,u)$ can be determined from the condition on the conformal Weil tensor to be
zero. Such a condition selects spaces which are conformally flat. Examples
of such selection are discussed in [166]. Based on just presented results,
it should be clear that, although techically permissible, parts of the
results of [152],[153] describing non null electromagnetic knots \ are
without physical justification.

\ The seminal work of Penrose [$179$], pages 271-275, entitled: "Any
space-time has a plane wave as a limit" explains the universal significance
of the metric given by Eq.(B7) which is also known in physics literature as
Penrose limit metrics. The significance of Theorem B.1. and appreciation of
its role in relating the conformal transformations to Huygens' principle
apparently, totally escaped the attention of physics community. In Appendix
D we provide \ solid evidence that the conformally flat space-times
supporting the Lie sphere geometry are indeed the boundaries of space-times
in which the AdS-CFT correspondence holds.\bigskip\ \medskip

\textbf{Appendix C\bigskip }

\bigskip\ 

\textbf{Dupin cyclides from \ Madelung's \ hydrodynamical reformulation}

\textbf{of the Schr\"{o}dinger equation\bigskip }

\ Using progressive waves method by Friedlander [70] we provided \ enough
evidence in\ sections 4 and 7 that Dupin cyclides \ can exist as solutions
of the Shr\"{o}dinger equation. In this appendix we shall approach the same
problem from yet another direction which is very illuminating in its own
right. \ In our study we were motivated by two papers by Ferapontov
[180],[181]. Ferapontov was interested in establishing the connection
between the dynamics of Hamiltonian systems of hydrodynamical type and its
realization in terms of differential geometry of hypersurfaces embedded in
Minkowski-type spacetimes [180]. \ Using general results developed in [180],
in [181] Ferapontov illustrated general results of [180] \ using
hydrodynamical Hamiltonian systems \ which do not possess \ Riemann
invariants\footnote{%
Riemann invariants are to be discussed in a separate publication.}. As
result, he demonstrated the duality between such Hamiltonian systems and
Dupin hypersurfaces. These are the hypersurfaces with constant principal
curvatures [133]. In 3 dimensions mathematicians call them as "Dupin
surfaces".

Almost immediately after Schr\"{o}dinger published the installment of his
foundational papers on quantum mechanics in Annalen der Physik in 1926 [1],
on \ 25th of October of 1926 Madelung submitted \ his paper \ entitled
"Quantentheorie in hydrodynamischer form" to the Zeitschrift fur Physik [$%
182 $] where it was published in 1927. Although equations (3) and (4) of his
paper contain typos, the key new equations ($3^{\prime }$),($3^{\prime
\prime }$) and ($4^{\prime }$) are correct. It is worth to reproduce these
equations in this appendix. To avoid any ambiguities, we shall use exactly
the same symbols as Madelung was using. He begins with the stationary Schr%
\"{o}dinger equation 
\begin{equation}
\Delta \psi _{0}+\frac{8\pi ^{2}m}{h^{2}}(W-U)\psi _{0}=0  \tag{C.1}
\end{equation}%
in which $W$ is an energy. Next, he writes $\psi =\psi _{0}e^{\frac{2\pi W}{h%
}t}$ and uses this result in the time-dependent version of the Schr\"{o}%
dinger equation 
\begin{equation}
\Delta \psi -\frac{8\pi ^{2}m}{h^{2}}U\psi -i\frac{4\pi m}{h}\frac{\partial
\psi }{\partial t}=0.  \tag{C.2}
\end{equation}%
Next, he was \ looking for a solution in the form $\psi =\alpha e^{i\beta },$
where he is saying that, in view of Eq.(C.1), \ it is sufficient to consider
only $\beta $ to be linearly-dependent upon $t$ while if we use Eq.(C.2),
then it makes sense to consider both $\alpha $ and $\beta $ to be time
-dependent. Surely, the consistency between equations (C.1) and (C.2)
requires $\beta $ to be linearly dependent upon $t.$ Substituting $\psi
=\alpha e^{i\beta }$ into Eq.(C.2) and separating the real part from
imaginary in resulting equation Madelung \ had obtained the following two
equations\footnote{%
We reproduce these equations without obvious typos which apparently were
overlooked by Madelung when he was proofreading the galleys of his paper.
His new key equations ($3^{\prime }$),($3^{\prime \prime }$) and ($4^{\prime
}$) are correct though.}%
\begin{equation}
\Delta \alpha -\alpha (grad\beta )^{2}-\frac{8\pi ^{2}m}{h^{2}}U\alpha +%
\frac{4\pi m}{h}\alpha \frac{\partial \beta }{\partial t}=0,  \tag{C.3}
\end{equation}%
\begin{equation}
\alpha \Delta \beta +2(grad\alpha \cdot grad\beta )-\frac{4\pi m}{h}\frac{%
\partial \alpha }{\partial t}=0.  \tag{C.4}
\end{equation}%
By introducing new notation $\varphi =-\beta \frac{h}{4\pi m}$ Eq.(C.4) is
converted into the continuity equation (this is Madelung's Eq.($4^{\prime
})) $%
\begin{equation}
div(\alpha ^{2}grad\varphi )+\frac{\partial \alpha ^{2}}{\partial t}=0. 
\tag{C.5}
\end{equation}%
Now, following Madelung, we introduce the velocity $\vec{u}=grad\varphi
\equiv \mathbf{u}$ . In terms of such defined velocity Eq.(C.3) acquires the
following form (this is Madelung's Eq.($3^{\prime }))$ 
\begin{equation}
\frac{\partial \varphi }{\partial t}+\frac{1}{2}(grad\varphi )^{2}+\frac{U}{m%
}-\frac{\Delta \alpha }{\alpha }\frac{h^{2}}{8\pi ^{2}m^{2}}=0.  \tag{C.6a}
\end{equation}%
This equation can be rewritten in recognizable hydrodynamical form by taking
into account that $\nabla \times \mathbf{u}=0$ and by applying the gradient
operator to Eq.(C.6a). In the end, we obtain: (Madelung's Eq.($3^{\prime
\prime }$)), 
\begin{equation}
\frac{\partial \mathbf{u}}{\partial t}+\frac{1}{2}grad(\mathbf{u})^{2}=\frac{%
d\mathbf{u}}{dt}=-\frac{1}{m}grad(U)+\frac{h^{2}}{8\pi ^{2}m^{2}}grad\frac{%
\Delta \alpha }{\alpha }.  \tag{C.6b}
\end{equation}

\textbf{Remark C.1.} Eq.(C.6b) along with the continuity Eq.(C.5) describes
the irrotational fluid moving under the action of conservative forces. In
our book [55], page 67, we noticed (and described in various places of the
book) the following chain of correspondences :

classical mechanics$\rightleftarrows $ thermodynamics$\rightleftarrows $%
electrodynamics$\rightleftarrows $

geometrical optics$\rightleftarrows $ hydrodynamics$\rightleftarrows $
magnetohydrodynamics$\rightleftarrows $

superconductivity$\rightleftarrows $ non Abelian gauge Yang-Mills theories.
\bigskip

These correspondences are all derivable from the \ formalism of contact
geometry and topology discussed in our book. Evidently, just described
result by Madelung is fully consistent with the gauge-theoretic Floer-type
description of Schr\"{o}dinger's quantum mechanics outlined in main text,
subsection 4.2.3.\bigskip

Now we recall that, according to Madelung, \ if we are only interested in
the description of the stationary Schr\"{o}dinger Eq.(C.1) the $t-$%
dependence of $\alpha $ can be omitted. In such a case using Eq.(C.3) we
obtain: 
\begin{equation}
\Delta \alpha =0  \tag{C.7}
\end{equation}%
and%
\begin{equation}
(grad\beta )^{2}=\frac{8\pi ^{2}m}{h^{2}}(W-U).  \tag{C.8}
\end{equation}%
Finally, Eq.(C.4) now acquires the following form:%
\begin{equation}
\alpha \Delta \beta +2(grad\alpha \cdot grad\beta )=0.  \tag{C.9}
\end{equation}%
\ A quick look at Eq.s (4.9 a-c) (producing Dupin's cyclides solutions ) and
comparing these with just obtained Eq.s( C.7)-(C.9) indicates that these two
sets of equations are the same when $\frac{8\pi ^{2}m}{h^{2}}(W-U)=1.$ But,
as we know, in general $\frac{8\pi ^{2}m}{h^{2}}(W-U)\neq 1$.

Following ingenious ideas by Luneburg [33], which we are about to describe,
just noticed difficulty can be resolved. By doing so, we are also going to
rederive the results of section 6 via entirely different set of arguments.

We begin with Eq.(C.8) which we formally rewrite as 
\begin{equation}
\beta _{x}^{2}+\beta _{y}^{2}+\beta _{y}^{2}=n^{2}(x,y,z).  \tag{C.10}
\end{equation}%
Consider now a set of wavefronts: $\beta (x,y,z)=const.$ An orthogonal
trajectory (a ray) through this wavefront at any point $x,y,z$ is normal to
the wavefront through this point. In analogy with Eq.(4.13), \ we introduce
the parameter $\tau $ \ along the ray trajectory so that this ray can be
described in terms of the set \{x($\tau $),y($\tau $),z($\tau $)\}. This
allows us to introduce the equations \ for these rays as follows:%
\begin{equation}
\frac{dx}{d\tau }=\lambda \beta _{x},\frac{dy}{d\tau }=\lambda \beta _{y},%
\frac{dz}{d\tau }=\lambda \beta _{z}.  \tag{C.11}
\end{equation}%
Here $\lambda =\lambda (x,y,z;\tau )>0$ is parameter describing various
choices of parametrization of the ray trajectory. This freedom of
reparametrization is the key element in achieving our goal-to bring Eq(C.10)
into the form of Eq.(4.9a). \ Toward this goal, using Eq.(C.11) \ we
consider the following chain of equalities%
\begin{eqnarray}
\frac{d}{d\tau }\left( \frac{1}{\lambda }\frac{dx}{d\tau }\right) &=&\beta
_{xx}\frac{dx}{d\tau }+\beta _{xy}\frac{dy}{d\tau }+\beta _{xz}\frac{dz}{%
d\tau }=\lambda (\beta _{xx}\beta _{x}+\beta _{xy}\beta _{y}+\beta
_{xz}\beta _{z})  \notag \\
&=&\frac{\lambda }{2}\frac{d}{dx}(\beta _{x}^{2}+\beta _{y}^{2}+\beta
_{y}^{2})=\frac{\lambda }{2}\frac{d}{dx}n^{2}  \TCItag{C.12a}
\end{eqnarray}%
Clearly,proceeding analogously, we obtain as well:%
\begin{equation}
\frac{d}{d\tau }\left( \frac{1}{\lambda }\frac{dy}{d\tau }\right) =\frac{%
\lambda }{2}\frac{d}{dy}n^{2},  \tag{C.12b}
\end{equation}%
and 
\begin{equation}
\frac{d}{d\tau }\left( \frac{1}{\lambda }\frac{dz}{d\tau }\right) =\frac{%
\lambda }{2}\frac{d}{dz}n^{2}.  \tag{C.12 c}
\end{equation}%
In equations (C.11) let $\lambda =\frac{1}{n}$ then, in view of Eq.(C.10),
we obtain%
\begin{equation}
\left( \frac{dx}{d\tau }\right) ^{2}+\left( \frac{dy}{d\tau }\right)
^{2}+\left( \frac{dz}{d\tau }\right) ^{2}=1.  \tag{C.13}
\end{equation}%
That is with such choice of the parameter $\lambda $ \ the yet arbitrary
parameter $\tau $ becomes a parameter describing natural parametrization
along the ray. We shall qualify it as "time". One still can do a better job,
though, by noticing that, say, 
\begin{equation}
\frac{1}{\lambda }\frac{d}{d\tau }\left( \frac{1}{\lambda }\frac{dx}{d\tau }%
\right) =\frac{d^{2}x}{d\sigma ^{2}}=\frac{1}{2}\frac{d}{dx}n^{2}. 
\tag{C.14}
\end{equation}%
Here d$\sigma =d(\lambda \tau ).$Evidently, Eq.s (C.12) become the Newtonian
equations of motion%
\begin{equation}
\frac{d^{2}}{d\sigma ^{2}}\mathbf{r=}\frac{1}{2}\mathbf{\nabla }n^{2}. 
\tag{C.16}
\end{equation}%
Here $\mathbf{r}=\{x,y,z\}.$If this is so, then \ in view of Eq.s(C.10) and
(C.11) we easily obtain\footnote{%
Since now $\dot{x}=\frac{dx}{d\sigma }$ and $\ \sigma =\lambda \tau $ and we
can use Eq.s(C.10) and (C.11).$.$} 
\begin{equation}
\dot{x}^{2}+\dot{y}^{2}+\dot{z}^{2}=n^{2}  \tag{C.18}
\end{equation}
This time, however, we can bring Eq.s(C.13) and (C.18) in correspondence
with each other by properly selecting "time". Introduce as well $\tilde{\beta%
}_{x}=\lambda \beta _{x},$ $\tilde{\beta}_{y}=\lambda \beta _{y},$ $\tilde{%
\beta}_{z}=\lambda \beta _{z}.$ Using this definition and selecting $\lambda
=\frac{1}{n}$we convert Eq.(C10) into 
\begin{equation}
\tilde{\beta}_{x}^{2}+\tilde{\beta}_{y}^{2}+\tilde{\beta}_{y}^{2}=1 
\tag{C.19}
\end{equation}%
easily recognizable as Eq.(4.9a) of the main text$.$ Now we multiply
Eq.(C.9) by $\lambda $ to obtain instead 
\begin{equation}
\alpha \lambda \Delta \beta +2(grad\alpha \cdot grad\tilde{\beta})=0. 
\tag{C.20}
\end{equation}%
Notice also that $\Delta \beta =\nabla _{x}\beta _{x}+\nabla _{y}\beta
_{y}+\nabla _{z}\beta _{z}.$ But $\tilde{\beta}_{x}/\lambda =\beta _{x},$%
etc. Since $\lambda =\frac{1}{n},$then $\Delta \beta =\nabla _{x}n\tilde{%
\beta}_{x}+\nabla _{y}n\tilde{\beta}_{y}+\nabla _{z}n\tilde{\beta}%
_{z}=n\Delta \tilde{\beta}+\tilde{\beta}_{x}\nabla _{x}n+\tilde{\beta}%
_{y}\nabla _{y}n+\tilde{\beta}_{z}\nabla _{z}n.$ In view of Eq.(C.20) and
taking again into account that $\lambda =\frac{1}{n},$ it only remains to
demonstrate that $\lambda ^{-1}\left( \tilde{\beta}_{x}\nabla _{x}n+\tilde{%
\beta}_{y}\nabla _{y}n+\tilde{\beta}_{z}\nabla _{z}n\right) =0.$Recall
Eq.(C.11) and present $\tilde{\beta}_{x}\nabla _{x}n+\tilde{\beta}_{y}\nabla
_{y}n+\tilde{\beta}_{z}\nabla _{z}n$ as $\frac{dx}{d\tau }\nabla _{x}n+\frac{%
dy}{d\tau }\nabla _{y}n+\frac{dz}{d\tau }\nabla _{z}n.$ Use equations of
motion, Eq.s(C.12), in order to write $\nabla _{x}n=\frac{d}{d\tau }(n\frac{%
dx}{d\tau }),$etc$.$ Finally, write $n\frac{dx}{d\tau }\frac{d}{d\tau }(n%
\frac{dx}{d\tau })=\frac{1}{2}\frac{d}{d\tau }(n\frac{dx}{d\tau })^{2}$ ,
etc. \ Using this result along with Eq.s(C.10),(C.11) we finally must prove
that $\frac{d}{d\tau }n^{2}=0.$ But $n^{2}=\frac{8\pi ^{2}m}{h^{2}}(W-U)$
and for the time -independent $U$ the desired result follows. Alternatively,
by comparing Eq.s(C.13) and (C.18) \ and by noticing that Eq.(C.18) is
converted to (C.13) for $\lambda =\frac{1}{n},$ we obtain: $n=1$ for $%
\lambda =\frac{1}{n}.$ Thus, we just demonstrated that the Dupin cyclides
generating set of equations, Eq.s (4.9), of the main text \ can be made to
coincide with the the set of equations obtained by Madelung [177].

\bigskip

\bigskip

\textbf{Appendix D\bigskip \medskip }

\textbf{Various models of hyperbolic and Anti- de Sitter spaces\bigskip }

\bigskip

D.1. \textsl{Hyperboloid model of hyperbolic space}\textbf{\ \medskip }

\ \ \ Following Danciger [159], consider \textbf{R}$^{n,1}$ denoting \textbf{%
R}$^{n+1}$ equipped with $(n,1)$ Minkowski-type metric tensor $g$:%
\begin{equation}
g=\left( 
\begin{array}{cc}
I_{n} & 0 \\ 
0 & -1%
\end{array}%
\right) .  \tag{D.1}
\end{equation}%
Using $g$ we define the hyperboloid of two sheets \ $x^{T}gx=-1.$ Sheets are
determined by the sign of the first(or last) coordinate $x_{1},$e.g. read
[105]or [159] for more details. Traditionally, the hyperbolic space \textbf{H%
}$^{n}$ is defined by the sheet for which $x_{1}>0.$ Alternatively, \ it can
be also defined as%
\begin{equation}
\mathbf{H}^{n}=\{x\in \mathbf{R}^{n+1}:x^{T}gx=-1\}/\{\pm I\}  \tag{D.2}
\end{equation}%
Taking a quotient $\{\pm I\}$ identifies two sheets. The hyperboloid $%
x^{T}gx=-1$ inherits the Riemannian metric of constant curvature $-1$ from \
that defined by Eq.(D.1). Isometries of $\mathbf{H}^{n}$ are defined by%
\begin{equation}
\text{Isom}(\mathbf{H}^{n})=\text{PO}(n.1):=\{A\in GL(n+1,\mathbf{R}%
):A^{T}gA=g\}/\{\pm I\}.  \tag{D.3}
\end{equation}%
This result defines \ the M\"{o}bius-type transformation in accord with
Definition 7.8. The orientation -preserving isometries are isometries lying
in the identity component of $PO(n,1)$: Isom$(\mathbf{H}^{n})=PO_{0}(n.1).$
For $n$ even (and this is our case) we have $PO_{0}(n.1)\simeq SO_{0}(n,1).$
If we had chosen to think about $\mathbf{H}^{n}$ as positive sheet of the
hyperboloid $x^{T}gx=-1,$then we should think about $PO(n,1)$as the subgroup
of $O(n,1)$ that preserves the positive sheet.\bigskip

D.2. \ \textsl{The projective model of hyperbolic space and Mobius geometry}%
\textbf{\ \medskip }

\ \ \ \ This model is easily obtainable from the hyperboloid model. Instead
of Eq.(D.2) we have now%
\begin{equation}
\mathbf{H}^{n}=\{x\in \mathbf{R}^{n+1}:x^{T}gx>0\}/\sim  \tag{D.4}
\end{equation}%
where the symbol $\sim $ denotes equivalence. That is $x\sim x^{\prime }$
whenever there is some nonzero $\lambda \in \mathbf{R}^{\ast }$ such that \ $%
x=\lambda x^{\prime }.$ Thus, every hyperbolic structure is also a
projective structure defined on some domain of \textbf{RP}$^{n}$ determined
by the equivalence relation. This fact then allows us to define the boundary
at infinity \ as 
\begin{equation}
\partial ^{\infty }\mathbf{H}^{n}=\{x\neq 0,x\in \mathbf{R}%
^{n+1}:x^{T}gx=0\}/\sim .  \tag{D.5}
\end{equation}%
We had encountered this equation already as Eq.(7.12). It can be interpreted
either as the projectivized light cone or as the condition for two spheres
to intersect transversally. $\ \partial ^{\infty }\mathbf{H}^{n}$ has
invariant flat conformal structure in the sense of Eq.(D.3). \ Since a
geodesic in $\mathbf{H}^{n}$ is determined by two distinct points on $%
\partial ^{\infty }\mathbf{H}^{n}$ this means that in this formalism every
geodesic in $\mathbf{H}^{n}$ is in one-to-one correspondence with the
orthogonal intersection of two spheres.\bigskip

D.3\textsl{.\ Hyperboloid model of Anti-de Sitter space\medskip }\textbf{\ }

\ \ \ \ \ \ \ Let \textbf{R}$^{n+1,2}$ denote \textbf{R}$^{n+3}$ equipped
with $(n+1,2)$ Minkowski-like metric tensor $g$%
\begin{equation}
g=\left( 
\begin{array}{ccc}
I_{n+1} & 0 & 0 \\ 
0 & -1 & 0 \\ 
0 & 0 & -1%
\end{array}%
\right) .  \tag{D.6}
\end{equation}%
Using $g,$ we define the hyperboloid $x^{T}gx=-1.$Instead of Eq.(D2)
defining the hyperbolic space $\mathbf{H}^{n}$ we are having now the
following definition of the Anti-de Sitter space (AdS)[$159$]%
\begin{equation}
AdS^{n+2}=\{x\in \mathbf{R}^{n+3}:x^{T}gx=-1\}/\{\pm I\}  \tag{D.7}
\end{equation}%
The hyperboloid $x^{T}gx=-1$ inherits a Lorentzian metric of constant
curvature $-1$ from \ the form defined by Eq.(D.6). Isometries of $AdS^{n+2}$
are defined as%
\begin{equation}
\text{Isom}(AdS^{n+2})=PO(n+1,2):=\{A\in GL(n+3,\mathbf{R}%
):A^{T}gA=g\}/\{\pm I\}.  \tag{D.8}
\end{equation}%
\bigskip

D.4. \ \textsl{The projective model of the Anti-de Sitter space and Lie
sphere geometry}\textbf{\ \medskip }

\ \ \ \ \ \ \ \ The projective model of AdS is easily obtainable from the
hyperboloid AdS model. Instead of Eq.(D.7) we have now%
\begin{equation}
AdS^{n+2}=\{x\in \mathbf{R}^{n+3}:x^{T}gx>0\}/\sim  \tag{D.9}
\end{equation}%
The boundary of AdS space is defined now by analogy with Eq.(D.5) \ that is%
\begin{equation}
\partial ^{\infty }AdS^{n+2}=\{x\neq 0,x\in \mathbf{R}^{n+3}:x^{T}gx=0\}/\sim
\tag{D.10}
\end{equation}%
In view of Remark 7.4. this result is easily recognizable as the Lie quadric 
\textbf{Q}$^{n+1,2}.$ It can be interpreted either as the projectivized
light cone or as a condition for two spheres to touch each other.
Accordingly, the geodesics in Anti-de Sitter space can be of three types:

a) A space-like geodesics is determined by two distinct points lying on

\ \ \ $\partial ^{\infty }AdS^{n+2},$that is on \textbf{Q}$^{n+1,2}.$ In
terms of the Lie sphere geometry such

\ \ \ a geodesics corresponds to two spheres touching each other.

b) A light-like geodesics has its both ends lying at the same point

\ \ \ \ of the quadric \textbf{Q}$^{n+1,2}.$ This situation is characterizes
the projectivized

\ \ \ \ light cone. In the language of the\ Lie

\ \ \ \ sphere geometry \ such situation corresponds to spheres participating

\ \ \ \ in the formation of canal \ surfaces \ other than Dupin cyclides as
explained

\ \ \ \ in section 7.5.1.

c) \ A time-like periodic geodesics which does not touch the boundary

\ \ \ $\partial ^{\infty }AdS^{n+2}.\bigskip $

\bigskip

The action of $PO(n+1,2)$ preserves $\partial ^{\infty }AdS^{n+2}$ in \
accord with Definition 7.6. and Corollary B.9. Furthermore, according to
[161], page 184, Einstein's space Ein$^{n,1}=$ $\partial ^{\infty
}AdS^{n+2}. $ Thus, Einstein's space is the same as the the space where Lie
sphere geometry acts and, using the discovered by Sophus Lie correspondence
between the Lie sphere and Pl\"{u}cker line geometries described in section
7.4., it is possible to map such defined Einstein's space into the space of
twistors. \ The connection with Einstein spaces (and with spaces where Lie
sphere geometry acts) and, therefore, with Penrose boundary can be easily
explained now. For this, following [161] and using Eq.(7.10) (or Eq.(6.41a))
of the main text, \ we relate the quadratic form%
\begin{equation}
<\mathbf{v},\mathbf{v}>=v_{1}^{2}+\cdot \cdot \cdot
+v_{n+1}^{2}-v_{n+2}^{2}-v_{n+3}^{2}  \tag{D.11a}
\end{equation}%
i to the nullcone as:%
\begin{equation}
v_{1}^{2}+\cdot \cdot \cdot +v_{n+1}^{2}=v_{n+2}^{2}+v_{n+3}^{2}. 
\tag{D.11b}
\end{equation}%
Since each of $v^{\prime }s$ is never zero, it is possible to divide both
sides of Eq.(D.11b) by the positive number $\sqrt{v_{n+2}^{2}+v_{n+3}^{2}}$
so that this equation can be rewritten as 
\begin{equation}
v_{1}^{2}+\cdot \cdot \cdot +v_{n+1}^{2}=1=v_{n+2}^{2}+v_{n+3}^{2}. 
\tag{D.12}
\end{equation}%
It describes \ the product $S^{n}\times S^{1}.$ For $n=3$ we obtain: $%
S^{3}\times S^{1}.$ This is an example of static Einstein spacetime with
compactified time axis [22]. In fact, $S^{3}\times S^{1}=$ \^{E}in$^{n,1}$
while Ein$^{n,1}=$\^{E}in$^{n,1}/\{\pm 1\}.$K\"{u}hnel and Rademacher had
classified \ all (pseudo) Riemannian Einstein spacetimes having local or
global conformal conformal groups [183]. These are surely including the pp
wave \ type Ricci-flat spaces which, based on results of Appendix B, can be
identified with Penrose limits.

\bigskip

$\bigskip $

\textbf{References}

$\mathbf{\bigskip }$

[1] \ \ E.Schr\"{o}dinger, Collected Papers on Wave Mechanics\textit{, }

\ \ \ \ \textit{\ \ }Chelsea Publ.Co, New York, 1978.

[2] \ \ R.Courant and D.Hilbert, Methods of Mathematical Physics,Vol.1,

\ \ \ \ \ \ Interscience Publishers, Inc., New York, 1953.

[3] \ \ B.Baker and E.Copson, The Mathematical Theory of Huygens'

\ \ \ \ \ \ \ Principle, Clarendon Press,Oxford, 1939.

[4] \ \ R.Feynman, Rev. Mod.Phys. 20 (1948) 367.

[5] \ \ M.Gutzwiller, Huygens' principle and the path integral, in

\ \ \ \ \ \ \ Path Summation: Achievements and Goals (Trieste, 1987), pp.
47--73,

\ \ \ \ \ \ \ L.Schulman editor, World Sci. Publishing, Singapore, 1988.

[6] \ \ R.Feynman, R.Leighton and M.Sands, The Feynman Lectures on

\ \ \ \ \ \ Physics, Vol.3, Addison -Wesley Publishing Co., New York, 1989.

[7] \ \ R.Feynman and A.Hibbs, Quantum Mechanics and Path Integrals,

\ \ \ \ \ \ McGraw -Hill Co, New York, 1965.

[8] \ \ M.Born and E.Wolf, Principles of Optics, Cambridge U.Press.

\ \ \ \ \ \ Cambridge, UK, 1980

[9] \ \ C. J\"{o}nsson, Zeitschrift f\"{u}r Physik, 161(1961) 454.

[10] \ M.Arndt, O.Nairz, J.Vos-Andreae, C.Keller, G.Zouw and A.Zellinger,

\ \ \ \ \ \ \ Nature, 401 (1999) 680.

[11] \ S.Eibenberger, S. Gerlich, M. Arndt, M. Mayor and J. T\"{u}xen, Phys.

\ \ \ \ \ \ \ Chem. Chem. Phys. 15 (2013) 14696.

[12] \ D.Bohm, Quantum Theory, Dover Publications Inc., New York, 1989.

[13] \ A. Sanz, and S. Miret-Art\'{e}s, A Trajectory Description of Quantum

\ \ \ \ \ \ \ Processes I. Fundamentals, Springer-Verlag, Berlin, 2012.

[14] \ A.Kholodenko, arXiv:1703.04674.

[15] \ U. Leonhardt, Measuring the Quantum State of Light, Cambridge U.

\ \ \ \ \ \ \ Press, Cambridge, UK, 1997.

[16] \ M. Fox, Quantum Optics: An Introduction, Oxford U. Press,

\ \ \ \ \ \ \ Oxford,UK, 2006.

[17] \ U.Niederer, Helv. Phys. Acta 45 (1972) 802.

[18] \ R.Littlejohn, Phys. Rep. 138 (1986) 193.

[19] \ B.Wybourne, Classical Groups for Physicists, Wiley-Interscience Publ.,

\ \ \ \ \ \ \ New York, 1974.

[20] M.Kibler, Found.Chem. 9 (2007) 221.

[21] R. Campoamor-Stursberg, J.of Phys.: Conference Series

\ \ \ \ \ \ 538 (2014) 012004.

[22] A.Keane and R.Barrett, Class.Quantum Grav. 17 (2000) 201.

[23] A. Bohm, Y. Ne'eman and A.Barut, Dynamical Groups and

\ \ \ \ \ \ Spectrum Generating Algebras, Vol's 1 \& 2, World Scientific,

\ \ \ \ \ \ Singapore, 1988.

[24]\ E.Schr\"{o}dinger, Expanding Universes,

\ \ \ \ \ \ Cambridge University Press, Cambrige, 1956.

[25] S.Huggett and K.Tod, An Introduction to Twistor Theory

\ \ \ \ \ \ Cambridge University Press, Cambridge, 1994.

[26] E.Schr\"{o}dinger, Science and the Human Temperament,

\ \ \ \ \ \ George Allen \& Unwin LTD, London, 1935.

[27] V. Kac and P.Cheung, Quantum Calculus, Springer-Verlag, Berlin, 2002.

[28] E.Schr\"{o}dinger, Phys.Rev.28 (1926) 1049.

[29] R.Courant and D.Hilbert, Methods of Mathematical Physics,Vol.2,

\ \ \ \ \ \ \ \ Interscience Publishers, Inc., New York, 1962.

[30] F.Cardin, Elementary Symplectic Topology and Mechanics,

\ \ \ \ \ \ \ Springer-Verlag, Heidelberg, 2015.

[31] D.Tannor, Introduction to Quantum Mechanics. A Time-Dependent

\ \ \ \ \ \ \ Perspective, University Science Books,Saustalito, Ca., 2007.

[32] M.Gosson and B.Hiley, Found.Phys.41 (2011) 1415.

[33] R.Luneburg, \ Mathematical Theory of Optics, U.of California Press,

\ \ \ \ \ \ \ Berkeley, 1966.

[34] A.Davydov, Quantum Mechanics, Pergamon Press, Oxford, 1976.

[35] L.De Broglie , An Introduction to the Study of Wave Mechanics,

\ \ \ \ \ \ Methuen \& Co. LTD., 36 Essex Street, W.C. , 1930.

[36] C.Adams, M.Siegel and J. Mlynek, Phys. Reports 240 (1994) 143.

[37] R.Sawant, J.Samuel,A.Sinha,S.Sinha and U.Sinha,

\ \ \ \ \ \ PRL 113 \ (2014) 120406.

[38] K.Ito and H. McKean Jr., Diffusion Processes and Their Sample Paths,

\ \ \ \ \ \ Springer-Verlag,Berlin, 1965.

[39] H.De\ Raedt, K.Michelsen and K.Hess, Phys.Rev.A 85 (2012) 012101.

[40] U.Sinha,C.Couteau, T.Jennewein,R.Laflamme and G.Weihs,

\ \ \ \ \ \ Science 329 (2010) 418.

[41] I. S\"{o}llner, B.Gsch\"{o}sser, P. Mai, B. Pressl, Z. V\"{o}r\"{o}s
and G. Weihs,

\ \ \ \ \ \ Found. Phys. 42 (2012) 742.

[42] A. Sommerfeld, Optics, Academic Press, NY, 1964.

[43] L.Landau and E.Lifshitz, Classical Theory of Fields,

\ \ \ \ \ \ Butterworth-Heinemann, Oxford, 2000.

[44] A.Pazy, Semigroups of Linear Operators and Applications to

\ \ \ \ \ \ Partial Differential Equations, Springer-Verlag,Berlin, 1982.

[45] L.Evans, Partial Differential Equations,

\ \ \ \ \ \ AMS Publishers, Providence, RI, 2010.

[46] J.Hadamard, \ Lectures on Cauchy's Problem in Linear Partial

\ \ \ \ \ \ Differential Equations, \ Nauka Co, Moscow, 1978

\ \ \ \ \ \ (Russian translation from French original).

[47] V.Arnol'd, Contact Geometry and Wave Propagation,

\ \ \ \ \ \ Enseign. Math. 36 (1990) 215.

[48] V. Arnol'd, Lectures on Partial Differential Equations,

\ \ \ \ \ \ Springer-Verlag, Berlin, 2004.

[49] V.Arnol'd, Mathematical Methods of Classical Mechanics,

\ \ \ \ \ \ Springer-Verlag, Heidelberg, 1978.

[50] V.Maslov and M. Fedoryuk, Semiclassical Approximation in Quantum

\ \ \ \ \ \ Mechanics. \ D. Reidel Publishing Co., Dordrecht-Boston, Mass.,
1981.

[51] M.Fedoryuk, Partial Differential equations V,

\ \ \ \ \ \ Springer-Verlag, Berlin, 1999.

[52] V.Arnol'd , Ordinary Differential Equations,

\ \ \ \ \ \ Springer-Verlag,Berlin, 1992.

[53] C.Chicone, Oerdinary Differential equations with Applications,

\ \ \ \ \ \ Springer-Verlag, Berlin, 1999.

[54] A.Fasano and S.Marmi, Analytical Mechanics,

\ \ \ \ \ \ Oxford U.Press, Oxford, 2006.

[55] A.Kholodenko, Applications of Contact Geometry and Topology in Physics,

\ \ \ \ \ \ World Scientific,Singapore, 2013.

[56] M.Gotay, JMP 40 (1999) 2017.

[57] V.Guilleming and S.Sternberg, Variations on the Theme by Kepler,

\ \ \ \ \ \ \ AMS Publishers, Providence, RI 1990.

[58] M. Giaquinta and S. Hildebrandt,

\ \ \ \ \ \ Calculus of variations II, Springer-Verlag, Berlin, 1996.

[59] S.Benenti, Hamiltonian Structures and Generating Families,

\ \ \ \ \ \ \ Springer-Verlag, Berlin, 2011.

[60] I.Gelfand and S.Fomin, Calculus of Variations, Prentice-Hall Inc.,

\ \ \ \ \ \ \ New Jersey, 1963.

[61] M.Mathisson, Acta Math. 71 (1939) 249.

[62] L. Asgeirsson, Comm. Pure Appl. Math. 9 (1956) 307.

[63] P.G\"{u}nther, Math.Intelligencer 13 (1991) 56.

[64] P.G\"{u}nther, Huygens' Principle and Hyperbolic Equations.

\ \ \ \ \ \ \ Academic Press, Inc., Boston, MA, 1988.

[65] F. Friedlander, The Wave Equations on a Curved Space-Time,

\ \ \ \ \ \ Cambridge U.Press, Cambridge, 1975.

[66] R.McLenaghan and J.Carminati, Ann.Inst.Henri Poincare 44 (1986) 115.

[67] R.Goldoni, J. Math. Phys. 18 (1977) 2125.

[68] N.Ibragimov, A.Oganesyan, Russian Math.Surveys 46 (1991) 137.

[69] M.Belger, R.Schimming and V.Wunsch, J.for Analysis and its

\ \ \ \ \ \ \ Applications 16 (1997) 9.

[70] F.Friedlander, Proc.Camb.Phil.Soc. 43(1946) 360.

[71] A.Kholodenko, Anal.Math.Phys. 6 (2016) 163.

[72] M.Audin and M.Damian, Morse Theory and Floer Homology,

\ \ \ \ \ \ Springer-Verlag, Berlin, 2014.

[73] D.Bohm and B.Hiley, Undivided Universe, Routledge,

\ \ \ \ \ \ New York and London, 1993.

[74] A.Sym, J.Nonlinear Math.Phys. 12, Supplement 1 (2005) 648.

[75] R.Pruss and A.Sym, Phys.Lett. A 336 (2005) 459.

[76] A.Sym and A.Szereszewski, SIGMA 7 (2011) 095.

[77] P.Broadbridge, C.Chanu and W.Miller Jr. SIGMA 8 (2012) 089.

[78] J. Alc\'{a}zar, H. Dahl and G. Muntingh, arXiv:1611.06768.

[79] T.Cecil and P.Ryan, Geometry of Hypersurfaces, Springer-Verlag,

\ \ \ \ \ \ Berlin , 2015.

[80] Y.Suris, in Sophus Lie and Felix Klein. The Erlangen Program

\ \ \ \ \ \ and its Impact in Mathematics and Physics, EMS 23 (2015).

[81] J.Ratcliffe, Foundations of Hyperbolic Manifolds, Springer-Verlag,

\ \ \ \ \ Berlin, 1994.

[82] C.Codirla and H.Osborn, Ann.Phys. 260 (1997) 91.

[83] I.Bars and J.Terning, Extra Dimensions in Space and Time,

\ \ \ \ \ \ Springer Science +Business Media, LLC 2010.

[84] P.Arvidsson, R.Marnelius, arXiv:hep-th/0612060.

[85] S.Brodsky, G.de Teramond, H. Dosch and J.Elich,

\ \ \ \ \ \ Phys.Reports 584 (2015) 1.

[86] A.Kolesnik and N.Ratanov, Telegraph Processes and Option Pricing,

\ \ \ \ \ \ Springer-Verlag, Berlin, 2013.

[87] W.Greiner, Relativistic Quantum Mechanics.Wave Equations,

\ \ \ \ \ \ \ Springer-Verlag, Berlin, 2000.

[88] \ V. Berestetskii, E. Lifshitz and L.Pitaevskii, Relativistic Quantum

\ \ \ \ \ \ \ \ Theory, Pergamon Press, Oxford, UK, 1971.

[89] \ N.Bogoliubov and D.Shirkov, Introduction to the Theory of Quantized

\ \ \ \ \ \ \ Fields, John Wiley\&Sons, New York,1976.

[90] \ M. Kac, Rocky Mountain J. of Math. 4 (1974) 497.

[91] \ B.Gaveau, T. Jacobson, M. Kac and L.Schulman, PRL 53 (1984) 419.

[92] \ C. Dewitt-Morette and P. Cartier, Functional Integration:

\ \ \ \ \ \ \ Action and Symmetries, Cambridge U.Press, Cambridge, 2006.

[93] \ A.Kolesnik and M. Pinsky, J.Stat.Phys.142 (2011) 828.

[94] \ E.Zauderer, Partal Differential Equations of Applied Mathematics,

\ \ \ \ \ \ \ Wiley-Interscience, New York, 1989.

[95] \ B.Gaveau and L.Schulman, Il Nuovo Cimento 11 (1989) 31.

\bigskip \lbrack 96] \ A. Kholodenko, Ann.Phys.201 (1990) 186.

[97] \ S. Deguchi and T.Suzuki, Phys.Lett.B 731 (2014) 337.

[98] \ S.Deguchi and S.Okano, Phys. Rev. D93 (2016) 045016.

[99] \ \ I. Bars, Phys.Rev.D 58 (1998) 066006.

[100] I.Araya and I.Bars, Phys.Rev.D 89 (2014) 066011.

[101] S. Singer, Lineraity,Symmetry and Prediction \ in the Hydrogen Atom,

\ \ \ \ \ \ \ \ \ Springer-Verlag, Berlin, 2005.

[102] V.Fock, Z.Phys. 98 (1935) 145.

[103] \ L.Landau and E. Lifshitz, Quantum Mechanics,

\ \ \ \ \ \ \ \ Elsevier Publ.Co, Amsterdam, 1981.

[104] H.Goldstein, Classical Mechanics,

\ \ \ \ \ \ \ Addison-Wesley Publ.Co., Reading, MA, 1980.

[105] I.Gelfand, R.Milnos and Z.Shapiro, Representations of the Rotation

\ \ \ \ \ \ \ and Lorentz Groups and Their Applications, Martino Fine Books,

\ \ \ \ \ \ \ P.O. Box 913, Eastford, CT, 2012. (Translation from the
original

\ \ \ \ \ \ \ 1958 Russian edition).

[106] J.Bros and G.Viano, Forum Math.8 (1996) 621.

[107] J.Bros and G.Viano, Forum Math.8 (1996) 659.

[108] J.Bros and G.Viano, Forum Math. 9 (1997) 165.

[109] D.Basu and S.Srinivasan, Czech.J.Phys.B 27 (1997) 635.

[110] J.Milnor, The American Mathematical Monthly 90 (1983), 353.

[111] T.de Laat, Regularization and quantization of the Kepler problem,

\ \ \ \ \ \ \ \ \ PhD Thesis, Department of Mathematics, Radboud University,

\ \ \ \ \ \ \ \ \ Nijmegen, Netheerlands, 2010.

[112] \ A.Fomenko, Symplectic Geometry, Gordon and Breach Co.,

\ \ \ \ \ \ \ \ New York, 1988.

[113] \ M.Dunajski, Solitons, Instantons and Twistors, Oxford U.Press,

\ \ \ \ \ \ \ \ Oxford,UK, 2010.

[114] R.Penrose and W. Rindler, Spinors and Space-Time,

\ \ \ \ \ \ \ \ Cambridge U.Press,Cambridge, UK, 1984.

[115] \ D.Sommerville, An Introduction to the Geometry of N Dimensions

\ \ \ \ \ \ \ \ \ Dower Publications , Inc., New York, 1958.

[116] \ H.Pottmann and J.Wallner, Computational Line Geometry,

\ \ \ \ \ \ \ \ \ Springer-Verlag, Berlin, 2010.

[117] \ R.Penrose, Relativistic symmetry groups, in

\ \ \ \ \ \ \ \ \ Group Theory and Nonlinear Problems, A.Barut Editor,
pp.1-58,

\ \ \ \ \ \ \ \ \ D.Reidel Publ.Co., Boston, 1974.

[118] M.Kriele, Spacetime, Springer-Verlag, Berlin, 1999.

[119] A.Kholodenko, Int'l J.Mod Phys.A 30 (2015) 1550189.

[120] A.Kholodenko, E.Ballard, Physica A 380 (2007) 115.

[121] M.Bander and C.Itzykson, Rev.Mod.Phys. 38 (1966) 346.

[122] I.Frenkel and M.Libine, Adv.Math. 218 (2008) 1806.

[123] A.Hurwitz and R.Courant, Complex Function Theory (in German)

\ \ \ \ \ \ \ \ \ Interscience Publishers, Inc., New York, 1944.

[124] R. Fueter, Comment. Math. Helv. 8 (1) (1935) 371.

[125] B.Cordani, The Kepler Problem, Birkh\"{a}user, Basel, 2003.

[126] I.Bars, Phys.Rev. D 58 (1998) 066006.

[127] I.Bars, C. Deliduman and O.Andreev, Phys. Rev. D \ 58 (1998) 066004.

[128] P.Kustaanheimo and E.Stiefel, J.Fur.Reine

\ \ \ \ \ \ \ \ und Angevante Math. 218 (1965), 204.

[129] F.Cornish, J.Phys.A 17 (1984) 323.

[130] A.Chen, Phys.Rev.A 22 (1980) 333.

[131] A.Chen, Phys.Rev.A 23 (1981) 1655.

[132] A.Chen,Phys.Rev.A 26 (1982) 669.

[133] G.Thorbergsson, Bull. London Math.Soc. 15 (1983) 493.

[134] E.Huhnen-Venedey, PhD Thesis, department of Mathematics,

\ \ \ \ \ \ \ Technical University,Berlin, 2007.

[135] E.Musso, G.Jensen and L.Nicolodi, Surfaces in Cassical Geometries,

\ \ \ \ \ \ \ \ Springer -Verlag, Berlin, 2016.

[136] C.Doran and A.Lasenby, Geometric Algebra,

\ \ \ \ \ \ \ Cambridge U.Press, Cambridge, 2004.

[137] A.Bobenko and Y.Suris, Discrete Differential Geometry,

\ \ \ \ \ \ \ \ AMS Publishers, Providence, RI, 2008.

[138] \ A.Bobenko and E.Huhnen-Venedey, Geom. Dedicata 159 (2012) 207.

[139] \ L.Drouton, L.Fuchs, L.Garnier and R.Langevin,

\ \ \ \ \ \ \ \ Adv. in Appl.Clifford Algebra 24 (2014) 515.

[140] L.Dorst, Math.Comput. Sci 10 (2016) 97.

[141] S.Helgason, in The Sophus Lie memorial conference (Oslo 1992),

\ \ \ \ \ \ \ pages 3-21, Scand.U.Press, Oslo, 1994.

[142] L.Mason and N.Woodhouse, Integrability, Self-Duality,

\ \ \ \ \ \ \ and Twistor Theory, Clarendon Press, Oxford, 1996.

[143] R.Ward, R.Wells,Jr. , Twistor Geometry and Field Theory,

\ \ \ \ \ \ \ \ Cambridge University Press, Cambridge, 1990.

[144] F.Klein, Vorlesungen \"{U}ber H\"{o}here Geometrie,

\ \ \ \ \ \ \ \ Spinger-Verlag, Berlin, 1926.

[145] W. Blaschke, Volresungen \"{U}ber Differential-Geometrie \ III,

\ \ \ \ \ \ \ \ Springer-Verlag , Berlin, 1929.

[146] M.Schrott and B.Odenhal, J. for Geometry and Graphics, 10 (2006)73.\ \
\ \ \ \ 

[147] A.Gray, Modern Differential Geometry of Curves and Surfaces

\ \ \ \ \ \ \ \ with Mathematica, Taylor\&Francis, Roca Baton, FL 2006.

[148] T. Ivey, AMS Proceedings 123 (1995) 865.

[149] P. Colapino, Articulating Space: Geometric Algebra for Parametric

\ \ \ \ \ \ \ \ Design-Symmetry Kinematics and Curvature, PhD Thesis,

\ \ \ \ \ \ \ U.of California, Santa Barbara, 2016.

[150] A.Bobenko and E.Huhnen -Venedey, Geom.Dedicata 159 (2012) 207.

[151] M.Lavicka and J.Vrsek, \ J.for Geometry and Graphics 13 (2009)145.

[152] M.Arrayas, D. Bouwmeester and J. Trueba, Phys.Reports 667 (2017), 1.

[153] C.Hoyos, N.Sicar and J.Sonnenschein, J.Phys.A 48 (2015) 255204.

[154] H.Nastase, Introduction to the AdS-CFT Correspondence, Cambridge

\ \ \ \ \ \ \ \ U.Press, Cambridge, UK, 2015.

[155] A.Kholodenko, J.Geom.Phys.35 (2000),193.

[156] A.Kholodenko, J.Geom.Phys. 38 (2001) 81.

[157] A.Kholodenko, J.Geom.Phys.43 (2005) 45.

[158] C.Frances, Comm.Math.Helv. 80 (2005) 883.

[159] J.Danciger, Geometric Transitions: From Hyperbolic to AdS Geometry,

\ \ \ \ \ \ \ \ PhD Thesis, Department of Mathematics, Stanford University,
2011.

[160] S.Takeuchi, Japanese J.of Appl.Physics 53 (2014) 030101.

[161] T.Barbot, V.Charette, T.Drumm, W.Goldman and K.Melnick, in

\ \ \ \ \ \ \ \ Recent Developments in Pseudo-Riemannian Geometry,

\ \ \ \ \ \ \ \ D.Alekseevsky and H.Baum Editors, pp 179-229,

\ \ \ \ \ \ \ \ ESI Lect. Math. Phys., Eur. Math. Soc., Z\"{u}rich, 2008.

[162] \ V. Mazya and T.Shaposhnikova, Jaques Hadamard,

\ \ \ \ \ \ \ \ A Universal Mathematician,

\ \ \ \ \ \ \ \ AMS Publishers, \ Providence, RI, 1998.

[163] Y.Kravtsov, Geometric Optics in Engineering Physics,

\ \ \ \ \ \ \ \ Alpha Sci.International Ltd., Harrow, UK. 2005.

[164] A.Petrov, Einstein Spaces, Pergamon Press, New York, 1969.

[165] A.Petrov, New Methods in General Theory of Relativity,

\ \ \ \ \ \ \ Nauka, Moscow, 1966 (in Russian)

[166] N.Ibragimov, Transformation Groups Applied to Mathematical Physics,

\ \ \ \ \ \ \ D.Reidel Publ.Co., Boston, MA, 1985.

[167] R.Ward, Class.Quant.Gravity 4 (1987) 775.

[168] J.Griffits and J.Podolsky, Exact Space-Times in Einstein's

\ \ \ \ \ \ \ \ General Relativity, Cambidge U.Press, Cambridge UK, 2009.

[169] \ H.Stephani, D.Kramer,M.MacCallum, C.Hoenselaers and E.Herlt,

\ \ \ \ \ \ \ \ Exact Solutions of Einstein's Field Equations,

\ \ \ \ \ \ \ \ Cambridge U.Press, Cambridge,UK, 2003.

[170] \ R.Geroch, Ann.Phys. 36 (1966) 147.

[171] \ C.Torre, Class.Quantum Grav. 31 (2014) 045022.

[172] \ G.Rainich, AMS Transactions 27 (1925) 106.

[173] Ch.Misner and J.Wheeler, Ann.Phys. 2 (1957) 525.

[174] N.Kuiper, Ann.Math. 30 (1949) 916.

[175] W.K\"{u}hnel \ and H-B. Rademacher, J.Math. Pures Appl. 88 (2007) 251.

[176] A.Aminova, Russ.Math.Surveys 50 (1995) 69.

[177] A.Keane and B.Tupper, Class.Quantum Grav.21 (2004) 2037.

[178] W.K\"{u}hnel \ and H-B. Rademacher,Geom.Dedicata 109 (2004) 175.

[179] M.Cahen and M.Flato, Differential Geometry and Relativity,

\ \ \ \ \ \ \ \ D.Reidel Publ.Co., Boston, MA.1976.

[180] \ E.Ferapontov, Soviet Jorn.Math. 55 (1991) 1970.

[181] \ E.Ferapontov, Differential Geom. and Applications 5 (1995) 121.

[182] \ E.Madelung, Zeit.f.Phys.40 (1927) 322.

[183] \ W. K\"{u}hnel \ and H-B. Rademacher, Result.Math. 56 (2009) 421.

\bigskip

\bigskip

\bigskip

\bigskip

\bigskip

\bigskip

\bigskip

\bigskip

\bigskip

\bigskip

\textsl{\bigskip }

\bigskip

\bigskip

\bigskip

\bigskip

\bigskip

\bigskip

\bigskip

\bigskip

\bigskip

\bigskip

\bigskip

\bigskip

\bigskip

\bigskip

\bigskip

\bigskip

\bigskip

\bigskip

\bigskip

\bigskip

\bigskip

\bigskip

\bigskip

\bigskip

\bigskip

\bigskip

\bigskip

\bigskip

\bigskip

\bigskip

\bigskip

\bigskip

\bigskip

\bigskip

\bigskip

\bigskip

\bigskip

\bigskip

\bigskip

\bigskip

\bigskip

\bigskip

\bigskip

\bigskip

\bigskip

\bigskip

\bigskip

\bigskip

\bigskip

\end{document}